\documentclass[a4paper,fleqn]{article}

\usepackage[numbers]{natbib}
\usepackage[acronym]{glossaries}
\usepackage{tikz}
\usetikzlibrary{positioning}
\usepackage{color}
\usepackage{siunitx}
\usepackage{pifont}
\usepackage{amsmath}
\usepackage{hyperref}
\usepackage[margin=2.5cm]{geometry}
\usepackage{amssymb}
\usepackage{graphicx}
\usepackage{changepage}

\newcommand{\cmark}{\ding{51}}%

\definecolor{lightblue}{RGB}{199,221,236}
\definecolor{orange}{RGB}{255, 127, 14}
\definecolor{lightorange}{RGB}{255,223,195}
\definecolor{darkblue}{RGB}{31, 119, 180}
\definecolor{lightred}{RGB}{255, 220, 220}
\definecolor{darkred}{RGB}{150, 20, 20}
\definecolor{lightgreen}{RGB}{220, 250, 220}
\definecolor{darkgreen}{RGB}{20, 150, 20}
\definecolor{emeraldgreen}{RGB}{44,160,44}

\makeglossaries

\newacronym{mor}{MOR}{Model Order Reduction}
\newacronym{pod}{POD}{Proper Orthogonal Decomposition}
\newacronym{deim}{DEIM}{Discrete Empirical Interpolation Method}
\newacronym{eim}{EIM}{Empirical Interpolation Method}
\newacronym{fem}{FEM}{Finite Element Method}
\newacronym{neuralodes}{Neural ODEs}{Neural Ordinary Differential Equations}
\newacronym{neuralode}{Neural ODE}{Neural Ordinary Differential Equation}
\newacronym{rom}{ROM}{Reduced Order Model}
\newacronym{fom}{FOM}{Full Order Model}
\newacronym{svd}{SVD}{Singular Value Decomposition}
\newacronym{flops}{FLOPs}{Floating-Point Operations}
\newacronym{ffnn}{FFNN}{Feed-Forward Neural Network}
\newacronym{mse}{MSE}{Mean Squared Error}
\newacronym{dae}{DAE}{Differential Algebraic Equation}
\newacronym{efie}{EFIE}{Electric Field Integral Equation}
\newacronym{peec}{PEEC}{Partial Equivalent Element Circuit}
\newacronym{hts}{HTS}{High Temperature Superconductors}
\newacronym{lts}{LTS}{Low Temperature Superconductors}
\newacronym{dofs}{DOFs}{Degrees of Freedom}
\newacronym{corc}{CORC\textsuperscript{\textregistered}}{Conductor on Round Core}
\newacronym{iem}{IEM}{Integral Equation Method}
\newacronym{rebco}{REBCO}{Rare-Earth Barium Copper Oxide}
\newacronym{ems}{EM}{electromagnetics}
\newacronym{nns}{NNs}{Neural Networks}
\newacronym{ml}{ML}{Machine Learning}
\newacronym{pinns}{PINNs}{Physics-Informed Neural Networks}
\newacronym{opinf}{OpInf}{Operator Inference}
\newacronym{sindy}{SINDy}{Sparse Identification of Nonlinear Dynamics}

\usepackage{fancyhdr}
\usepackage{lipsum} 

\fancypagestyle{plain}{
  \fancyhf{}
}

\begin{document}
\let\WriteBookmarks\relax
\def\floatpagepagefraction{1}
\def\textpagefraction{.001}

\title{A Structured Neural ODE Approach for Real-Time Evaluation of AC Losses in 3D Superconducting Tapes}                      

\author{R. Basei \thanks{\texttt{riccardo.basei@phd.unipd.it} (Corresponding Author)} \thanks{Department of Industrial Engineering, University of Padua}
\and
F. Pase \thanks{Newtwen}
\and
F. Lucchini \footnotemark[2]
\and 
F. Toso  \footnotemark[3]
\and
R. Torchio  \footnotemark[2] \thanks{Department of Information Engineering, University of Padua}}

\date{}

\maketitle

\begin{abstract}
Efficient modeling of \gls{hts} is crucial for real-time quench monitoring; however, full-order electromagnetic simulations remain prohibitively costly due to the strong nonlinearities. Conventional projection-based reduced-order modeling pipelines for nonlinear problems, such as \gls{pod}–\gls{deim}, alleviate this cost but often require intrusive access to the \gls{fom} operators and a substantial number of interpolation points for hyperreduction.
This work investigates reduced-order strategies for \gls{iem} of \gls{hts} systems. We present the first application of \gls{pod}–\gls{deim} to \gls{iem}-based \gls{hts} models, and introduce a Structured \gls{neuralode} approach that learns nonlinear dynamics directly in the reduced space. The benchmark results show that \gls{neuralode} outperforms \gls{pod}–\gls{deim} both in efficiency and accuracy, highlighting its potential for real-time simulations of superconductors.
\end{abstract}

\begin{adjustwidth}{2.5em}{2.5em}
\end{adjustwidth}

\section{Introduction}

The second generation (2G) of \gls{rebco}-based \gls{hts} has emerged as a promising technology for a wide range of electrical engineering applications. These include power distribution networks, high-field magnets for experimental magnetic confinement fusion devices \cite{9756879}, and particle accelerators \cite{ROSSI20231354360}. \gls{hts} tapes can be configured into several cable architectures, such as the twisted stacked-tape conductor \cite{7393501}, the ROEBEL cable \cite{Goldacker_2014}, and the \gls{corc} cable \cite{van_der_Laan_2019}. Among these, the \gls{corc} design has attracted significant attention due to its advantageous mechanical strength and electromagnetic performance \cite{van_der_Laan_2019}. 

Although \gls{hts} offer the major advantage of operating at significantly higher temperatures than conventional \gls{lts}, monitoring the quench phenomenon remains a critical challenge. This difficulty arises from the relatively slow reduction of the critical current $I_c$ with increasing temperature \cite{9765384,instruments5030027}. The sudden transition from the superconducting state to the normal resistive state generates localized hot spots through the Joule effect, which, if not properly controlled, can lead to irreversible damage of the conductor. Reliable quench detection and mitigation are therefore essential for the safe and continuous operation of future fusion demonstrator power plants, such as the European DEMO \cite{MERRILL20152196}. 

Modeling quench phenomena requires solving multiphysics problems that couple \gls{ems}, thermal behavior, and fluid dynamics \cite{Zappatore_2024}. While lumped-element models have been proposed, capturing the underlying physics with sufficient accuracy typically demands finite element analysis. For superconducting structures, the \gls{ems} problem can be addressed using several numerical formulations, most notably the \gls{fem} with magnetic field ($H$) formulations~\cite{9097858}, or the $T$–$A$ approach \cite{Huber_2022}. An alternative is \glsfirst{iem}, which is traditionally used in open-domain problems such as antenna radiation, but has also proven effective in computing AC losses in superconductors~\cite{fast_J_phi,10897790}.

Due to the strong nonlinearities in the resistivity model, both \gls{fem} and \gls{iem} simulations require substantial computational resources to capture the transient, nonlinear dynamics of realistic 3D superconducting structures. This high computational cost renders them impractical for real-time monitoring. To address this limitation, several studies have explored the use of surrogate models to accelerate the \gls{ems} solution, particularly in the context of estimating AC loss \cite{11143214,10966202}. A first attempt to apply \gls{pod} to \gls{hts} systems within the \gls{iem} framework was reported in~\cite{10830004}. However, in this work, no hyperreduction strategy was adopted, as the authors argued that the sparsity of the operator $\mathbf{R}$ kept the assembly cost manageable. This is true for relatively small problems, but as the mesh dimension increases, the assembly quickly becomes considerably more demanding. In such cases, \gls{eim}~\cite{barrault_empirical_2004} and its discrete counterpart~\cite{chaturantabut_discrete_2009} can be applied to alleviate this cost, and these techniques have already been successfully used in electromagnetic problems \cite{GUO2025114352}. Nevertheless, \gls{deim} suffers from two critical drawbacks in the present context: the strong nonlinearity and spatial locality of superconducting material laws require a large number of interpolation points, which reduces efficiency, and its intrusive nature demands full access to the high-fidelity model assembly routines, which limits practical applicability. Furthermore, since the governing equations are highly nonlinear, the reduced-order model must still be solved with an implicit time-integration scheme, which entails a large number of Newton iterations and repeated Jacobian evaluations. As a result, even with \gls{deim}, the simulation cost remains significant.

These challenges have motivated a growing interest in \gls{ml}-based modeling strategies, which are typically non-intrusive and data-driven. Here, we provide only a brief overview of the main approaches, without aiming to be exhaustive; several recent reviews give a comprehensive account of this rapidly evolving area~\cite{JMLR:v24:21-1524, benner_survey_2015, brunton_data-driven_2021}.

Neural Operators such as DeepONet~\cite{lu_learning_2021}, Fourier Neural Operator~\cite{li2021fourier}, and graph neural operator variants~\cite{anandkumar2019neural} aim to learn mappings from function spaces, rather than between finite-dimensional vectors. This formulation enables them to generalize across families of input functions, achieving mesh-independent predictions and fast inference once trained. Their flexibility and non-intrusivity make them particularly attractive for parametric problems and large-scale simulations. However, they also present drawbacks: training requires substantial data, the networks can be very large, and their ability to handle complex geometries is limited. Graph-based neural operators alleviate this to some extent, but at the price of predicting the full-order field directly, which implies very large network outputs and nontrivial computational costs.

\gls{pinns}~\cite{RAISSI2019686} incorporate the residual of the physical equation into the training loss, enabling the solution of forward and inverse problems even when data are scarce. \gls{pinns} are especially powerful for inverse problems and for settings where measurements are sparse. However, they tend to struggle with complex geometries, and their formulation is not well-suited for input-driven or control problems: a PINN is typically trained for a fixed set of inputs or excitations, and adapting it to new ones generally requires retraining the network. Additionally, their implementation requires formulating the governing equations as differentiable residuals, which can be cumbersome when the full-order model is complex or not available in closed form.

\gls{opinf}~\cite{peherstorfer_data-driven_2016,BENNER2020113433} offers a physics-aware, non-intrusive alternative. Instead of learning black-box surrogates, \gls{opinf} identifies reduced operators directly from trajectory data projected onto a reduced basis. This approach has the advantage of being parsimonious, data-efficient, and interpretable, while remaining consistent with the structure of the \gls{rom}. Nevertheless, \gls{opinf} typically assumes that the reduced operators have a polynomial structure, which can be restrictive in cases where the dynamics involve more complex nonlinear interactions.

Latent Dynamics modeling frameworks~\cite{bonneville2024comprehensivereviewlatentspace,FARENGA2025107146} take yet another route. The central idea is that the solution trajectories of high-dimensional problems evolve on a low-dimensional manifold, which can be identified either with linear projections (e.g. \gls{pod}) or with nonlinear encoders (e.g., autoencoders~\cite{lusch_deep_2017,kramer_nonlinear_2019}). Once the manifold is fixed, the latent dynamics can be learned in different ways. \glspl{neuralode}~\cite{chen_neural_2019} model continuous-time dynamics by learning the time derivative as a neural network. \gls{sindy}~\cite{1517384113} also parametrizes the derivative in continuous time, but restricts it to a sparse linear combination of predefined candidate functions, leading to interpretable governing equations. More recent developments further combine physics-based structure and data-driven learning for parametric reduced-order modeling~\cite{10.1063/5.0308144}, and explore advanced nonlinear manifold-learning strategies such as probabilistic manifold decomposition~\cite{guo2025nonlinearmodelreductionprobabilistic}, as well as autoencoder-based \glspl{rom} enhanced through self-attention mechanisms~\cite{FuAnonlinear}.

In the \gls{hts} field as well, machine learning–based approaches have recently been applied to accelerate quench detection and AC loss estimation~\cite{10412643, 10896621}. However, their application to reduced-order modeling of nonlinear electromagnetic simulations remains largely unexplored.

In this work, we build on the latent dynamics paradigm and propose a novel neural architecture for nonlinear electromagnetic modeling that operates directly in the reduced-order space. The starting point is a discretized integral equation model of \gls{hts} conductors, which is compressed through \gls{pod}-based dimensionality reduction. Instead of relying on intrusive hyper-reduction techniques such as \gls{deim}, we design a Structured \gls{neuralode} that learns only the nonlinear component of the reduced dynamics, while the linear part is inherited directly from the full-order model. This idea of embedding problem structure into \glspl{neuralode} architectures has also been explored in recent works on Structured \glspl{neuralode}~\cite{LINOT2023111838, SERINO2025114090, loya2025structurepreservingneuralordinarydifferential}, albeit motivated primarily by stiffness and time-integration considerations. In this way, physical knowledge is embedded into the reduced-order evolution, ensuring parsimony, interpretability, and efficiency. The approach is validated on a \gls{hts} \gls{corc} cable case study and benchmarked against the standard \gls{pod}–\gls{deim} method, demonstrating the potential of physics-informed latent dynamics modeling for challenging electromagnetic simulation.
The results demonstrate improved accuracy and significantly lower computational cost, highlighting the potential of the proposed strategy for real-time, industrially relevant superconducting applications.

The main contributions of this work are twofold:
\begin{itemize}
    \item We demonstrate, for the first time, the application of the \gls{pod}-\gls{deim} methodology to \gls{iem}-based models of \gls{hts} systems;
    \item We introduce and validate a Structured \gls{neuralode} approach that directly learns the nonlinear dynamics in the reduced-order space, offering a less-intrusive alternative to traditional hyperreduction techniques.
\end{itemize}

The remainder of this paper is organized as follows. Section~\ref{sec:model} describes the physical system under study, including the mesh and material properties. Section~\ref{sec:formulation} introduces the \gls{iem} formulation. Section~\ref{sec:poddeim} discusses the application of the \gls{pod}-\gls{deim} method, while Section~\ref{sec:neuralode} presents the proposed \gls{neuralode} approach. Section~\ref{sec:results} reports the numerical results and Section~\ref{sec:discussion} discusses them. Finally, Section~\ref{sec:conclusion} summarizes the conclusions and outlines directions for future research.

\section{Model Description}\label{sec:model}
\begin{figure}[!t]
\centering

{\includegraphics[trim=0cm 0cm 0cm 0cm, clip, width=0.225\columnwidth]{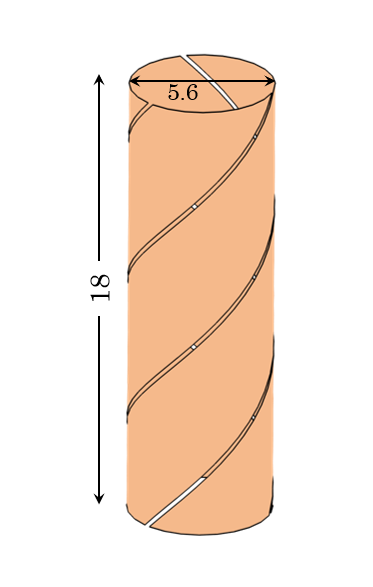}}
{\includegraphics[trim=4cm 0cm 3cm 0cm, clip, width=0.225\columnwidth]{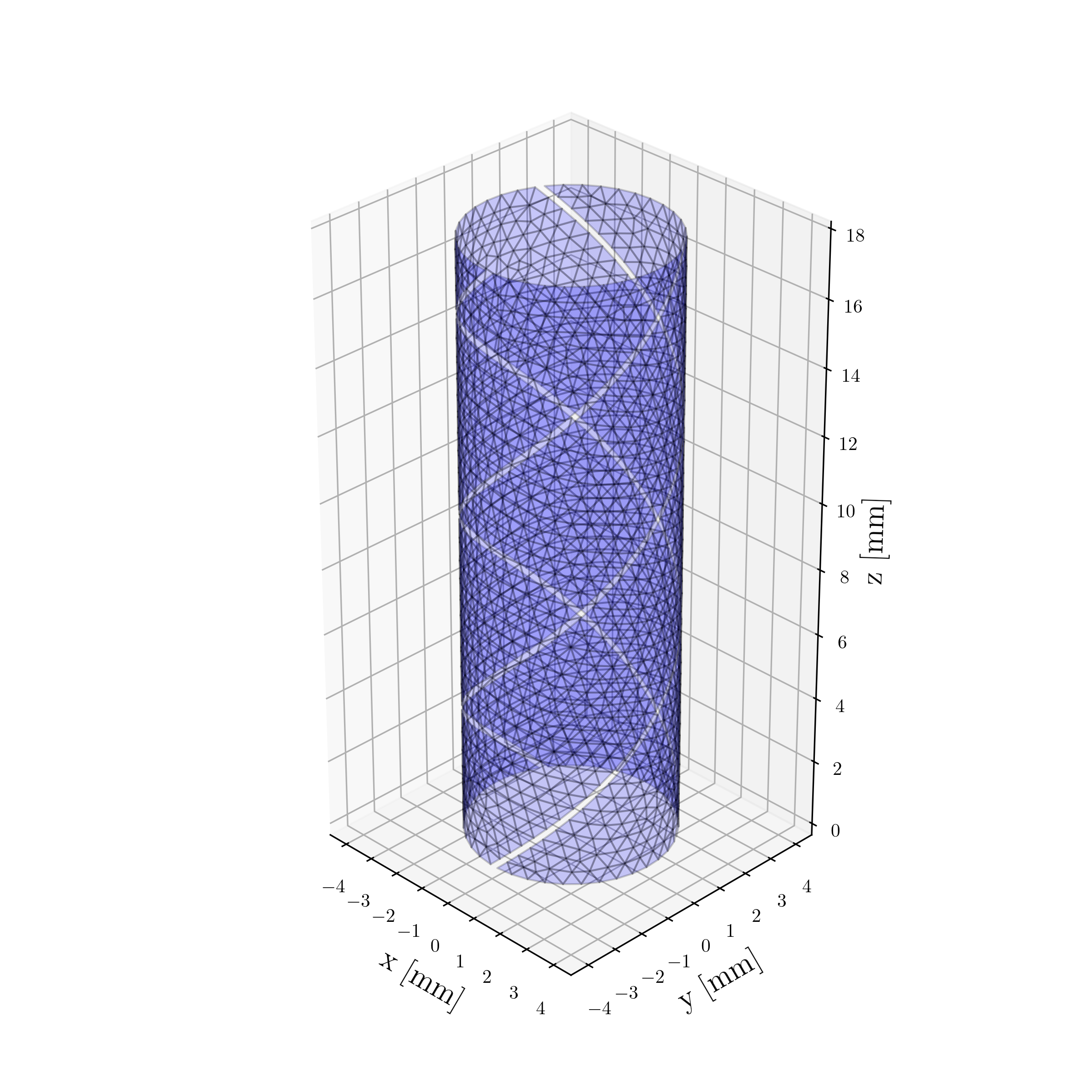}}

\caption{Geometrical model of \gls{hts} tapes under analysis (units in mm), and 2D surface mesh.}
\label{fig:mesh}
\end{figure}

The \gls{corc} cables are realized by winding \gls{hts} tapes made of \gls{rebco}, arranged in a helical geometry, around a thick core made of Copper \cite{9462395}. The cable is subjected to various AC loss mechanisms, including magnetization losses and coupling losses. From the EM modeling perspective, a common practice is to focus only on the \gls{hts} tapes and determine the AC losses due to the current loops therein \cite{Wang_2019,9693316,10591441,11220970}. Moreover, the \gls{rebco} tapes with a thickess of approximately \SI{1}{\micro\meter}, represent a small portion of the whole thickess of the \gls{hts} tapes ($\approx$ \SI{100}{\micro\meter}) \cite{Wang_2019}. Owing to the extremely high aspect ratio of the superconducting layer, the tapes are modeled as two-dimensional surfaces embedded in three-dimensional space, using surface formulations, e.g, exploiting the $T-A$ approach \cite{10897790}.
Following \cite{Wang_2019}, the \gls{hts} tapes analyzed in this work 
are extracted from a \gls{corc} conductor made of 3 tapes in a single layer, arranged in a helical geometry with a winding angle of 40°, representing an 18 mm-long segment of the cable. The diameter of the Copper former is 5.6 mm. The model representation is shown in Figure~\ref{fig:mesh}.

Within the superconducting domain $\Omega_{sc}\subseteq\Omega$, the resistivity depends on the local current density. In this work, we adopt a power-law constitutive relation~\cite{sirios_comparison_2019}:
\begin{equation}
    \mathbf{E}(\mathbf{J}) = \frac{E_c}{J_c}\bigg(\frac{\lVert\mathbf{J}\rVert}{J_c}\bigg)^{n-1}\mathbf{J},
\end{equation}
where the electric field is assumed to be parallel to the current density. The corresponding resistivity can thus be expressed as
\begin{equation}
    \rho_{sc}(\mathbf{J}) = \frac{E_c}{J_c}\bigg(\frac{\lVert\mathbf{J}\rVert}{J_c}\bigg)^{n-1}.
\end{equation}
Here $E_c=10^{-4}$~V/m is the critical electric field, $J_c=236$~MA/m$^{2}$ is the critical current density, and $n=25$ is the power-law exponent characterizing the superconducting material. In general, both $n$ and $J_c$ depend on temperature and magnetic field; here, they are assumed constant for simplicity. Being this constitutive equation highly nonlinear, the resulting problem is also highly nonlinear, which represents a significant challenge when solving it numerically. 

We investigate the current density distribution induced by an external time-varying magnetic field. Specifically, the magnetic field is assumed to be oriented in the $y$-direction.
This external field is modeled as:
\begin{equation}
    \mathbf{B}_\text{ext}(t)=B_y(t)\hat{\mathbf{y}}=B_0\sin(2\pi ft)\hat{\mathbf{y}},
\end{equation}
where $B_0$ is the magnetic field amplitude
, and $f=50$~Hz is the frequency. According to Faraday’s law of induction, this time-varying magnetic field induces an electric field:
\begin{equation}
    \nabla\times\mathbf{E}_\text{ext} = -\frac{\partial\mathbf{B}}{\partial t}.
\end{equation}
For the given magnetic field orientation, the induced electric field can be expressed as
\begin{equation}
    \mathbf{E}_\text{ext}(x,y,z)=\begin{bmatrix}-\frac{\partial B_y}{\partial t}z \\ 0 \\ \frac{\partial B_y}{\partial t}x \end{bmatrix}.
\end{equation}
This induced electric field acts as the external excitation driving the current distribution in the superconducting tapes.

\section{$J-\varphi_e$ Formulation}\label{sec:formulation} 

The $J-\varphi_e$ formulation~\cite{fast_J_phi} for solving electromagnetic problems is based on the \gls{efie}, which serves as its starting point. We consider an electromagnetic analysis over a bounded domain $\Omega\subset\mathbb{R}^3$, where the \gls{efie} is coupled with the current continuity equation:
\begin{equation}
\begin{aligned}
    \mathbf{E}(\mathbf{r}) &= -\frac{\partial\mathbf{A}(\mathbf{r})}{\partial t}-\nabla\varphi_e(\mathbf{r})+\mathbf{E}_\text{ext}(\mathbf{r}),\\\
    \nabla\cdot\mathbf{J}(\mathbf{r})&=0,
    \label{eq:efie}
    \end{aligned}
\end{equation}
where $\mathbf{E}(\mathbf{r})$ is the electric field, $\mathbf{A}(\mathbf{r})$ is the magnetic vector potential, $\varphi_e(\mathbf{r})$ is the scalar electric potential, $\mathbf{J}(\mathbf{r})$ is the current density, and $\mathbf{E}_\text{ext}(\mathbf{r})$ is the external applied electric field. The constitutive relation between electric field and current density is given by Ohm’s law:
\begin{equation}
    \mathbf{E}=\rho(\mathbf{J})\mathbf{J},
\end{equation}
with $\rho(\mathbf{J})$ denoting the electrical resistivity. The magnetic vector potential $\mathbf{A}$ can be expressed in terms of $\mathbf{J}$ via the integral relation:
\begin{equation}
    \mathbf{A}(\mathbf{r})=\mu_0\int_\Omega g(\mathbf{r},\mathbf{r}')\mathbf{J}(\mathbf{r}')d\Omega,
\end{equation}
where
\begin{equation}
    g(\mathbf{r},\mathbf{r}')=\frac{1}{4\pi\lVert\mathbf{r}-\mathbf{r}'\lVert}
\end{equation}
is the static Green's function.

To solve \eqref{eq:efie} numerically, we adopt the \gls{peec} methodology. The computational domain $\Omega$ is discretized into triangular elements. The current density and electric potential are expanded using vector basis functions $\mathbf{w}_h(\mathbf{r})$ (associated with edges), and piecewise constant basis functions $p_h(\mathbf{r})$ (associated with faces):
\begin{equation}
    \mathbf{J}(\mathbf{r})=\sum_{h=1}^{n_e}i_h\mathbf{w}_h(\mathbf{r}), \quad \varphi_e(\mathbf{r})=\sum_{h=1}^{n_f}\varphi_jp_h(\mathbf{r}),
\end{equation}
where $n_e$ is the number of edges and $n_f$ the number of faces. Applying Galerkin-testing to \eqref{eq:efie} yields a system of nonlinear \gls{dae}:
\begin{equation}
    \begin{bmatrix}
        \mathbf{R}(\mathbf{i}) & \mathbf{G}^T\\
        \mathbf{G} & \mathbf{0}
    \end{bmatrix}\begin{bmatrix}
        \mathbf{i} \\ \mathbf{\varphi}
    \end{bmatrix}+\begin{bmatrix}
        \mathbf{L} & \mathbf{0} \\
        \mathbf{0} & \mathbf{0}
    \end{bmatrix} \frac{d}{dt}\begin{bmatrix}
        \mathbf{i} \\ \mathbf{\varphi}
    \end{bmatrix}=\begin{bmatrix}
        \mathbf{e}_s(t) \\ \mathbf{0}
    \end{bmatrix}.
\end{equation}
Here $\mathbf{i}\in\mathbb{R}^{n_e}$ collects the unknown edge currents, and $\mathbf{\varphi}\in\mathbb{R}^{n_f}$ contains the face-associated scalar potentials. The matrix $\mathbf{L}\in\mathbb{R}^{n_e\times n_e}$ is the inductance matrix, $\mathbf{R}(\mathbf{i}):\mathbb{R}^{n_e}\rightarrow\mathbb{R}^{n_e\times n_e}$ is the current-dependent resistance operator, and $\mathbf{G}$ is the faces-to-edges incidence matrix, acting as discrete counterpart of the divergence operator. The source term $\mathbf{e}_s(t)$ arises from the discretization of the external electric field.

Although this formulation leads to non-sparse matrices, it is advantageous because only the regions carrying current--namely, the superconducting domains--need to be meshed. This avoids the need to mesh the surrounding air, unlike the $H$ and $H-\varphi_e$ formulations, which require solving for air elements as well. As a result, the overall model size is significantly reduced.

\section{\gls{pod}-\gls{deim}}\label{sec:poddeim}

\subsection{Structure-Preserving \gls{pod}}

The \gls{pod} method~\cite{Chatterjee_intro_00} is used to derive a reduced-order basis for both current and potential fields while preserving the structure of the underlying equations~\cite{10686274}. The procedure is as follows:
\begin{enumerate}
    \item Using \gls{fom} simulations, snapshots of the current and potential are collected:
    \begin{equation}
        \mathbf{I} = \begin{bmatrix} | & & |\\ \mathbf{i}^{(1)} & \dots & \mathbf{i}^{(n_s)} \\| & & |\end{bmatrix}\in\mathbb{R}^{n_e\times n_s}, \quad
    \mathbf{\Phi} = \begin{bmatrix}| & & |\\ \mathbf{\varphi}^{(1)} & \dots & \mathbf{\varphi}^{(n_s)}\\| & & | \end{bmatrix}\in\mathbb{R}^{n_f\times n_s},
    \end{equation}
    where $n_e$, $n_f$, and $n_s$ are the number of edges, faces, and snapshot instances, respectively. Each column of the snapshot matrices corresponds to the system state at a given time instant, and the variability across snapshots arises from both temporal evolution and changes in the simulation setup, such as different inputs or initial conditions, depending on the specific application.
    \item Apply \gls{svd} separately to the current and potential snapshot matrices:
    \begin{equation}
        \mathbf{V}_i\mathbf{\Sigma}_i\mathbf{U}_i^T =\mathbf{I}, \quad \mathbf{V}_\varphi\mathbf{\Sigma}_\varphi\mathbf{U}_\varphi^T =\mathbf{\Phi}.
    \end{equation}
    The columns of $\mathbf{V}_i$ and $\mathbf{V}_\varphi$ define orthonormal basis vectors for the current and potential, respectively;
    \item Retain $r_i$ and $r_\varphi$ dominant modes, forming the reduced basis
    \begin{equation}
        \mathbf{V}_{i,r}\in \mathbb{R}^{n_e\times r_i}, \quad \mathbf{V}_{\varphi,r} \in \mathbb{R}^{n_f\times r_\varphi}.
    \end{equation}
    The state variables are projected onto the reduced space as:
    \begin{equation}
    \mathbf{i}_r=\mathbf{V}_{i,r}^T\mathbf{i}, \quad \mathbf{\varphi}_r=\mathbf{V}_{\varphi,r}^T\mathbf{\varphi},
    \end{equation}
    and the system matrices are modified accordingly:
    \begin{equation}
        \mathbf{R}_r(\mathbf{i}) = \mathbf{V}_{i,r}^T\mathbf{R}(\mathbf{i})\mathbf{V}_{i,r},\quad \mathbf{L}_r=\mathbf{V}_{i,r}\mathbf{L}\mathbf{V}_{i,r},\quad \mathbf{G}_r = \mathbf{V}_{\varphi,r}\mathbf{G}\mathbf{V}_{i,r}, \quad \mathbf{e}_{s,r}(t)=\mathbf{V}_{i,r}^T\mathbf{e}_s(t).
    \end{equation}
\end{enumerate}
This structure-preserving \gls{pod} approach ensures that the \gls{rom} captures the dominant dynamics of both the current and potential while maintaining the original physical structure of the equations.

An a priori estimate of the error introduced by truncating the \gls{pod} modes is given by the retained energy ratio:
\begin{equation}
    \epsilon = \frac{\sum_{k=1}^r \sigma_k}{\sum_{k=1}^{n_s} \sigma_k},
\end{equation}
where $\sigma_k$ are the singular values of $\mathbf{I}$ or $\mathbf{\Phi}$, stored in descending order in the diagonal of $\mathbf{\Sigma}_i$ or $\mathbf{\Sigma}_\varphi$, respectively. A higher value of $\epsilon$ indicates a more accurate \gls{rom} relative to the full model.

\subsection{\gls{deim}}

The \gls{deim}~\cite{chaturantabut_discrete_2009} efficiently approximates nonlinear functions in \glspl{rom} by evaluating them at a small number of carefully selected spatial points. We define the nonlinear function:
\begin{equation}
    \mathbf{f}(\mathbf{i}) := \mathbf{R}(\mathbf{i})\mathbf{i}.
\label{eq:f_def}
\end{equation}
\gls{deim} proceeds by constructing a low-rank approximation of $\mathbf{f}$ through the following steps:
\begin{enumerate}
    \item \textit{snapshots} of $\mathbf{f}$ are generated under various input conditions by evaluating \eqref{eq:f_def}, using the \gls{fom} simulations performed for \gls{pod}. These snapshots are assembled into a matrix:
    \begin{equation}
    \mathbf{F}=
        \begin{bmatrix} | & & |\\ \mathbf{f}^{(1)} & \dots & \mathbf{f}^{(n_s)} \\| & & |\end{bmatrix}\in\mathbb{R}^{n_e\times n_s}.
    \end{equation}
    \item as in \gls{pod}, a \gls{svd} is applied to the snapshot matrix:
    \begin{equation}
       \mathbf{F}=\mathbf{V}_f\mathbf{\Sigma}_f\mathbf{U}_f^T,
    \end{equation}
    and the reduced basis $\mathbf{V}_{f,r} \in \mathbb{R}^{n_e \times r_{deim}}$ is formed from the first $r_{deim} \ll n$ columns of $\mathbf{V}_f$;
    \item a set of $n_p$ optimal interpolation points is determined using a greedy algorithm~\cite{barrault_empirical_2004}. These points are encoded in a selection matrix $\mathbf{P}\in\mathbb{R}^{n_e \times n_p}$.
\end{enumerate}
The number of modes retained, $r_{\text{DEIM}}$, can be chosen based on the retained energy criterion, similar to \gls{pod}.

Using \gls{deim}, the nonlinear function is approximated as:
\begin{equation}
    \mathbf{f}(\mathbf{i}) \approx \mathbf{V}_{f,r}(\mathbf{P}^T\mathbf{V}_{f,r})^{-1}\mathbf{P}^T\mathbf{f}(\mathbf{i}).
\end{equation}
This formulation allows for evaluating the nonlinear term $\mathbf{f}(\mathbf{i})$ only at the selected spatial points determined by $\mathbf{P}$, significantly reducing computational effort.

\subsection{\gls{pod}-\gls{deim} \gls{rom} Simulation}\label{subsec:rom_sim}

We have described how the system is assembled. To simulate its dynamics, we must discretize the system of ODEs in time and solve it step by step. We consider the time step interval $[t_n, t_{n+1}] \subset \mathbb{R}_{\geq 0}$ of length $\Delta t=t_{n+1}-t_n$. Given the initial values of the current and potential, $\mathbf{i}_r^{(n)} = \mathbf{i}_r(t_n)$ and $\mathbf{\varphi}_r^{(n)} = \mathbf{\varphi}_r(t_n)$, our goal is to compute the next-step values $\mathbf{i}_r^{(n+1)} = \mathbf{i}_r(t_{n+1})$ and $\mathbf{\varphi}_r^{(n+1)} = \mathbf{\varphi}_r(t_{n+1})$.\\
To this end, we approximate the time derivatives over $[t_n, t_{n+1}]$ using a first-order backward finite difference, namely
\begin{equation}
\frac{d\mathbf{i}}{dt}\approx\frac{\mathbf{i}_r^{(n+1)}-\mathbf{i}_r^{(n)}}{\Delta t},
\quad
\frac{d\mathbf{\varphi}}{dt}\approx\frac{\mathbf{\varphi}_r^{(n+1)}-\mathbf{\varphi}_r^{(n)}}{\Delta t}.
\end{equation}
Using this approximation, we discretize the system with a backward Euler scheme:
\begin{equation}
\begin{bmatrix}
\Delta t\mathbf{R}_{deim}\big(\mathbf{i}_r^{(n+1)}\big)+\mathbf{L}_r & \Delta t\mathbf{G}_r^T\\
\Delta t\mathbf{G}_r & \mathbf{0}
\end{bmatrix}
\begin{bmatrix}
\mathbf{i}_r^{(n+1)} \\ \mathbf{\varphi}_r^{(n+1)}
\end{bmatrix}
=
\begin{bmatrix}
\mathbf{L}_r \mathbf{0} \\
\mathbf{0} & \mathbf{0}
\end{bmatrix}
\begin{bmatrix}
\mathbf{i}_r^{(n)} \\ \mathbf{\varphi}_r^{(n)}
\end{bmatrix}
+
\begin{bmatrix}
\Delta t\mathbf{e}_{s,r}^{(n+1)} \\ \mathbf{0}
\end{bmatrix},
\end{equation}
with $\mathbf{R}_{deim}$ is assembled using only the rows selected by the \gls{deim} procedure.
This discretization leads to a nonlinear system, since $\mathbf{R}_{deim}$ depends nonlinearly on $\mathbf{i}_r^{(n+1)}$. Solving it requires a Newton solver, which in turn involves repeated Jacobian evaluations. This process is computationally expensive.
To reduce the cost, we can, in principle, introduce an approximation: we evaluate $\mathbf{R}_{deim}$ at the current from the previous time step $\mathbf{i}_r^{(n)}$, rather than at the unknown current $\mathbf{i}_r^{(n+1)}$:
\begin{equation}
\begin{bmatrix}
\Delta t\mathbf{R}_{deim}\big(\mathbf{i}_r^{(n)}\big)+\mathbf{L}_r & \Delta t\mathbf{G}_r^T\\
\Delta t\mathbf{G}_r & \mathbf{0}
\end{bmatrix}
\begin{bmatrix}
\mathbf{i}_r^{(n+1)} \\ \mathbf{\varphi}_r^{(n+1)}
\end{bmatrix}
=
\begin{bmatrix}
\mathbf{L}_r & \mathbf{0} \\
\mathbf{0} & \mathbf{0}
\end{bmatrix}
\begin{bmatrix}
\mathbf{i}_r^{(n)} \\ \mathbf{\varphi}_r^{(n)}
\end{bmatrix}
+
\begin{bmatrix}
\Delta t\mathbf{e}_{s,r}^{(n+1)} \\ \mathbf{0}
\end{bmatrix},
\end{equation}
where $\mathbf{e}_{s,r}^{(n+1)}=\mathbf{e}_{s,r}(t_{n+1})$, and $\mathbf{R}_{deim}$ again assembled from the \gls{deim}-selected rows.
This approximation linearizes the system in the unknowns, so advancing the state reduces to solving a linear system--no Newton iterations or Jacobian evaluations are required. However, care must be taken: if the problem is strongly nonlinear, this approximation may introduce large errors, since even a small error in the current can significantly affect $\mathbf{R}_{deim}$, and hence the computed next-state current. For the highly nonlinear problem addressed in this paper, this approach proves infeasible, as demonstrated in Section~\ref{subsec:ts}.

\section{Structured \gls{neuralode}}\label{sec:neuralode}

\subsection{Model Definition}

We aim at constructing a non-intrusive \gls{rom} that avoids hyperreduction of the nonlinear operator. 
Starting from the \gls{pod}-reduced formulation, the reduced dynamics can be written in the form
\begin{equation}
    \begin{aligned}
    \begin{bmatrix}
        \mathbf{R}_r(\mathbf{i_r}) & \mathbf{G}_r^T\\
        \mathbf{G}_r & \mathbf{0}
    \end{bmatrix}\begin{bmatrix}
        \mathbf{i}_r \\ \mathbf{\varphi}_r
    \end{bmatrix} +
    \begin{bmatrix}
        \mathbf{L}_r & \mathbf{0} \\
        \mathbf{0} & \mathbf{0}
    \end{bmatrix}&\frac{d}{dt}\begin{bmatrix}
        \mathbf{i}_r\\ \mathbf{\varphi}_r
    \end{bmatrix}=\begin{bmatrix}
        \mathbf{e}_{s,r}(t) \\ \mathbf{0}
    \end{bmatrix}.
    \end{aligned}
    \label{eq:reduced_equation}
\end{equation}
where the linear operators $\mathbf{L}_r$ and $\mathbf{G}_r$ are inherited from \gls{pod} projection of the \gls{fom}, while the only remaining nonlinear contribution is the reduced resistance operator $\mathbf{R}_r(\mathbf{i}_r)$:
\begin{equation}
\begin{aligned}
    \mathbf{R}_r:\mathbb{R}^{r_i}&\longrightarrow\mathbb{R}^{r_i\times r_i},\\
    \mathbf{i}_r&\longmapsto\mathbf{R}_r(\mathbf{i}_r).
\end{aligned}
\end{equation}
Instead of approximating $\mathbf{R}_r(\mathbf{i}_r)$ via \gls{pod}-\gls{deim}, we introduce a neural-network surrogate that operates entirely in the reduced current space. Specifically, we approximate the reduced resistance operator with a feed-forward neural network
\begin{equation}
\widehat{\mathbf{R}}_r(\mathbf{i}_r;\psi) = \mathrm{NN}_\psi(\mathbf{i}_r),
\qquad 
\mathrm{NN}_\psi:\mathbb{R}^{r_i}\rightarrow\mathbb{R}^{r_i\times r_i},
\label{eq:NN_Rr}
\end{equation}
where $\psi$ denotes the trainable parameters (weights and biases). For ease of notation, in the remainder of the manuscript, we denote this neural surrogate simply by $\mathbf{R}_\psi(\mathbf{i}_r)$. The network takes as input the reduced current degrees of freedom $\mathbf{i}_r$ and outputs the entries of the reduced resistance operator. Substituting $\mathbf{R}_\psi(\mathbf{i}_r)$ into \eqref{eq:reduced_equation} yields a reduced dynamical system parametrized by $\psi$; in this sense, the neural network defines a learnable vector field and the resulting model falls within the \gls{neuralode} paradigm.

\subsection{Training}

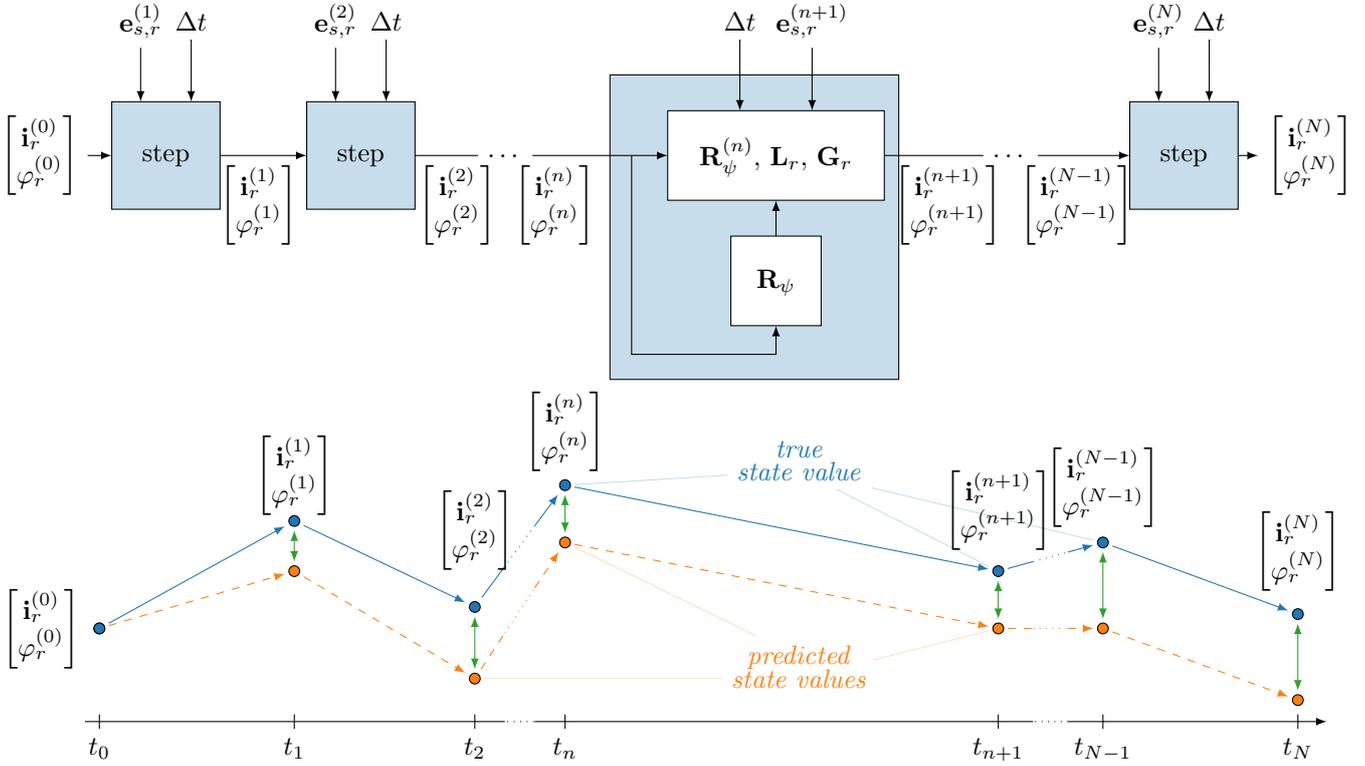
\begin{figure}[!htbp]
\centering

\begin{tikzpicture}[scale=0.85][font=\small]
    \node (x0) at (-1.2,2.875) {$\begin{bmatrix}\mathbf{i}_r^{(0)}\\\mathbf{\varphi}_r^{(0)} \end{bmatrix}$};
    \draw[-latex] (x0) -- (-0.2,2.875);
    \draw[fill=lightblue] (-0.2,2.125)--(-0.2,3.625)--(1.3,3.625)--(1.3,2.125)--cycle;
    \node at (0.55,2.875) {step};
    \draw[-latex] (1.3,2.875) -- (2.5,2.875) node[below left] {$\begin{bmatrix}\mathbf{i}_r^{(1)}\\ \mathbf{\varphi}_r^{(1)}\end{bmatrix}$};
    \node (u1) at (0.2,4.75) {$\mathbf{e}_{s,r}^{(1)}$};
    \node (dt1) at (0.9,4.75) {$\Delta t$};
    \draw[-latex] (u1)--(0.2,3.625);
    \draw[-latex] (dt1)--(0.9,3.625);

    \draw[fill=lightblue] (2.5,2.125)--(2.5,3.625)--(4,3.625)--(4,2.125)--cycle;
    \node at (3.25,2.875) {step};
    \node (u2) at (2.9,4.75) {$\mathbf{e}_{s,r}^{(2)}$};
    \node (dt2) at (3.6,4.75) {$\Delta t$};
    \draw[-latex] (u2)--(2.9,3.625);
    \draw[-latex] (dt2)--(3.6,3.625);

    \node (ppp) at (5.25,2.875) {$\dots$};
    \draw (4,2.875) -- (ppp) node[below left] {$\begin{bmatrix}\mathbf{i}_r^{(2)}\\ \mathbf{\varphi}_r^{(2)}\end{bmatrix}$};

    \draw[fill=lightblue] (6.7,-0.25) -- (10.7,-0.25) -- (10.7,4) -- (6.7,4) -- cycle;
    \draw[latex-]  (7.5,2.875) -- (ppp) node[below right] {$\begin{bmatrix}\mathbf{i}_r^{(n)}\\ \mathbf{\varphi}_r^{(n)}\end{bmatrix}$};
    \draw[fill=white] (8.375,0.5) -- (9.625,0.5) -- (9.625,1.75) -- (8.375,1.75) -- cycle;
    \node at (9,1.1) {$\mathbf{R}_\psi$};
    \draw[fill=white] (7.5,3.5) -- (10.5,3.5) -- (10.5,2.25) -- (7.5,2.25) -- cycle;
    \node at (9,2.85) {$\mathbf{R}_\psi^{(n)}$, $\mathbf{L}_r$, $\mathbf{G}_r$};
    \draw[-latex] (7,2.875) -- (7,0.1) -- (9,0.1) -- (9,0.5);
    \draw[-latex] (9,1.75) -- (9,2.25);
    \node (dt) at (8.5,4.75) {$\Delta t$};
    \node (u) at (9.5,4.75) {$\mathbf{e}_{s,r}^{(n+1)}$};
    \draw[-latex] (dt) -- (8.5,3.5);
    \draw[-latex] (u) -- (9.5,3.5);

    \node (ppp1) at (12.25,2.875) {$\dots$};
    \node (pppp1) at (12.3,2.875) {};
    \draw (10.5,2.875) -- (ppp1);
    \draw[latex-]  (13.9,2.875) -- (ppp1);
    \node at (13.1,2.18) {$\begin{bmatrix}\mathbf{i}_r^{(N-1)}\\ \mathbf{\varphi}_r^{(N-1)}\end{bmatrix}$};
    \node at (11.45,2.18) {$\begin{bmatrix}\mathbf{i}_r^{(n+1)}\\ \mathbf{\varphi}_r^{(n+1)}\end{bmatrix}$};

    \draw[fill=lightblue] (13.9,2.125)--(13.9,3.625)--(15.4,3.625)--(15.4,2.125)--cycle;
    \node at (14.65,2.875) {step};
    \node (uT) at (14.3,4.75) {$\mathbf{e}_{s,r}^{(N)}$};
    \node (dtT) at (15,4.75) {$\Delta t$};
    \draw[-latex] (uT)--(14.3,3.625);
    \draw[-latex] (dtT)--(15,3.625);

    \node (tdt) at (16.4,2.875) {$\begin{bmatrix}\mathbf{i}_r^{(N)}\\ \mathbf{\varphi}_r^{(N)}\end{bmatrix}$};
    \draw[-latex] (15.4,2.875) -- (tdt);
    
    \end{tikzpicture}
    \begin{tikzpicture}[scale=0.85][font=\small]
            \node (x0) at (-0.2,0) {};
            \node (x1) at (2.5,0.8) {};
            \node (x2) at (5,-0.7) {};
            \node (xn) at (6.25,1.2) {};
            \node (xn1) at (12.25,0) {};
            \node (xN1) at (13.7,0) {};
            \node (xN) at (16.4,-1) {};
            
            \draw[fill=orange] (x0) circle (0.075cm);
            \draw[fill=orange] (x1) circle (0.075cm);
            \draw[fill=orange] (x2) circle (0.075cm);
            \draw[fill=orange] (xn) circle (0.075cm);
            \draw[fill=orange] (xn1) circle (0.075cm);
            \draw[fill=orange] (xN1) circle (0.075cm);
            \draw[fill=orange] (xN) circle (0.075cm);

            \draw[-latex,dashed,orange] (x0)--(x1);
            \draw[-latex,dashed,orange] (x1)--(x2);
            \draw[dashed,orange] (x2)--(5.416,-0.066);
            \draw[dotted,orange] (5.416,-0.066)--(5.832,0.566);
            \draw[-latex,dashed,orange] (5.83,0.566)--(xn);
            \draw[-latex,dashed,orange] (xn)--(xn1);
            \draw[dashed,orange] (xn1)--(12.733,0);
            \draw[dotted,orange] (12.733,0)--(13.217,0);
            \draw[-latex,dashed,orange] (13.217,0)--(xN1);
            \draw[-latex,dashed,orange] (xN1)--(xN);

            \node (x1t) at (2.5,1.5) {};
            \node (x2t) at (5,0.3) {};
            \node (xnt) at (6.25,2) {};
            \node (xn1t) at (12.25,0.8) {};
            \node (xN1t) at (13.7,1.2) {};
            \node (xNt) at (16.4,0.2) {};

            \draw[fill=darkblue] (x0) circle (0.075cm);
            \draw[fill=darkblue] (x1t) circle (0.075cm);
            \draw[fill=darkblue] (x2t) circle (0.075cm);
            \draw[fill=darkblue] (xnt) circle (0.075cm);
            \draw[fill=darkblue] (xn1t) circle (0.075cm);
            \draw[fill=darkblue] (xN1t) circle (0.075cm);
            \draw[fill=darkblue] (xNt) circle (0.075cm);

            \draw[-latex,darkblue] (x0)--(x1t);
            \draw[-latex,darkblue] (x1t)--(x2t);
            \draw[darkblue] (x2t)--(5.416,0.867);
            \draw[-latex,darkblue] (5.832,1.434)--(xnt);
            \draw[dotted,darkblue] (5.832,1.434)--(5.416,0.867);
            \draw[-latex,darkblue] (xnt)--(xn1t);
            \draw[darkblue] (xn1t)--(12.733,0.933);
            \draw[darkblue,dotted] (12.733,0.933)--(13.217,1.066);
            \draw[-latex,darkblue] (13.217,1.066)--(xN1t);
            \draw[-latex,darkblue] (xN1t)--(xNt);

            \draw[latex-latex,emeraldgreen] (x1)--(x1t);
            \draw[latex-latex,emeraldgreen] (x2)--(x2t);
            \draw[latex-latex,emeraldgreen] (xn)--(xnt);
            \draw[latex-latex,emeraldgreen] (xn1)--(xn1t);
            \draw[latex-latex,emeraldgreen] (xN1)--(xN1t);
            \draw[latex-latex,emeraldgreen] (xN)--(xNt);

            \node at (-0.8,-2) {};

            \draw(-0.4,-1.3)--(5.416,-1.3);
            \draw[dotted](5.45,-1.3)--(5.82,-1.3);
            \draw(5.83,-1.3)--(12.733,-1.3);
            \draw[dotted] (12.76,-1.3)--(13.217,-1.3);
            \draw[-latex] (13.217,-1.3)--(16.8,-1.3);

            \draw (-0.2,-1.2)--(-0.2,-1.4) node[below] {$t_0$};
            \draw (2.5,-1.2)--(2.5,-1.4) node[below] {$t_1$};
            \draw (5,-1.2)--(5,-1.4) node[below] {$t_2$};
            \draw (6.25,-1.2)--(6.25,-1.4) node[below] {$t_n$};
            \draw (12.25,-1.2)--(12.25,-1.4) node[below] {$t_{n+1}$};
            \draw (13.7,-1.2)--(13.7,-1.4) node[below] {$t_{N-1}$};
            \draw (16.4,-1.2)--(16.4,-1.4) node[below] {$t_N$};

            \node at (9.5,2.5) {\textcolor{darkblue}{\textit{true}}};
            \node (true) at (9.5,2.2) {\textcolor{darkblue}{\textit{state value}}};
            \draw[lightblue,very thin] (true)--(xnt);
            \draw[lightblue,very thin] (true)--(xn1t);
            \draw[lightblue,very thin] (true)--(xN1t);

            \node (pred) at (9.5,-0.4) {\textcolor{orange}{\textit{predicted}}};
            \node (pred2) at (9.5,-0.7) {\textcolor{orange}{\textit{state values}}};
            \draw[lightorange,very thin] (pred)--(xn);
            \draw[lightorange,very thin] (pred2)--(xn1);
            \draw[lightorange,very thin] (pred2)--(x2);

            \node [left=0.01 of x0]{\textcolor{black}{$\begin{bmatrix}\mathbf{i}_r^{(0)}\\\mathbf{\varphi}_r^{(0)}\end{bmatrix}$}};
            \node [above=0.5 of x1]{\textcolor{black}{$\begin{bmatrix}\mathbf{i}_r^{(1)}\\\mathbf{\varphi}_r^{(1)}\end{bmatrix}$}};
            \node [above=1.2 of x2]{\textcolor{black}{$\begin{bmatrix}\mathbf{i}_r^{(2)}\\\mathbf{\varphi}_r^{(2)}\end{bmatrix}$}};
            \node [above=0.7 of xn]{\textcolor{black}{$\begin{bmatrix}\mathbf{i}_r^{(n)}\\\mathbf{\varphi}_r^{(n)}\end{bmatrix}$}};
            \node [above=0.8 of xn1]{\textcolor{black}{$\begin{bmatrix}\mathbf{i}_r^{(n+1)}\\\mathbf{\varphi}_r^{(n+1)}\end{bmatrix}$}};
            \node [above=1.1 of xN1]{\textcolor{black}{$\begin{bmatrix}\mathbf{i}_r^{(N-1)}\\\mathbf{\varphi}_r^{(N-1)}\end{bmatrix}$}};
            \node [above=1.2 of xN]{\textcolor{black}{$\begin{bmatrix}\mathbf{i}_r^{(N)}\\\mathbf{\varphi}_r^{(N)}\end{bmatrix}$}};
            
    \end{tikzpicture}
    \caption{Graphical representation of the training process. Green arrows illustrate the discrepancy between the true state values and the predicted values, which the optimization procedure seeks to minimize.}
    \label{fig:training}
\end{figure}

Once the approximated operator $\mathbf{R}_\psi$ has been defined, the next step is to train it. A naive approach would be to treat the problem as a standard regression task: one would simulate the \gls{fom}, project the resulting current snapshots onto the \gls{pod} basis to obtain reduced states $\big\{\mathbf{i}_r^{(n)}\big\}$, evaluate (and, if needed, project) the corresponding nonlinear operators to form the reduced resistance matrices $\big\{\mathbf{R}_r(\mathbf{i}_r^{(n)})\big\}$, and then train the network by pointwise regression of the mapping $\mathbf{i}_r \mapsto \mathbf{R}_r(\mathbf{i}_r)$. However, this strategy is (i) intrusive, since it requires access to snapshots of the nonlinearity, which is often infeasible when dealing with commercial solvers, and (ii) less effective than training on entire trajectories. In fact, as demonstrated in \cite{MELCHERS202394}, trajectory-based training is superior to local derivative fitting in terms of robustness and generalization capabilities.

For this reason, we adopt a \textit{discretize-then-optimize} approach, where the time discretization is embedded into the learning process, and the parameters are updated by backpropagation through the solver. The procedure to advance the reduced system from $t_n$ to $t_{n+1}$ is as follows:
\begin{enumerate}
    \item \textit{Construction of the approximated operator:} evaluate the reduced-order operator
    \begin{equation}
        \mathbf{R}_\psi^{(n)}=\mathbf{R}_\psi\big(\mathbf{i}_r^{(n)}\big),
    \end{equation} using the reduced current at the previous time step.
    \item \textit{Time discretization}:
    apply the Backward Euler scheme to the dynamical system in reduced coordinates.
    \item \textit{Solution of the linear system}: solve the resulting linear system:
    \begin{equation}
    \begin{bmatrix}
        \Delta t\mathbf{R}_\psi^{(n)}+\mathbf{L}_r & \Delta t\mathbf{G}_r^T\\
        \mathbf{G}_r & \mathbf{0}
    \end{bmatrix}\begin{bmatrix}
        \mathbf{i}_r^{(n+1)} \\ \mathbf{\varphi}_r^{(n+1)}
    \end{bmatrix} =
    \begin{bmatrix}
        \mathbf{L}_r & \mathbf{0} \\
        \mathbf{0} & \mathbf{0}
    \end{bmatrix}\begin{bmatrix}
        \mathbf{i}_r^{(n)} \\ \mathbf{\varphi}_r^{(n)}
    \end{bmatrix}+\begin{bmatrix}
        \mathbf{e}_{s,r}^{(n+1)} \\ \mathbf{0}
    \end{bmatrix}.
    \label{eq:node_advance}
\end{equation}
Since $\mathbf{R}_\psi$ is evaluated at the previous time step, the resulting system is linear in $\mathbf{i}_r^{(n+1)}$ and can be solved without Newton iterations. In this case, considering the previous time current during training allows the network to implicitly account for the resulting lag. Therefore, unlike \gls{deim}, this approximation error is explicitly embedded into the training process: the network learns to compensate for it, thereby reducing the mismatch and improving accuracy.
\end{enumerate}
Repeating this procedure for all time steps $n=0,\dots,N-1$, with $N\Delta t=T$, produced the time evolution of the reduced system. The predicted trajectories $\{\mathbf{i}_r^{(n)}\}^{N-1}_{n=0}$, $\{\mathbf{\varphi}_r^{(n)}\}^{N-1}_{n=0}$ are then compared with the reference solutions obtained from the \gls{fom}, and the parameters $\psi$ are optimized by minimizing the deviation between the reduced and full-order trajectories over the entire time horizon.

This training procedure follows the \gls{neuralode}~\cite{chen_neural_2019} paradigm, where the dynamics are unrolled across the temporal domain and the parameters are optimized end-to-end. The loss is defined as the \gls{mse} by the predicted states and the true states obtained by simulating the \gls{fom}. Figure~\ref{fig:training} provides a graphical interpretation of this training process.\\
Importantly, inference must employ the same integration scheme used in training, since the learned operator is biased toward that discretization.

\section{Results}\label{sec:results}

This section presents the numerical results obtained for the \gls{hts} \gls{corc}. We evaluate the performance of both the classical \gls{pod}-\gls{deim} approach and the proposed Structured \gls{neuralode} model in terms of accuracy and computational efficiency.

\subsection{Data Generation}\label{sec:data}

The resulting model features $n_e=4518$ edges and $n_f=3140$ faces, yielding a total of 7658 \gls{dofs}. To generate the dataset used for \gls{pod}-\gls{deim} reduction and for training the Structured \gls{neuralode}, five transient simulations were performed. Each transient corresponds to a sinusoidal excitation at 50~Hz with different amplitudes. This setup reflects the physical scenario under consideration, where the external magnetic field induced by surrounding cables is sinusoidal at 50~Hz, and its amplitude varies with both current magnitude and cable distance. To evaluate the generalization capability of the proposed approach, additional transient simulations were conducted with varying amplitudes--both within and outside the training distribution--as well as with different excitation frequencies. The corresponding excitation parameters defining the training and testing datasets are reported in Table~\ref{tab:dataset}.

\begin{table*}[!htbp]
\small
\centering
\renewcommand{\arraystretch}{1.05}
\begin{tabular}{c c c c c c}
\hline
\textbf{Freq.} & \textbf{Ampl.} & \textbf{Training} & \textbf{Within-the-distr. val.} & \textbf{Outside-the-distr. val.} & \textbf{Different freq. val.}\\
\hline
50 Hz & 13 mT & \cmark &   &   &  \\
50 Hz & 15 mT & \cmark &   &   &  \\
50 Hz & 18 mT & \cmark &   &   &  \\
50 Hz & 22 mT & \cmark &   &   &  \\
50 Hz & 24 mT & \cmark &   &   &  \\
\hline
50 Hz & 20 mT &   & \cmark &   &  \\
\hline
50 Hz & 10 mT &   &   & \cmark &  \\
50 Hz & 30 mT &   &   & \cmark &  \\
\hline
40 Hz & 20 mT &   &   &   & \cmark\\
60 Hz & 20 mT &   &   &   & \cmark\\
100 Hz & 20 mT &   &   &   & \cmark\\
\hline
\end{tabular}
\renewcommand{\arraystretch}{1}
\caption{Overview of the excitation parameters adopted for dataset generation.}
\label{tab:dataset}
\end{table*}

Due to the pronounced nonlinearities of the model, time integration was carried out using a fully implicit Backward Euler scheme, with a Newton–Raphson solver applied at each time step. The simulations span a time window from 0 to 35~ms, discretized into 560 steps of duration 0.0625~ms. These runs are computationally expensive: the system is both high-dimensional and nonlinear, requiring many Newton iterations, each involving the solution of a large, dense linear system. All computations were performed on a 13th Gen Intel\textsuperscript{\textregistered} Core\textsuperscript{\texttrademark} i7-13700 CPU @2.1~GHz, with 24 cores, equipped with 64~GB of RAM. On this hardware, a single transient required roughly three hours to complete, resulting in about 15 hours for the training set and an additional 18 hours for validation. These times should be understood as indicative rather than absolute, as they depend on the hardware characteristics and implementation details. Importantly, this computational effort belongs to the offline stage of the methodology and must be carried out only once. In practice, the relevant cost is that of repeatedly simulating the \gls{rom}, which is several orders of magnitude lower and therefore well suited to design and control tasks.

\subsection{\gls{pod}-\gls{deim} Practical Implementation}

\begin{figure}[!htbp]
\centering

{\includegraphics[trim=0cm 0cm 1cm 0cm, clip, width=0.32\textwidth]{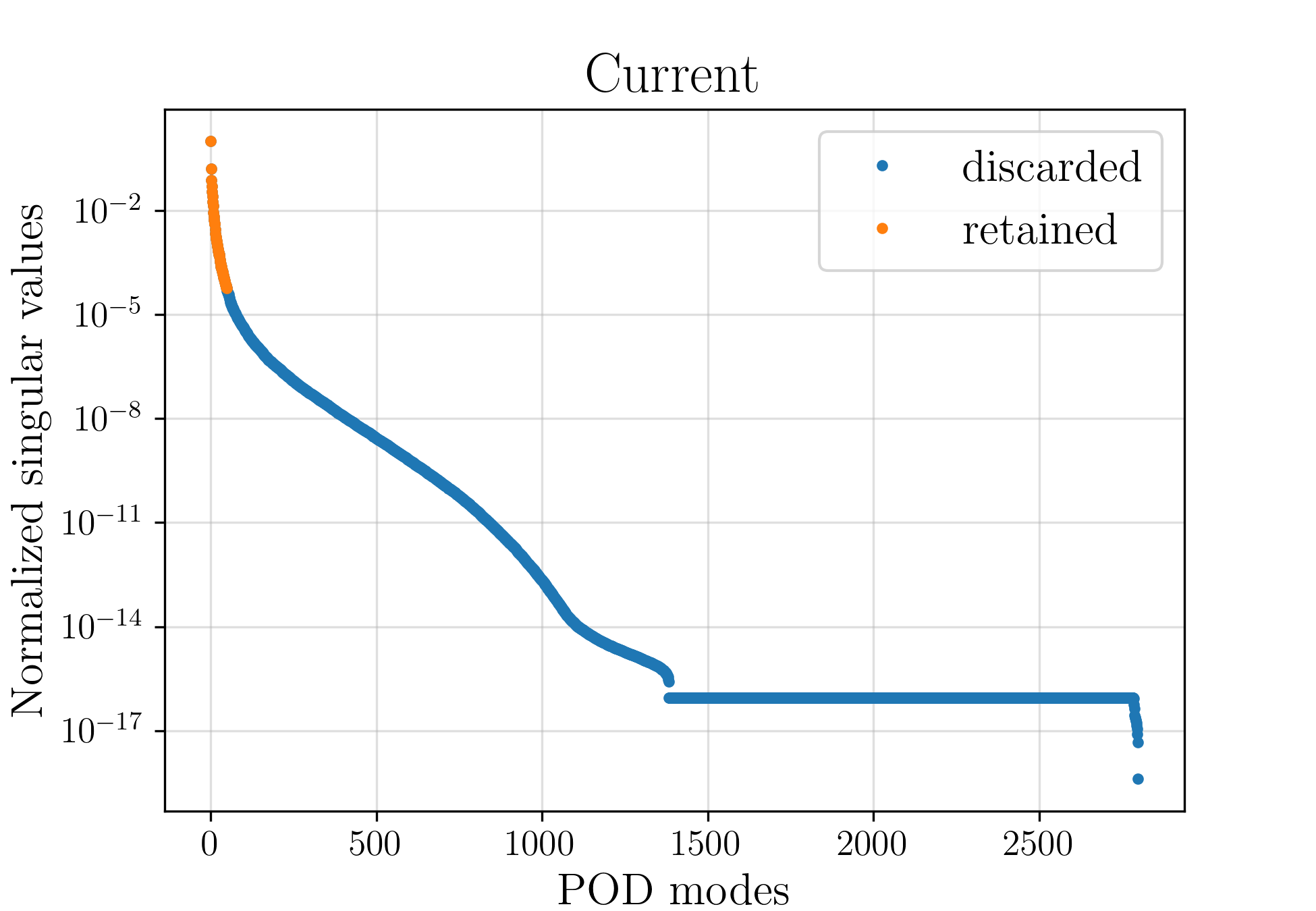}}
{\includegraphics[trim=0cm 0cm 1cm 0cm, clip, width=0.32\textwidth]{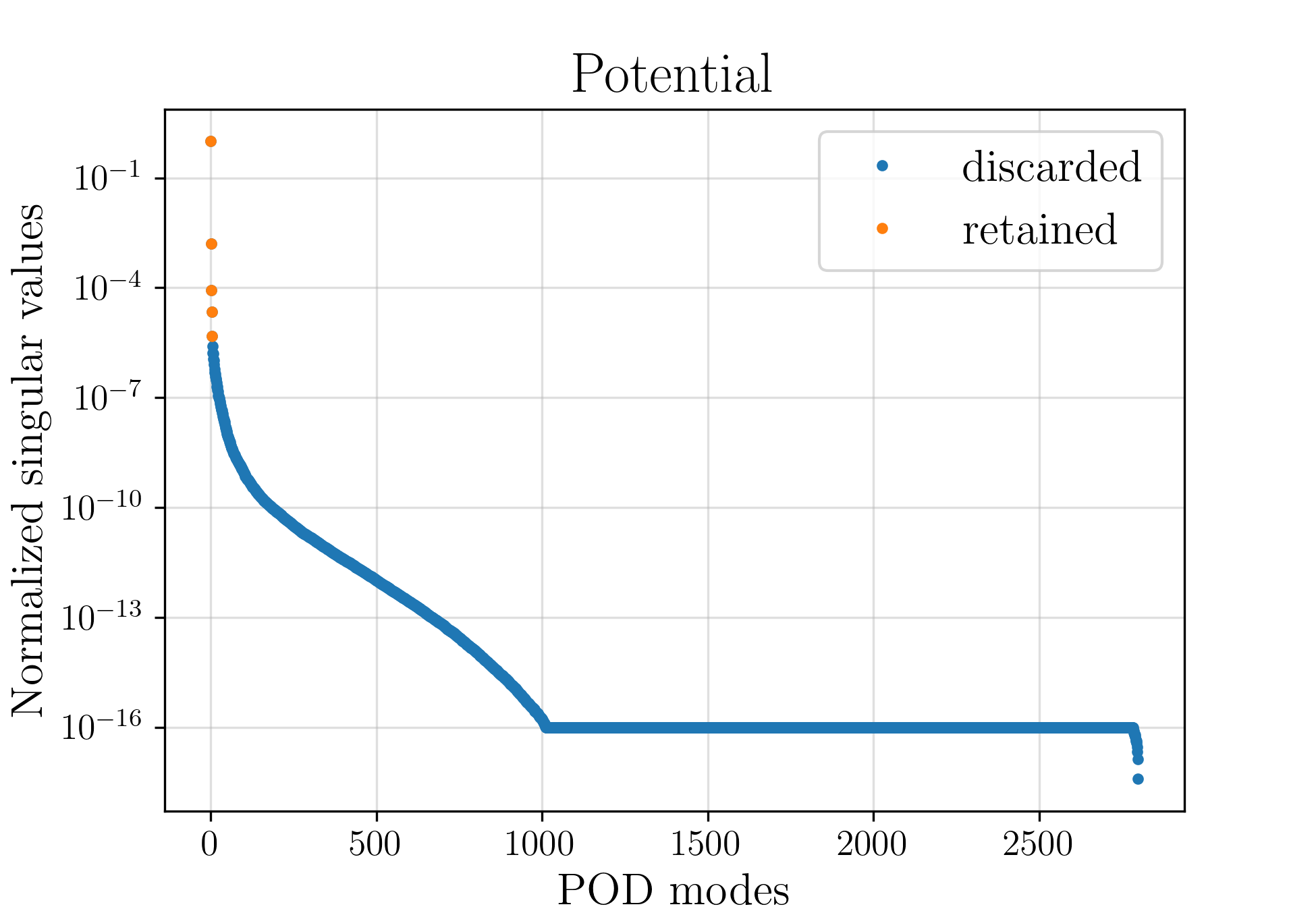}}
{\includegraphics[trim=0cm 0cm 1cm 0cm, clip, width=0.32\textwidth]{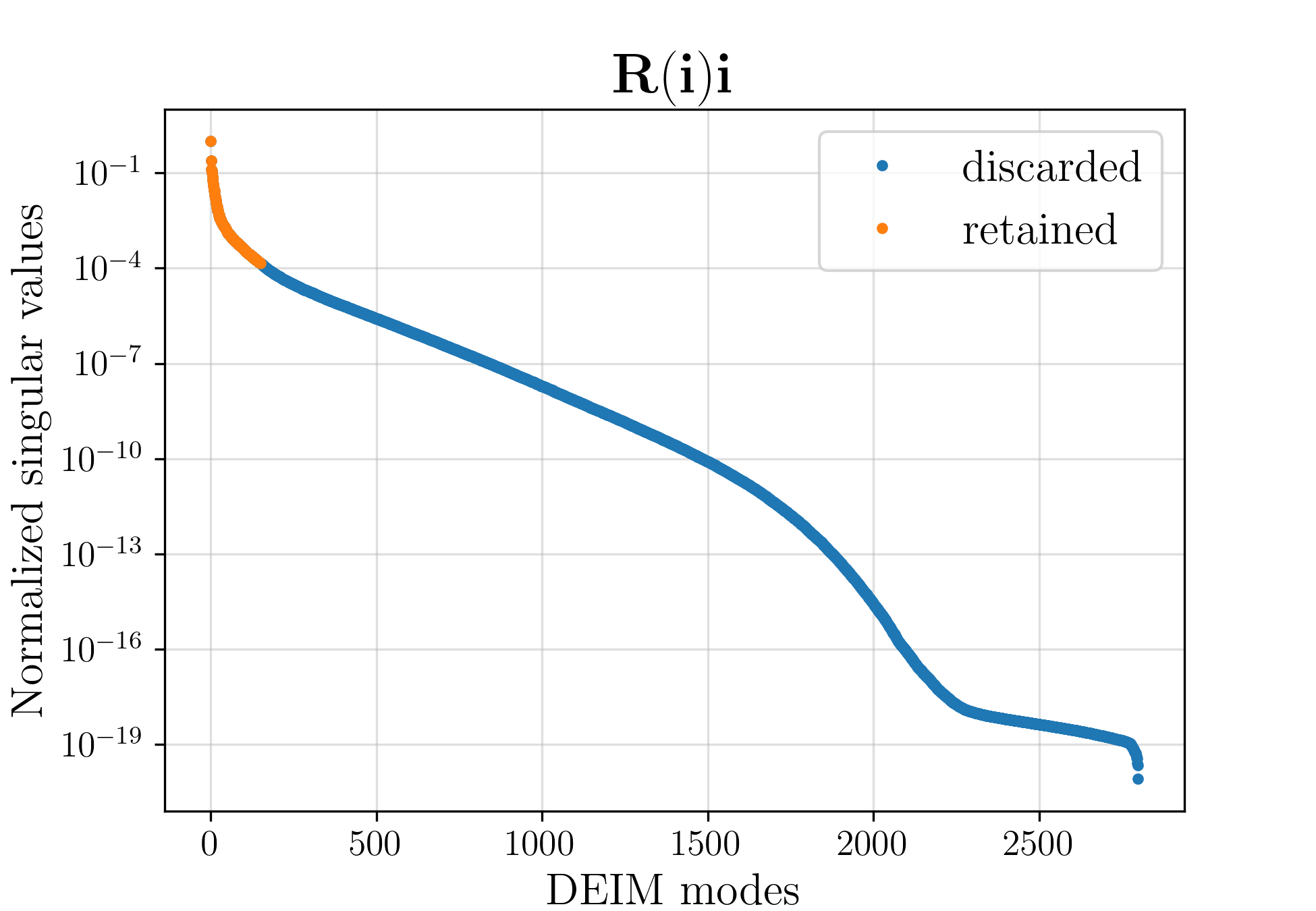}}

\caption{Singular value decay of the current, potential, and nonlinear term $\mathbf{R}(\mathbf{i})\mathbf{i}$ snapshot matrices. Retained modes are shown in orange, discarded modes in blue.}
\label{fig:poddeim_sv_decay}
\end{figure}

In the implementation of the proposed reduction framework, the training snapshots were used for constructing the snapshot matrices. This procedure yielded a total of $n_s=2800$ snapshots for the current, potential, and nonlinear term. For the current field, the dominant modes were selected based on the energy retained, resulting in $r_i=48$ retained modes. Also for the potential field, the energy retention criterion was applied, leading to the selection of the first five modes, i.e., $r_\varphi=5$. The singular values decay for the two variables are shown in Figure~\ref{fig:poddeim_sv_decay}.

For the \gls{deim} approximation of the nonlinear term, the number of interpolation points $n_p$ was chosen to be equal to the number of retained modes. In this case, 150 modes were retained. The relatively large number of \gls{deim} modes is motivated by the strong nonlinearities present in the system: accurately capturing localized variations in the current density requires a correspondingly dense set of interpolation points. The singular values decay for the nonlinear term, with the details of retained modes, are shown in Figure~\ref{fig:poddeim_sv_decay}.

\subsection{Structured \gls{neuralode} Practical Implementation}

The neural network $\mathbf{R}_\psi$ is implemented as a fully connected feed-forward architecture with $L_h=4$ hidden layers, each consisting of 140 neurons. As a preprocessing step, the absolute value of the reduced current $|\mathbf{i}_r|$ is fed to the network, reflecting the fact that resistivity--and therefore the $\mathbf{R}_\psi$--depends on the magnitude of the current rather than its sign. All the layers employ \textit{selu}~\cite{klambauer_self-normalizing_2017} activation functions.
The training dataset considers the very same transients used for constructing the \gls{pod}-\gls{deim} snapshot matrix. This dataset is split into fixed-length sequences that are organized into mini-batches, and the model is trained using the Adam optimizer~\cite{kingma_adam_2017}.

\subsection{Accuracy}

The accuracy of each reduced-order model was assessed by comparing the predicted current density with the reference solution obtained from the \gls{fom}. Specifically, the instantaneous absolute error was computed as the absolute difference between the \gls{rom}-predicted current and the corresponding \gls{fom} value. In the time-dependent plots, the mean, 95th-percentile, and maximum errors were evaluated across all current states at each time instant, thereby illustrating the temporal evolution of the error. Conversely, in the summary tables, the reported mean, 95th-percentile, and maximum values were computed over all current states and all time instants of the transient simulations. Moreover, the last row of the accuracy tables reports the relative discrepancy between the predicted and reference current trajectories, measured through the Frobenius norm:
\begin{equation}
\text{err}_{\lVert\cdot\rVert_F} = \frac{\lVert \mathbf{I}-\mathbf{I}_{\text{true}}\rVert_F}{\lVert \mathbf{I}_{\text{true}}\rVert_F},
\end{equation}
where $\mathbf{I}$ denotes the matrix collecting the full-order current states at all time instants reconstructed from the \gls{rom} solution (i.e., obtained by projecting the reduced currents back to the full space), and $\mathbf{I}_{true}$ is defined analogously using the \gls{fom} solution. This quantity provides a global measure of relative error over the entire transient. We note that a pointwise relative error is not well defined in this context, since the current may vanish at certain spatial locations and time instants.

\subsubsection{\textit{Within-the-distribution} Validation}

\begin{figure}[!htbp]
\centering

{\includegraphics[trim=1cm 0cm 2.5cm 0cm, clip, width=0.49\textwidth]{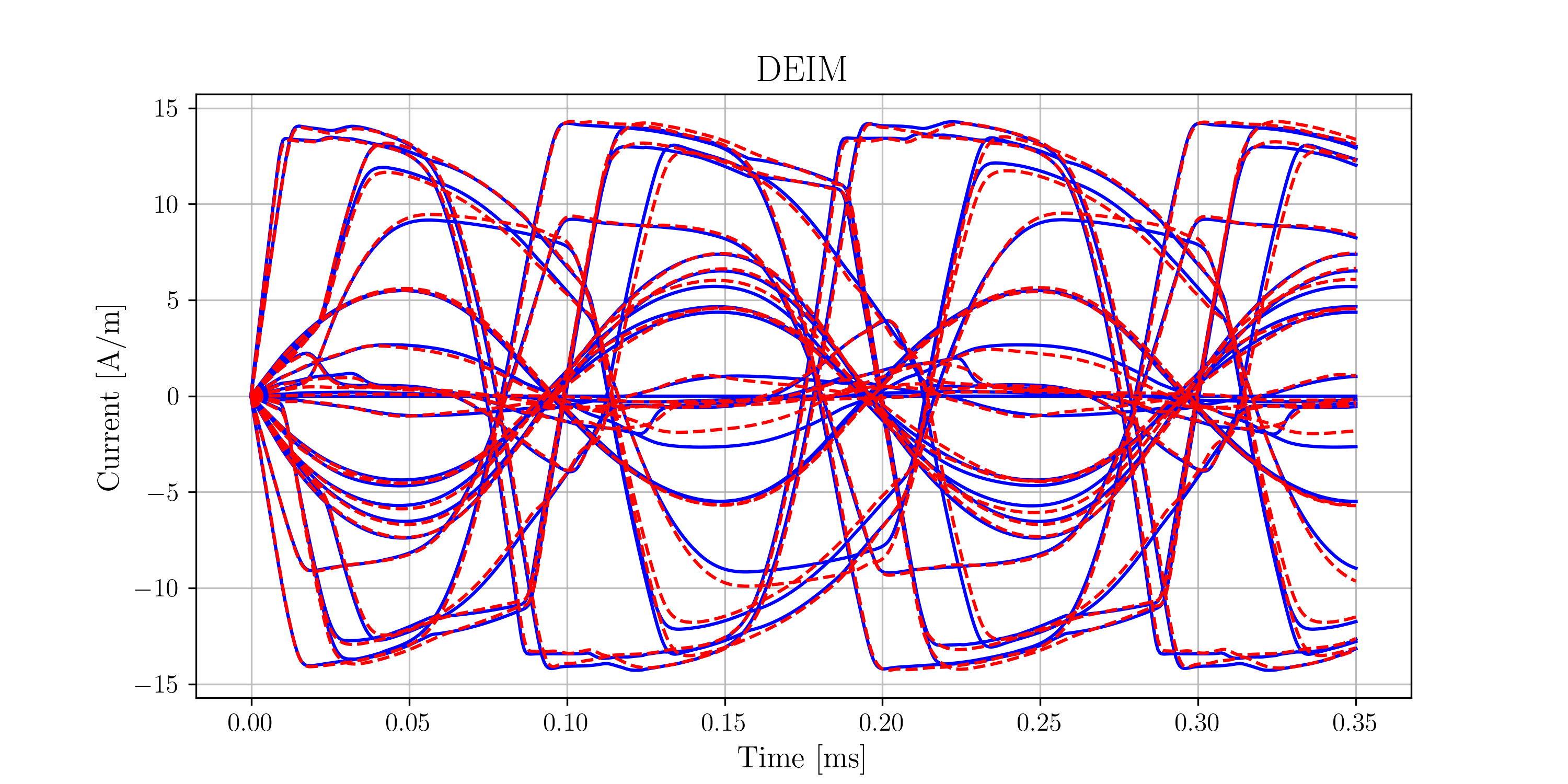}}
{\includegraphics[trim=1cm 0cm 2.5cm 0cm, clip, width=0.49\textwidth]{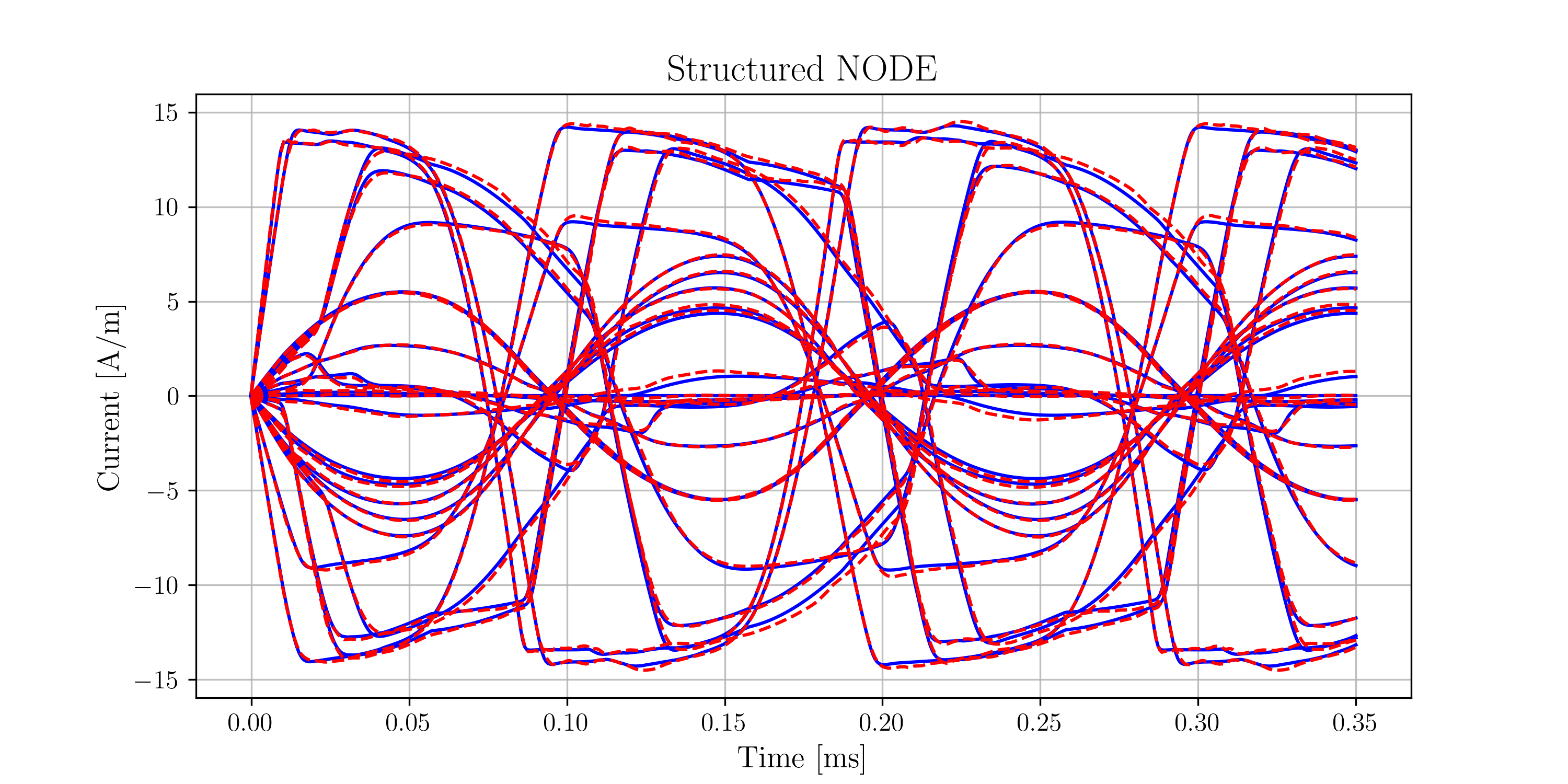}}

{\includegraphics[trim=1cm 0cm 2.5cm 0cm, clip, width=0.49\textwidth]{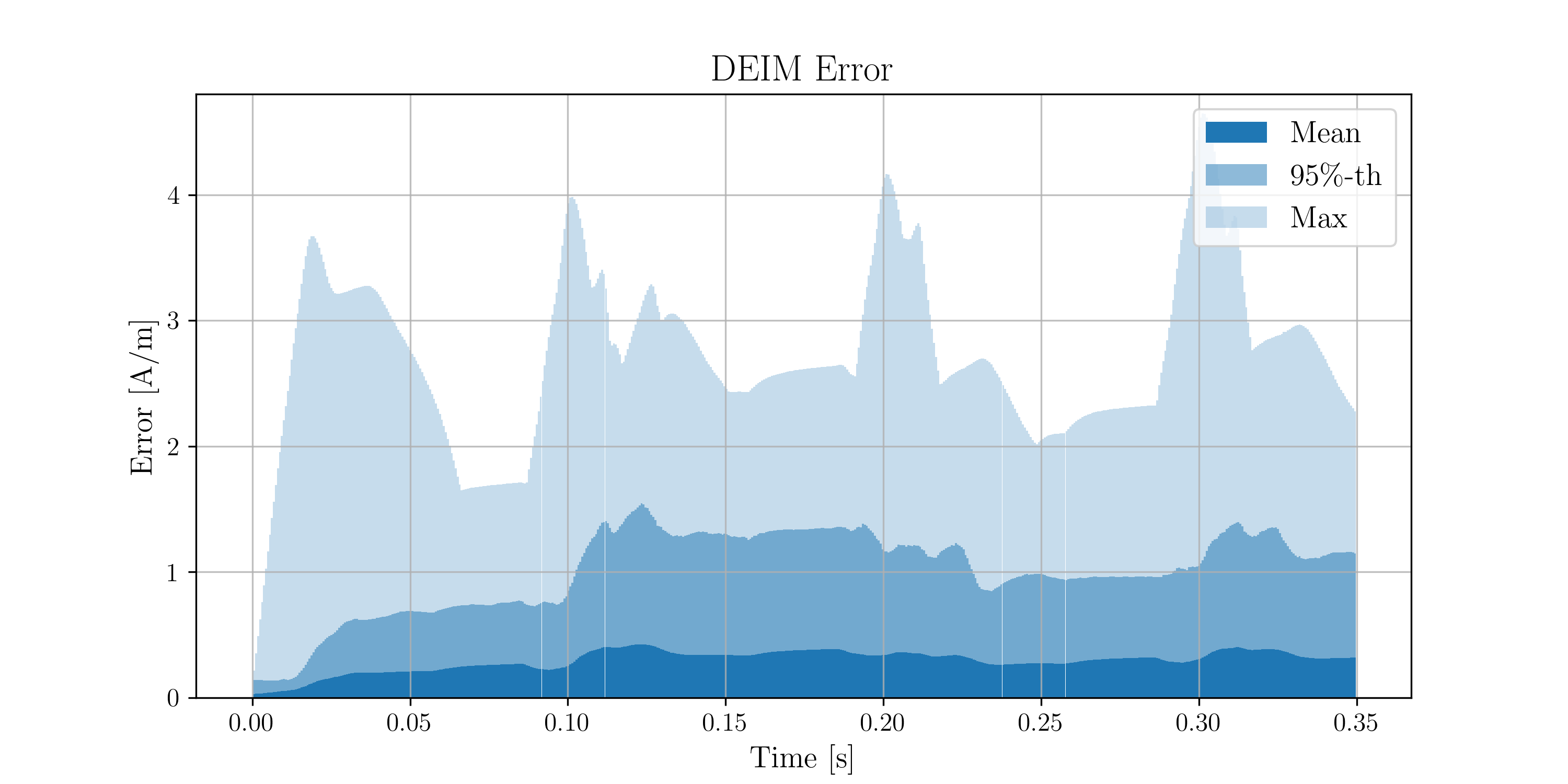}}
{\includegraphics[trim=1cm 0cm 2.5cm 0cm, clip, width=0.49\textwidth]{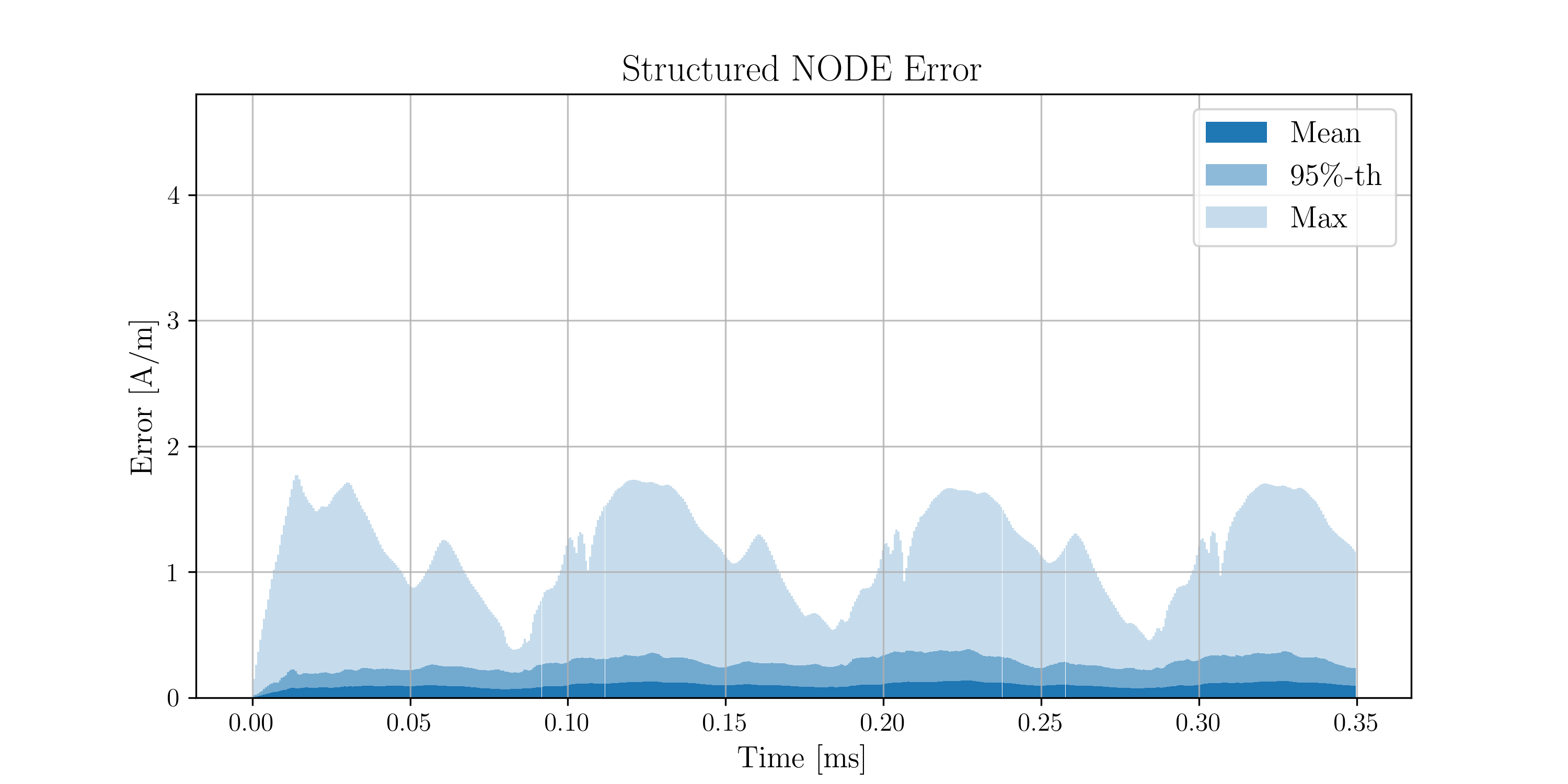}}

\caption{Top: Evolution of selected current states in the  \textit{within-the-distribution} validation transient for the \gls{rom} simulations (dashed red) compared to the \gls{fom} (solid blue). Bottom: Corresponding mean, 95th-percentile, and maximum errors between the \gls{rom}s and the \gls{fom}.}
\label{fig:current_snap}
\end{figure}

\begin{figure}[!htbp]
\centering

{\includegraphics[trim=4cm 0cm 3cm 0cm, clip, width=0.194\textwidth]{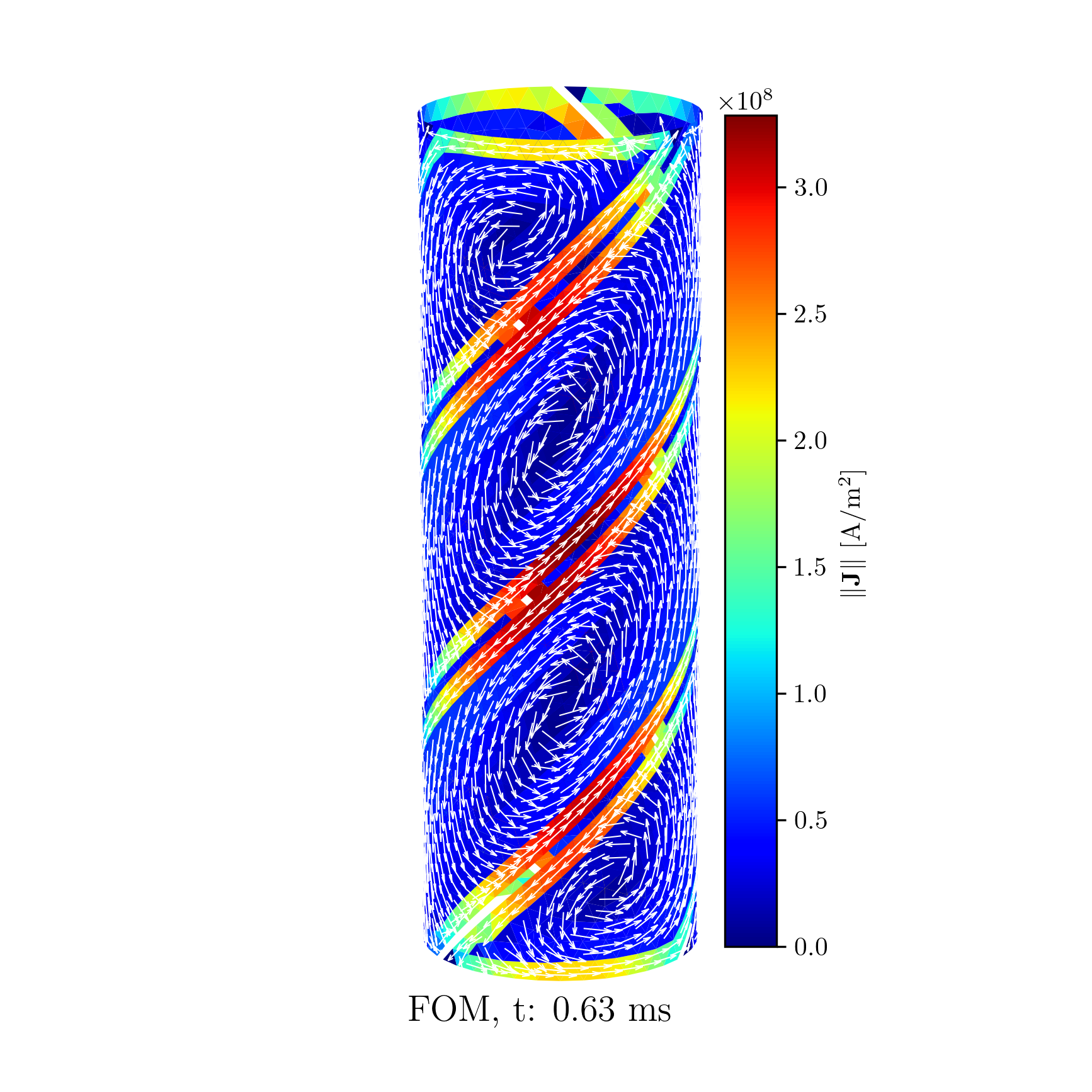}}
{\includegraphics[trim=4cm 0cm 3cm 0cm, clip, width=0.194\textwidth]{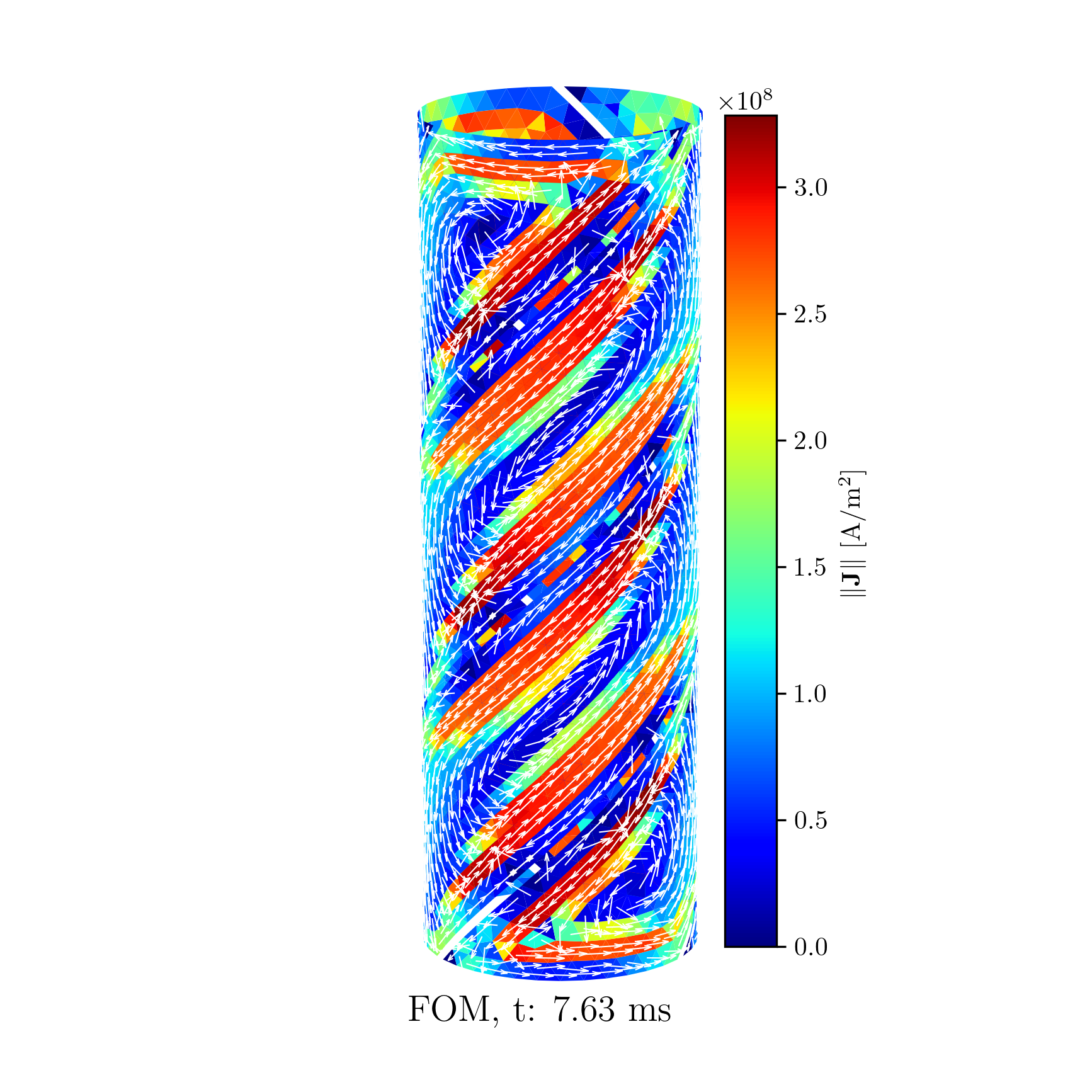}}
{\includegraphics[trim=4cm 0cm 3cm 0cm, clip, width=0.194\textwidth]{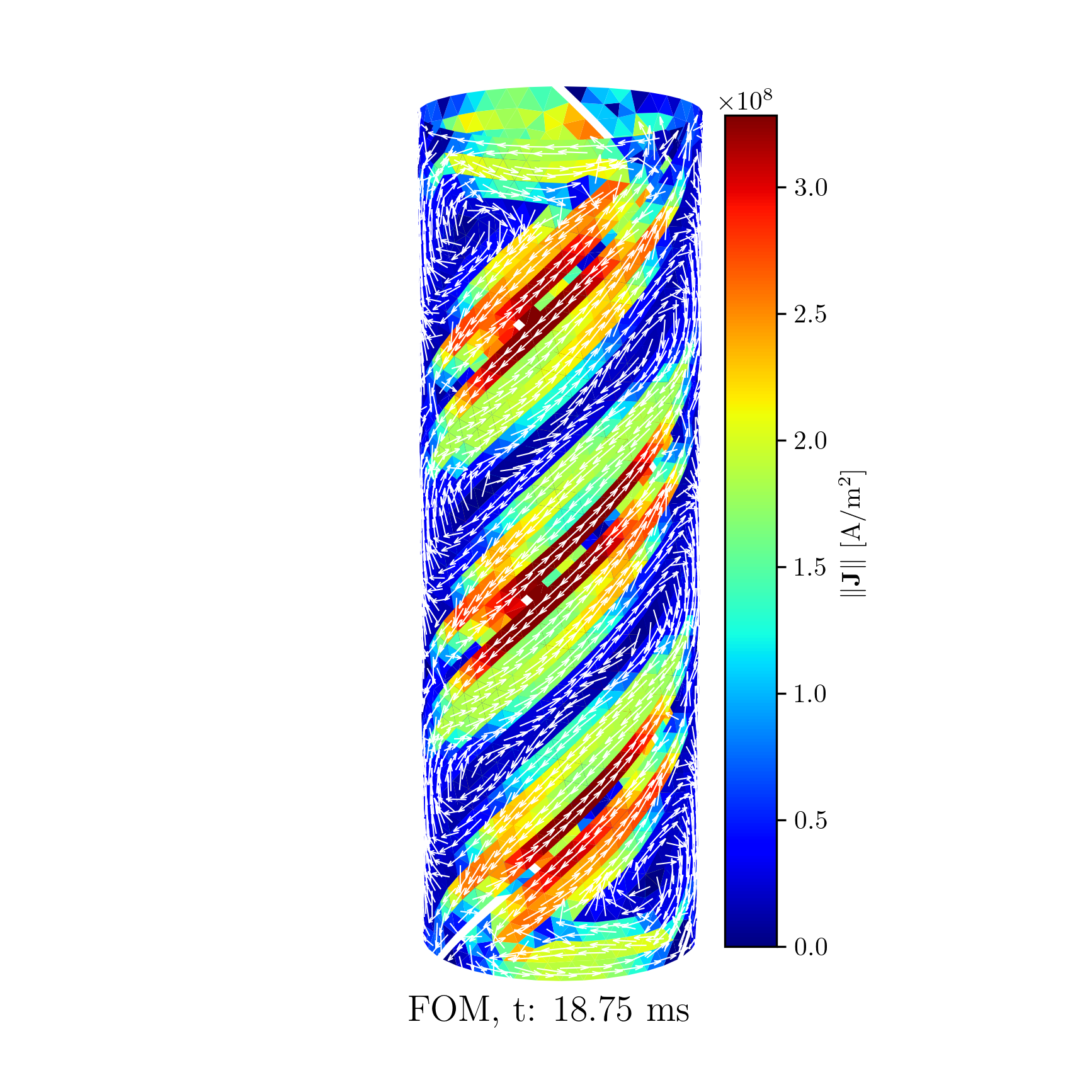}}
{\includegraphics[trim=4cm 0cm 3cm 0cm, clip, width=0.194\textwidth]{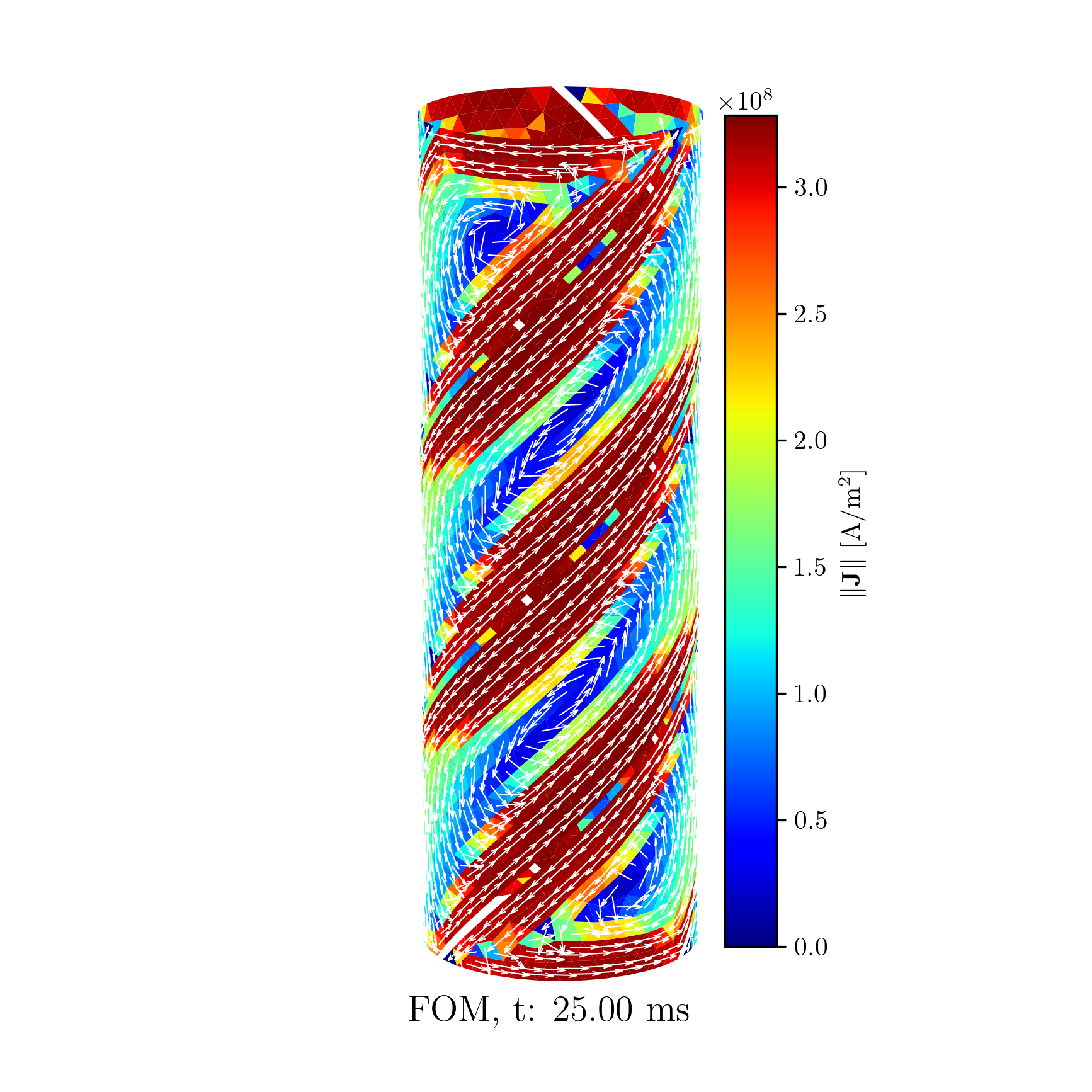}}
{\includegraphics[trim=4cm 0cm 3cm 0cm, clip, width=0.194\textwidth]{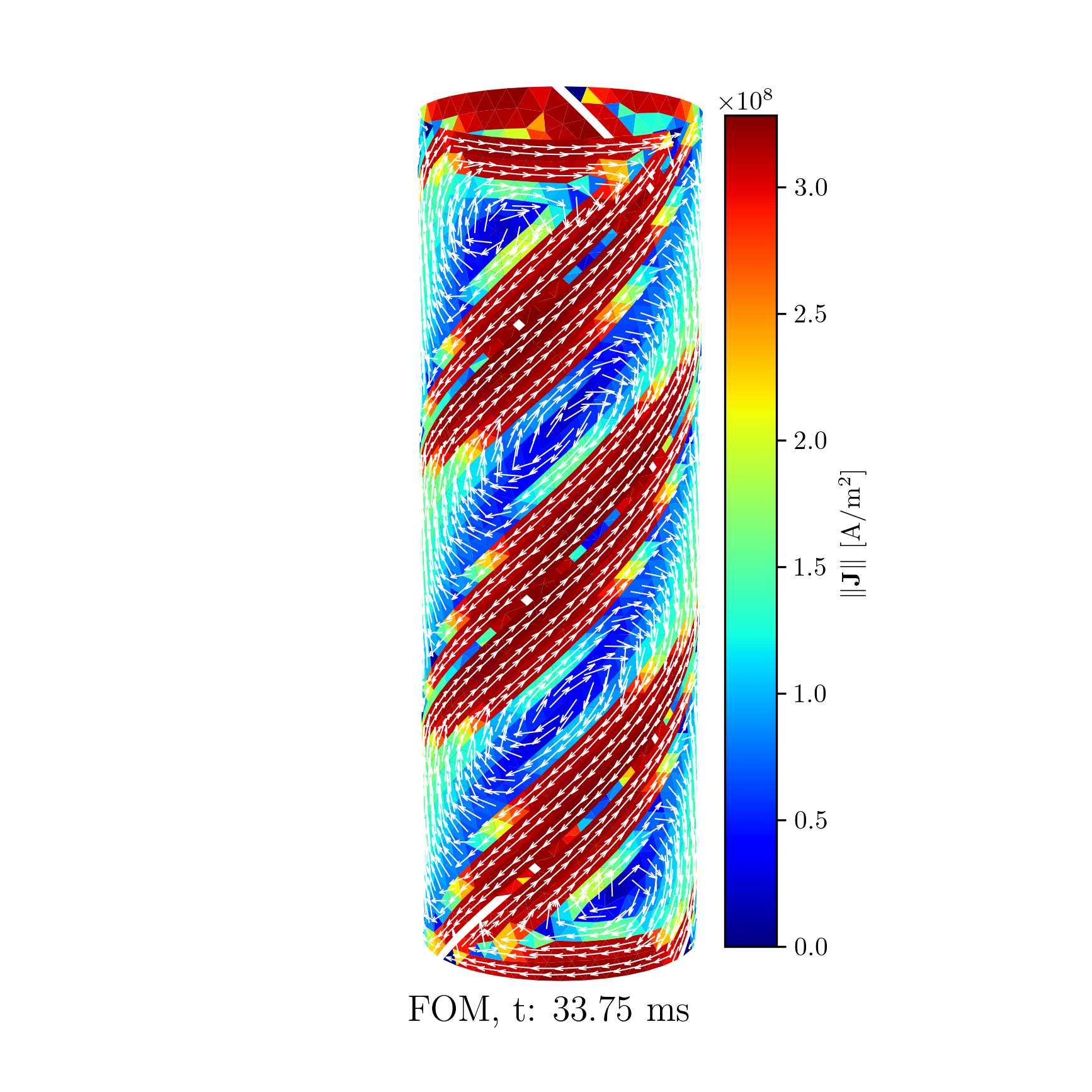}}

{\includegraphics[trim=4cm 0cm 3cm 0cm, clip, width=0.194\textwidth]{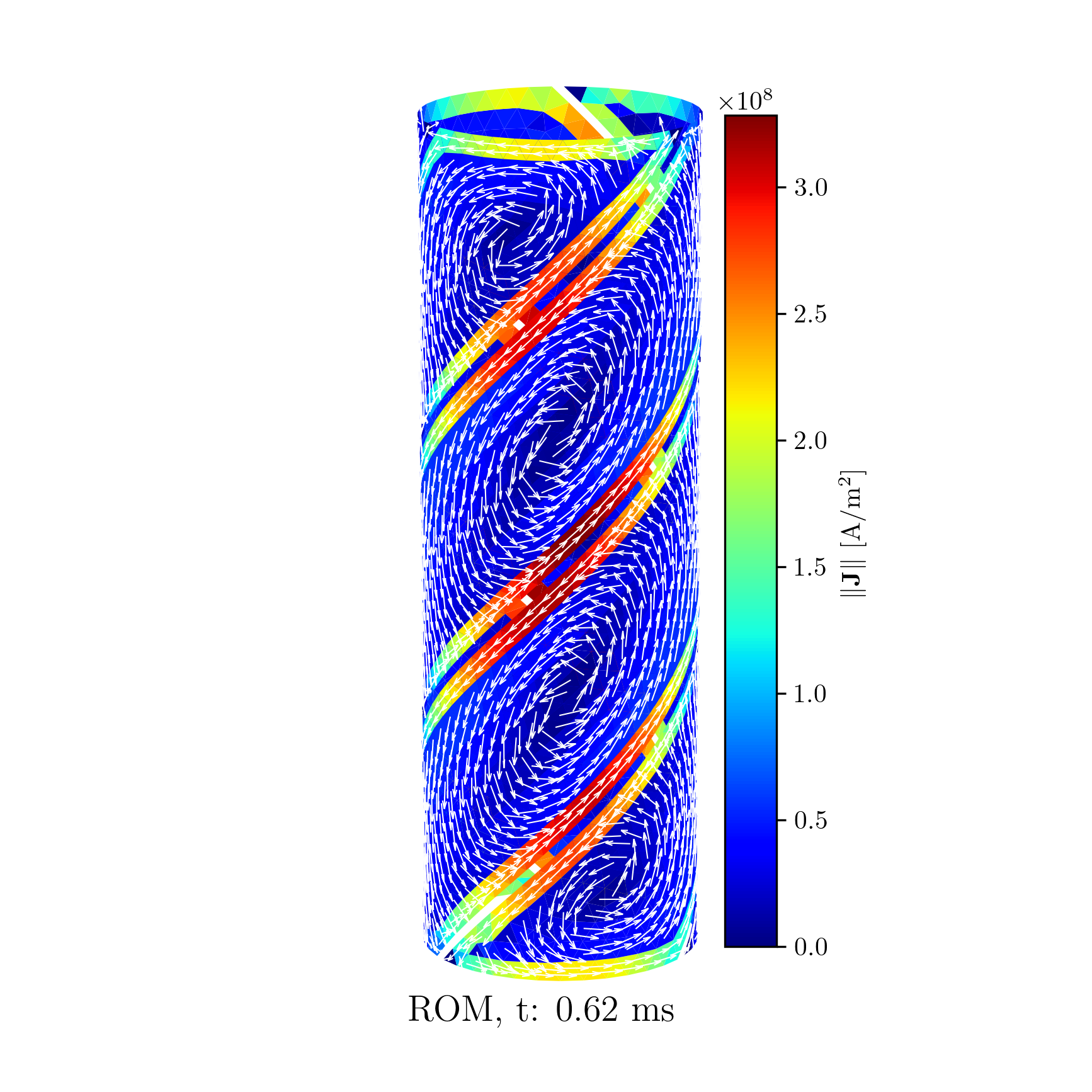}}
{\includegraphics[trim=4cm 0cm 3cm 0cm, clip, width=0.194\textwidth]{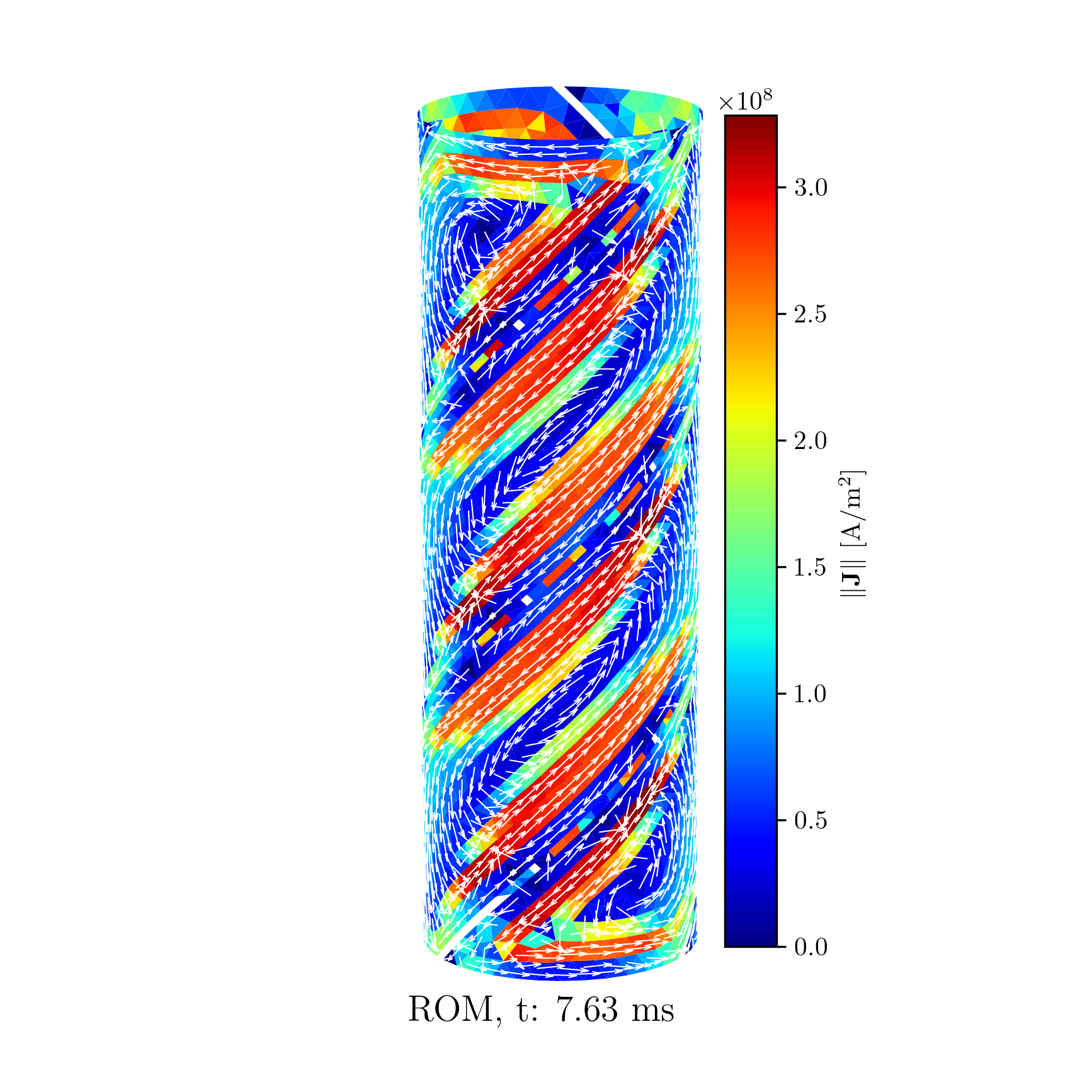}}
{\includegraphics[trim=4cm 0cm 3cm 0cm, clip, width=0.194\textwidth]{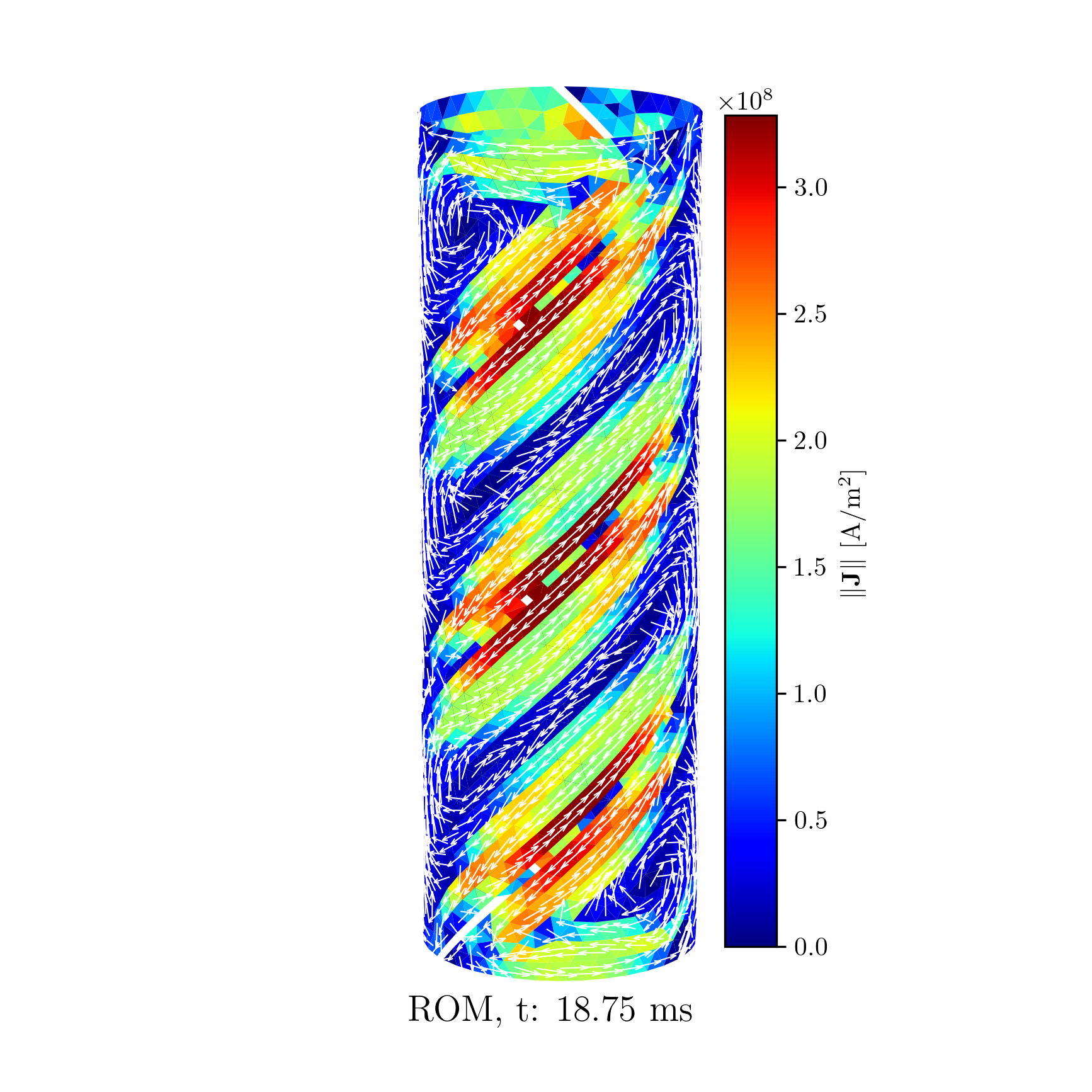}}
{\includegraphics[trim=4cm 0cm 3cm 0cm, clip, width=0.194\textwidth]{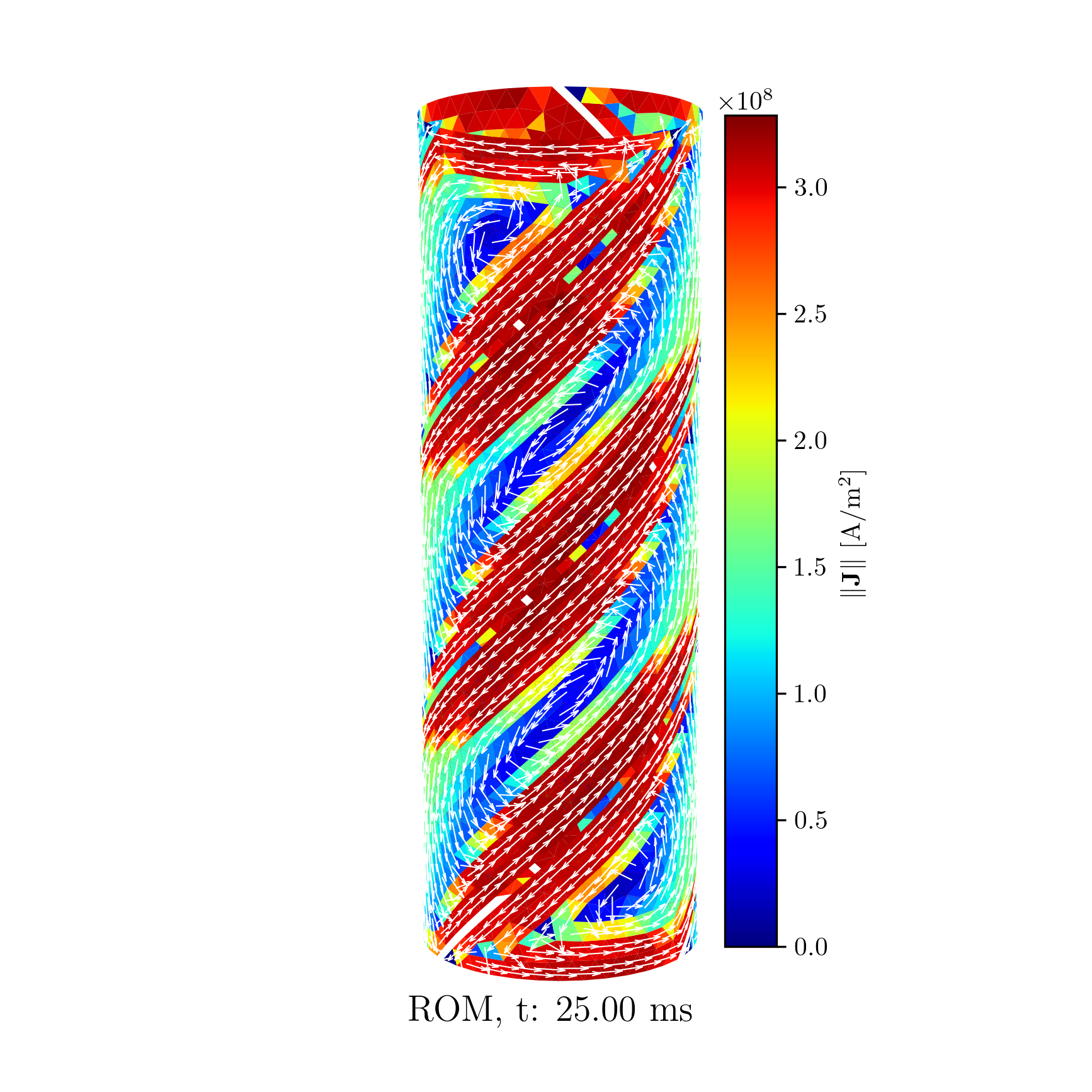}}
{\includegraphics[trim=4cm 0cm 3cm 0cm, clip, width=0.194\textwidth]{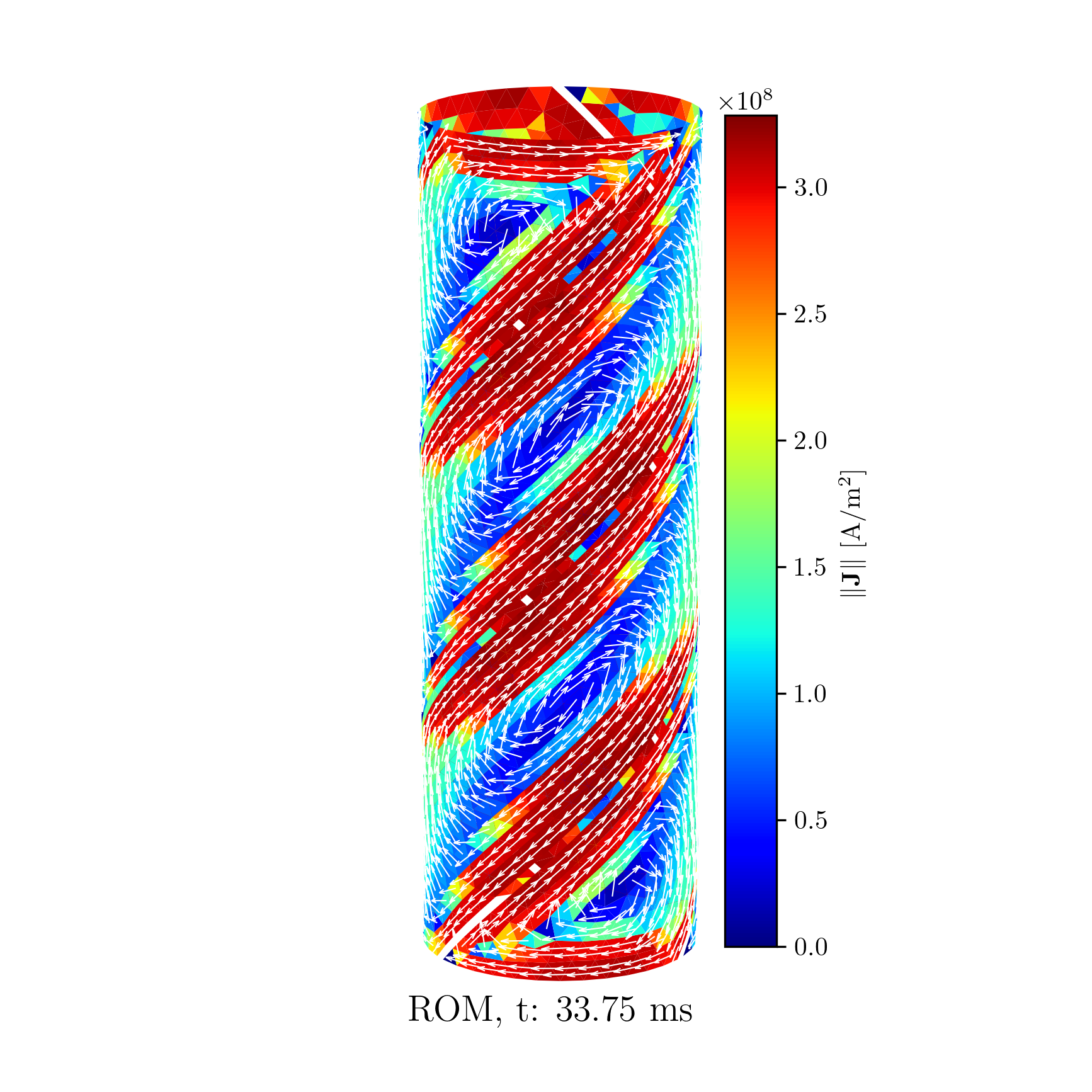}}

{\includegraphics[trim=4cm 0cm 3cm 0cm, clip, width=0.194\textwidth]{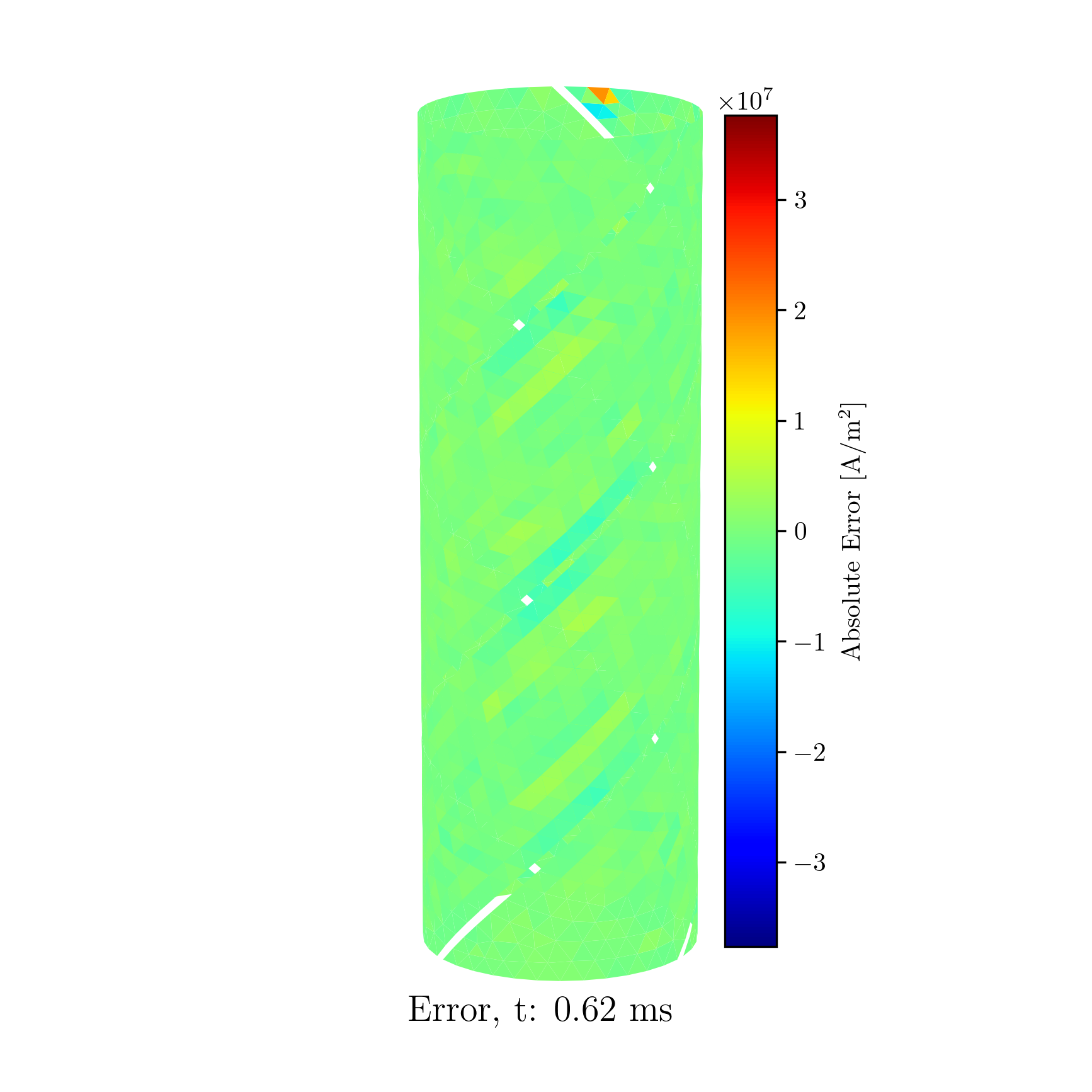}}
{\includegraphics[trim=4cm 0cm 3cm 0cm, clip, width=0.194\textwidth]{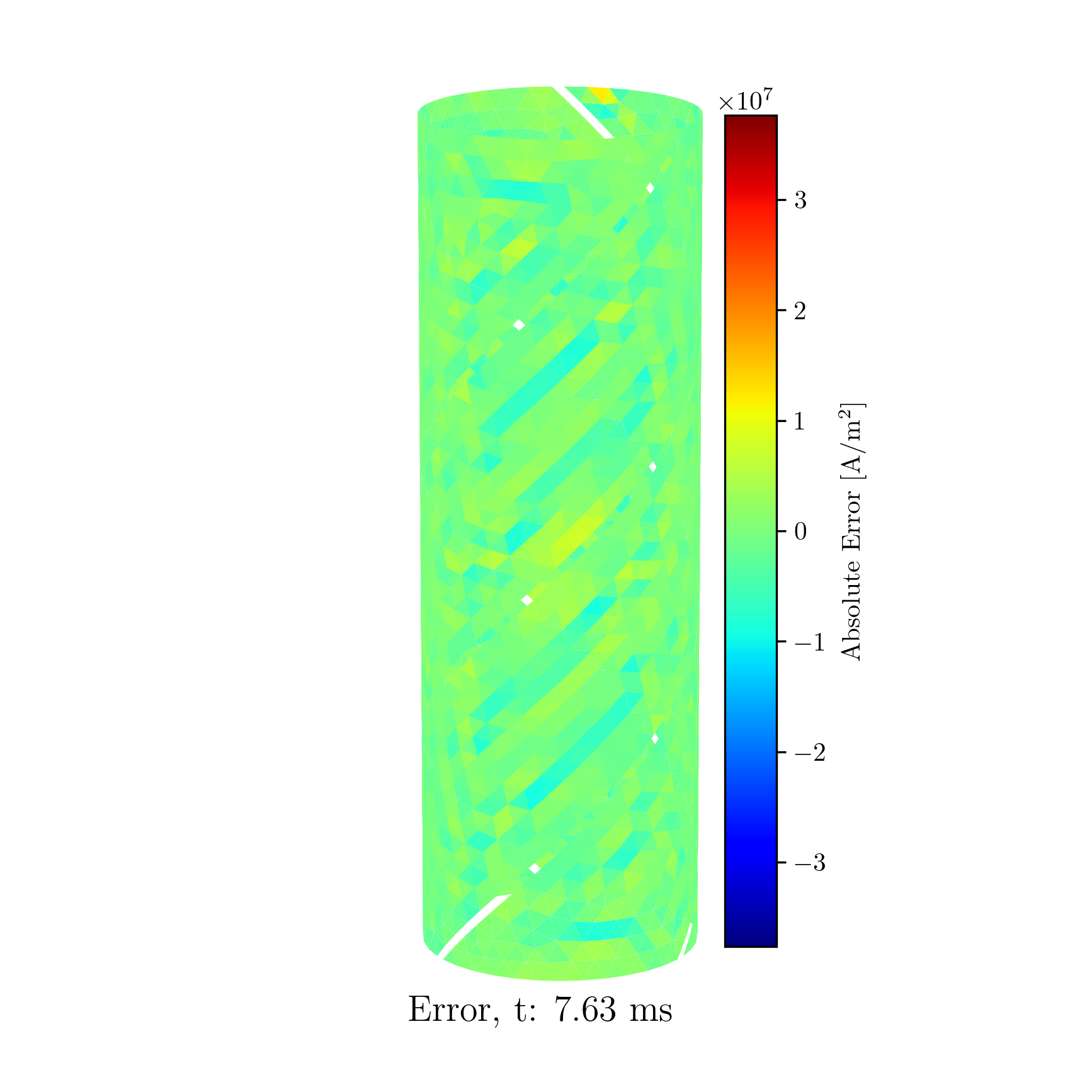}}
{\includegraphics[trim=4cm 0cm 3cm 0cm, clip, width=0.194\textwidth]{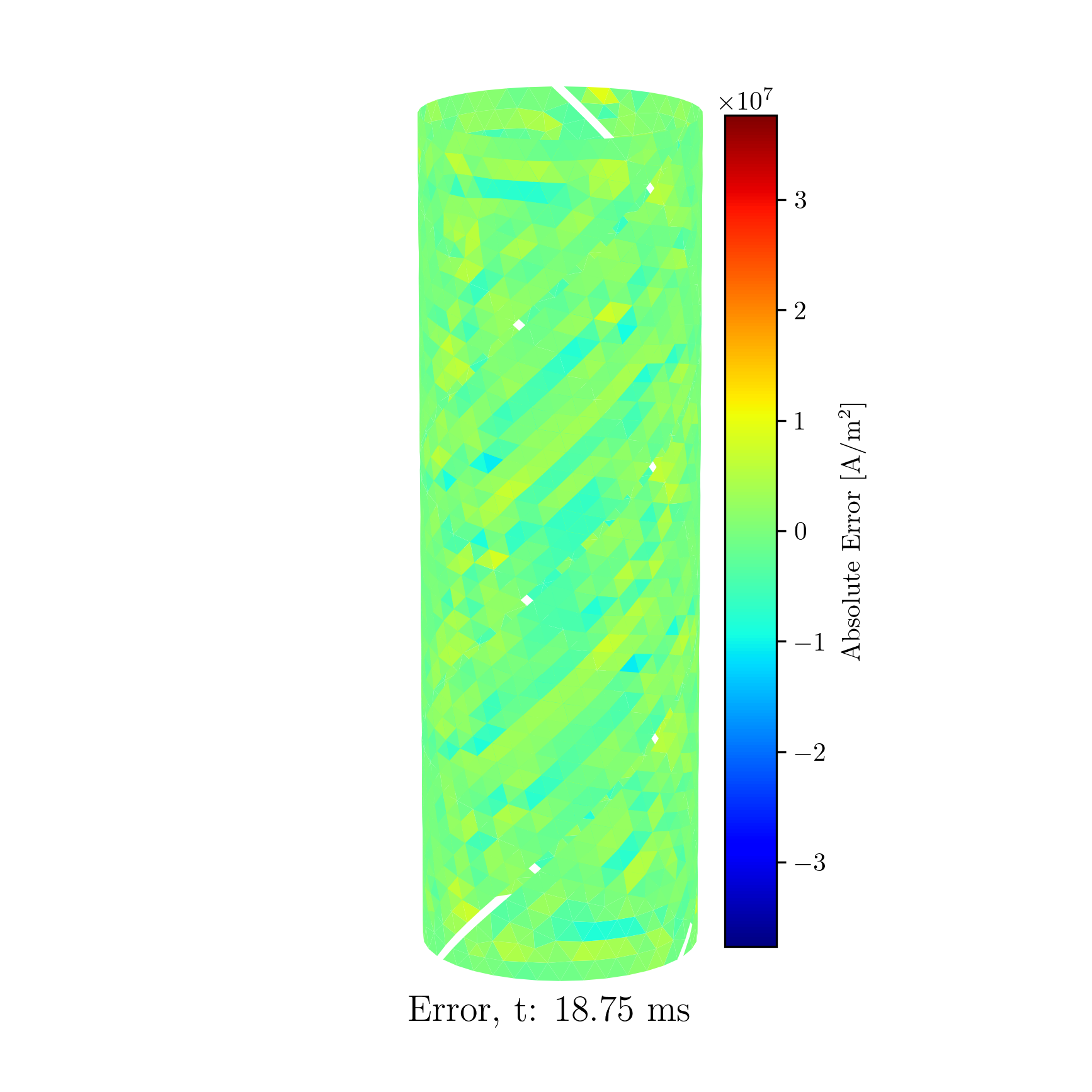}}
{\includegraphics[trim=4cm 0cm 3cm 0cm, clip, width=0.194\textwidth]{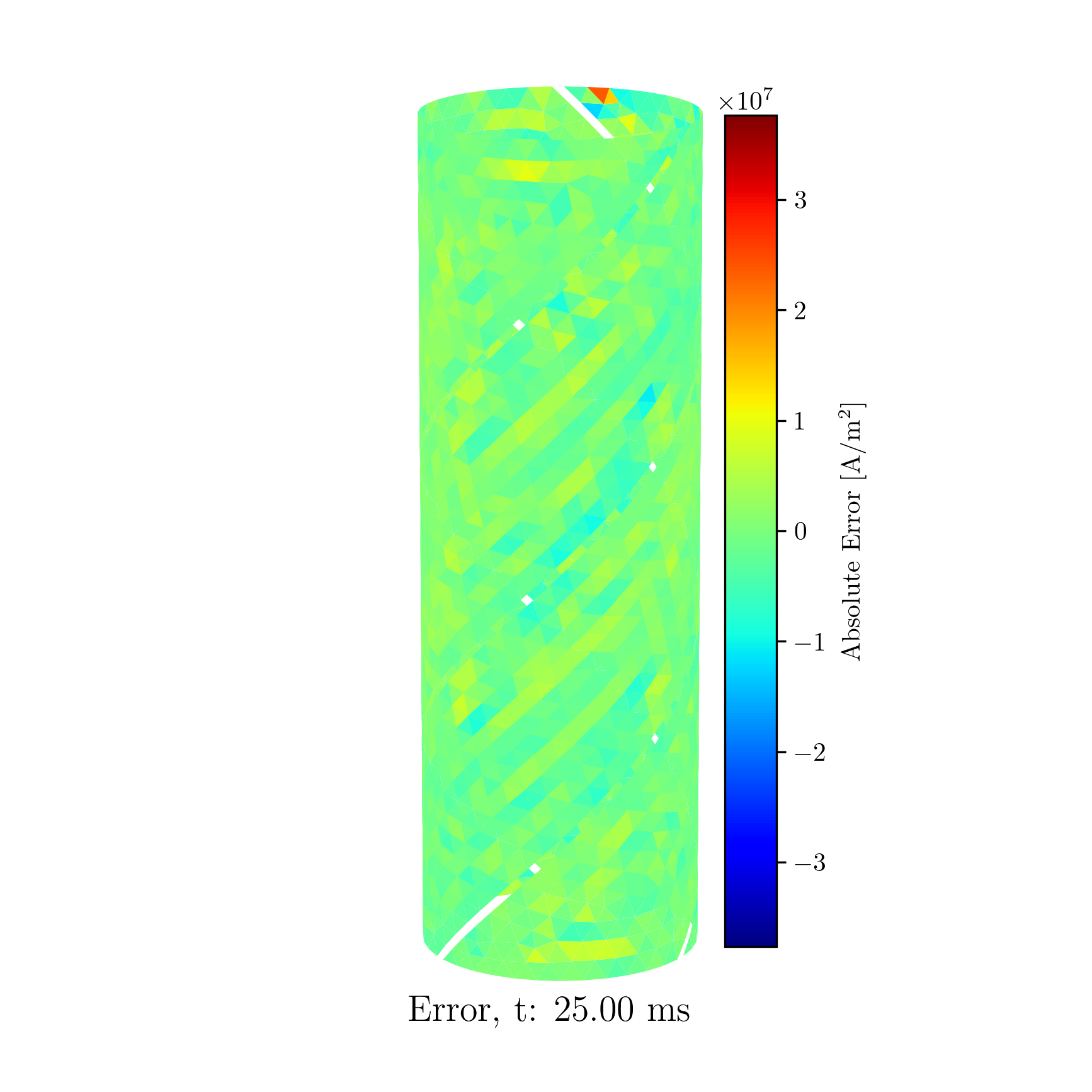}}
{\includegraphics[trim=4cm 0cm 3cm 0cm, clip, width=0.194\textwidth]{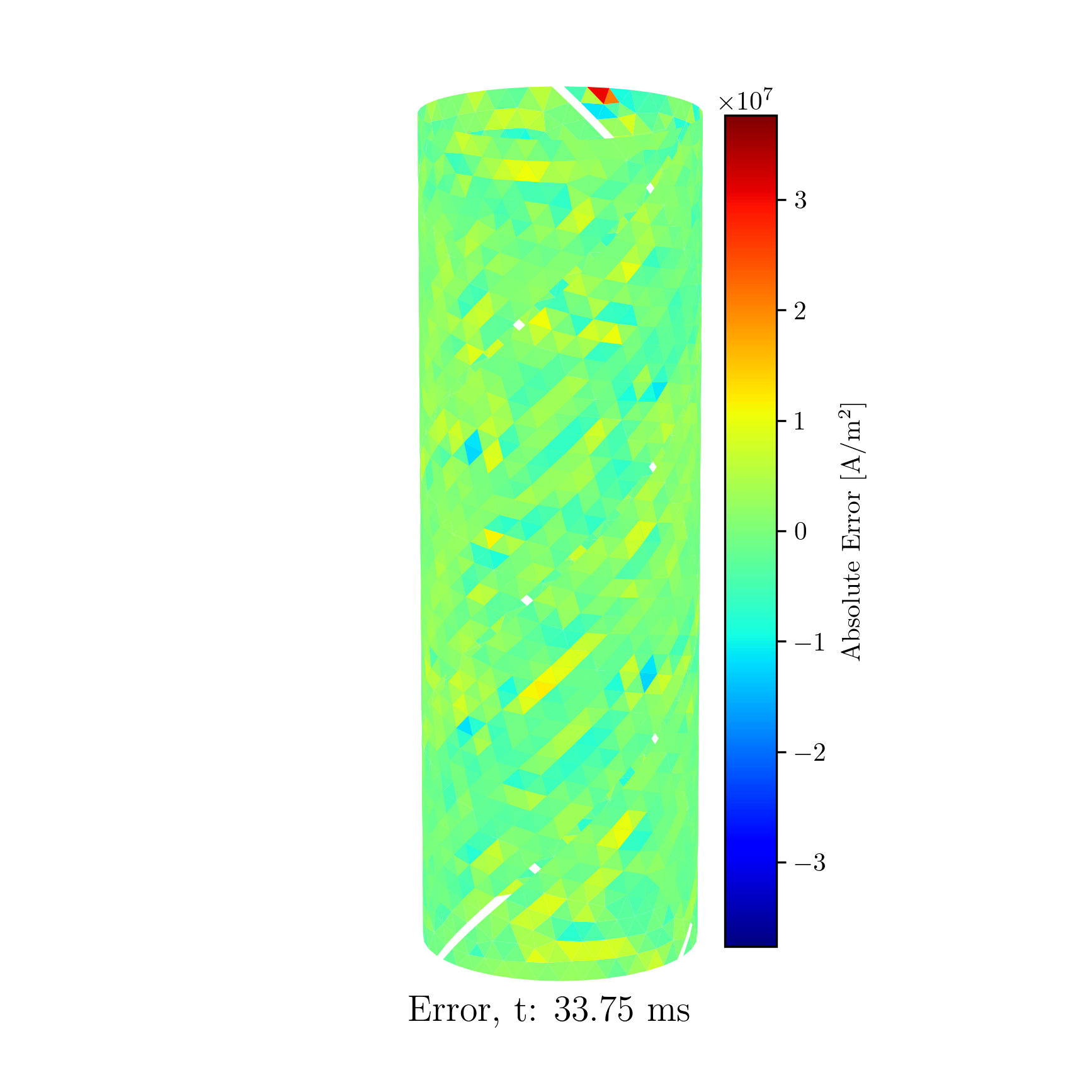}}

\caption{Comparison between the current density map of the \gls{fom} simulation and the \gls{neuralode}-based \gls{rom} on the  \textit{within-the-distribution} validation transient. The results demonstrate high accuracy.}
\label{fig:snap}
\end{figure}

To assess the accuracy of the proposed approach within the training distribution, the models were evaluated using sinusoidal inputs whose amplitude and frequency fall inside the range of the training data ($f=50$~Hz, $B_0=20$~mT). The \gls{pod}-\gls{deim} \gls{rom} was simulated with the same discretization strategy employed for the \gls{fom}, whereas the Structured \gls{neuralode} \gls{rom} was advanced according to \eqref{eq:node_advance}. As illustrated qualitatively at the top of Figure~\ref{fig:current_snap}, both reduced-order models reproduce the state evolution with good accuracy.\\
A quantitative comparison with the \gls{fom} simulation further supports this result: the bottom part of Figure~\ref{fig:current_snap} shows the mean, 95th-percentile, and maximum error among the states as functions of time, while the corresponding numerical values are summarized in Table~\ref{tab:IOD_error_tab}. These results indicate that the errors are of the same order of magnitude, with the Structured \gls{neuralode} yielding slightly lower values.
\begin{table}[!t]
\small
\centering
\renewcommand{\arraystretch}{1.3}
\begin{tabular}{c c c}
\hline
 & \textbf{\gls{deim}} & \textbf{\gls{neuralode}} \\
\hline
Mean Error & 0.1709~A/m &  0.1164~A/m\\
95th-percentile & 0.7921~A/m & 0.3398~A/m\\
Max Error & 4.5680~A/m & 1.7775~A/m\\
$\text{err}_{\lVert\cdot\rVert_F}$ & $7.5194 \cdot 10^{-2}$ & $2.3042 \cdot 10^{-2}$ \\
\hline
\end{tabular}
\renewcommand{\arraystretch}{1}
\caption{Mean, 95-th percentile and max absolute error for the \gls{deim} \gls{rom} and the \gls{neuralode} \gls{rom} in the  \textit{within-the-distribution} validation transient.}
\label{tab:IOD_error_tab}
\end{table}
Moreover, the current density map obtained with the proposed \gls{neuralode} approach is consistent with that from the \gls{fom} simulation, as shown in Figure~\ref{fig:snap}, which depicts the current distribution and the associated error at selected time instants.

\subsubsection{\textit{Outside-the-distribution} Validation}

To evaluate the performance of the \gls{rom}s outside the training distribution, two additional transient simulations were performed at 50 Hz with excitation amplitudes of 10 mT and 30 mT--values lying below and above the training range (13–24 mT), respectively. The corresponding quantitative results, summarized in Table~\ref{tab:OOD_error_tab}, indicate that the accuracy of the Structured \gls{neuralode} decreases noticeably, whereas the \gls{pod}-\gls{deim} \gls{rom} maintains performance comparable to the within-distribution case.

\begin{table}[!htbp]
\small
\centering
\renewcommand{\arraystretch}{1.3}
\begin{tabular}{c c c}
\hline
 & \textbf{\gls{deim}} & \textbf{\gls{neuralode}} \\
\hline
Mean Error & 0.1021~A/m &  0.4566~A/m\\
95th-percentile & 0.3860~A/m & 2.0990~A/m\\
Max Error & 6.0196~A/m & 6.7252~A/m\\
$\text{err}_{\lVert\cdot\rVert_F}$ & $7.6155 \cdot 10^{-2}$ & $1.8361\cdot10^{-1}$ \\
\hline
\end{tabular}
\renewcommand{\arraystretch}{1}
\caption{Mean, 95-th percentile and max absolute error for the \gls{deim} \gls{rom} and the \gls{neuralode} \gls{rom} in the  \textit{outside-the-distribution} validation transients.}
\label{tab:OOD_error_tab}
\end{table}

This outcome is consistent with expectations: data-driven models such as the Structured \gls{neuralode} are inherently constrained by the range of the training data. When the excitation amplitude extends beyond this range, the resulting state trajectories explore regions of the state space not represented during training, leading to a loss in predictive accuracy.

\subsubsection{\textit{Different Frequency} Validation}

To further assess the generalization capability of the proposed approach, additional simulations were carried out at an excitation amplitude of 20 mT--within the training range--but with frequencies of 40, 60, and 100 Hz, differing from the 50 Hz training frequency. The quantitative results, summarized in Table~\ref{tab:DF_error_tab}, demonstrate that the Structured \gls{neuralode} achieves very good accuracy across all tested frequencies, with errors of the same order of magnitude as in the 50 Hz validation and consistently lower than those of the \gls{pod}-\gls{deim} \gls{rom}.

\begin{table}[!htbp]
\small
\centering
\renewcommand{\arraystretch}{1.3}
\begin{tabular}{c c c}
\hline
 & \textbf{\gls{deim}} & \textbf{\gls{neuralode}} \\
\hline
Mean Error & 0.1729~A/m &  0.1484~A/m\\
95th-percentile & 0.7854~A/m & 0.4595~A/m\\
Max Error & 4.6427~A/m & 2.7591~A/m\\
$\text{err}_{\lVert\cdot\rVert_F}$ & $7.5948\cdot10^{-2}$ & $3.6895\cdot10^{-2}$ \\
\hline
\end{tabular}
\renewcommand{\arraystretch}{1}
\caption{Mean, 95-th percentile and max absolute error for the \gls{deim} \gls{rom} and the \gls{neuralode} \gls{rom} in the  \textit{different frequency} validation transients.}
\label{tab:DF_error_tab}
\end{table}

This strong performance can be attributed to the intrinsic structure of the proposed model: since the \gls{neuralode} component depends solely on the state variables, the learned dynamics remain accurate as long as the system evolution explores regions of the state space already represented in the training data. Although the model implicitly captures some temporal dependencies through the state update formulation, these do not significantly affect the accuracy, explaining the robust behavior observed across different excitation frequencies.

\subsection{Computational Cost}

To compare the computational cost of the two \glspl{rom}, we evaluate the number of \gls{flops} required to advance the system by one time step. The time-stepping scheme used to simulate the \gls{pod}-\gls{deim} \gls{rom} employs Newton iterations at each step, with residual and Jacobian evaluations at each iteration--an approach necessitated by the strong nonlinearity of the problem. In both cases, an optimal implementation scenario is assumed.

\subsubsection{\gls{pod}-\gls{deim} \gls{rom} Computational Cost}

The computational cost of the \gls{pod}-\gls{deim} reduced-order model involves mapping the reduced solution to the full current density $\mathbf{J}$, evaluating the resistivity $\rho$ and its derivative $\rho'$, assembling the \gls{deim}-selected rows of $\mathbf{R}$ and its Jacobian, performing matrix-vector products and projections, and solving the linear system within each Newton-Raphson iteration. Contributions from the right-hand side, which are independent of the solution, are computed once per time step.

\subsubsection{Structure \gls{neuralode} \gls{rom} Computational Cost}

The computational cost for advancing the state in the Structured \gls{neuralode} \gls{rom} includes evaluating the neural network, assembling the left- and right-hand sides through matrix-vector operations, and solving the resulting linear system.

\subsubsection{Results of the Evaluations}

For a single time step, the \gls{pod}-\gls{deim} \gls{rom} requires approximately $4.5\times10^7$ operations, whereas the Structured \gls{neuralode} \gls{rom} requires only $2.6\times10^5$, corresponding to a nominal \gls{flops}-based speed-up of $\times176$. This estimate reflects the algorithmic cost of advancing the reduced state under the fully implicit scheme employed in the \gls{pod}-\gls{deim} \gls{rom}, which requires repeated residual and Jacobian evaluations within Newton-Raphson iterations. In our experiments, the \gls{pod}-\gls{deim} \gls{rom} required an average of 19 Newton iterations per time-step, and the dominant cost arises from repeated assembly of the \gls{deim}-reconstructed nonlinear term and its Jacobian. By contrast, the Structured \gls{neuralode} \gls{rom} advances the reduced state by evaluating the neural-network once per step and solving a single reduced linear system, thereby avoiding Newton iterations and the associated repeated nonlinear assembly.

If, instead of considering operation counts, we measure computational cost in terms of wall-clock time for the full transient simulation, the observed speed is significantly larger. While the Structured \gls{neuralode} can be implemented in a highly optimized manner using only efficient neural-network evaluation and basic algebraic operations, the \gls{pod}-\gls{deim} \gls{rom} requires repeated operator assembly, and these assembly routines are more difficult to optimize in practice. As a result, the overhead associated with repeated assembly and Newton iterations dominates the execution time of the \gls{pod}-\gls{deim} \gls{rom}. In our implementation, this leads to a practical speed-up of approximately $\times1500$, reducing the runtime for a full transient simulation from 5 minutes to just 0.2 seconds. These results are summarized in Table~\ref{tab:comp_cost}.

\begin{table}[!t]
\small
\centering
\renewcommand{\arraystretch}{1.3}
\begin{tabular}{c c c}
\hline
\textbf{Approach} & \textbf{\# \gls{flops}} & \textbf{Eval. Time} \\
\hline
DEIM & 45,330,701 ($\times176$) & 334 s ($\times1500$) \\
\gls{neuralode} & 257,395 ($\times1$) & 0.2 s ($\times1$) \\
\hline
\end{tabular}
\renewcommand{\arraystretch}{1}
\caption{Number of \gls{flops} per time step and total evaluation time for reconstructing a full transient.}
\label{tab:comp_cost}
\end{table}

\subsection{Time-step sensitivity analysis}
\label{sec:dt_sensitivity}

\begin{figure}[!b]
    \centering
    \includegraphics[trim=0cm 0cm 0cm 0cm, clip, width=0.5\columnwidth]{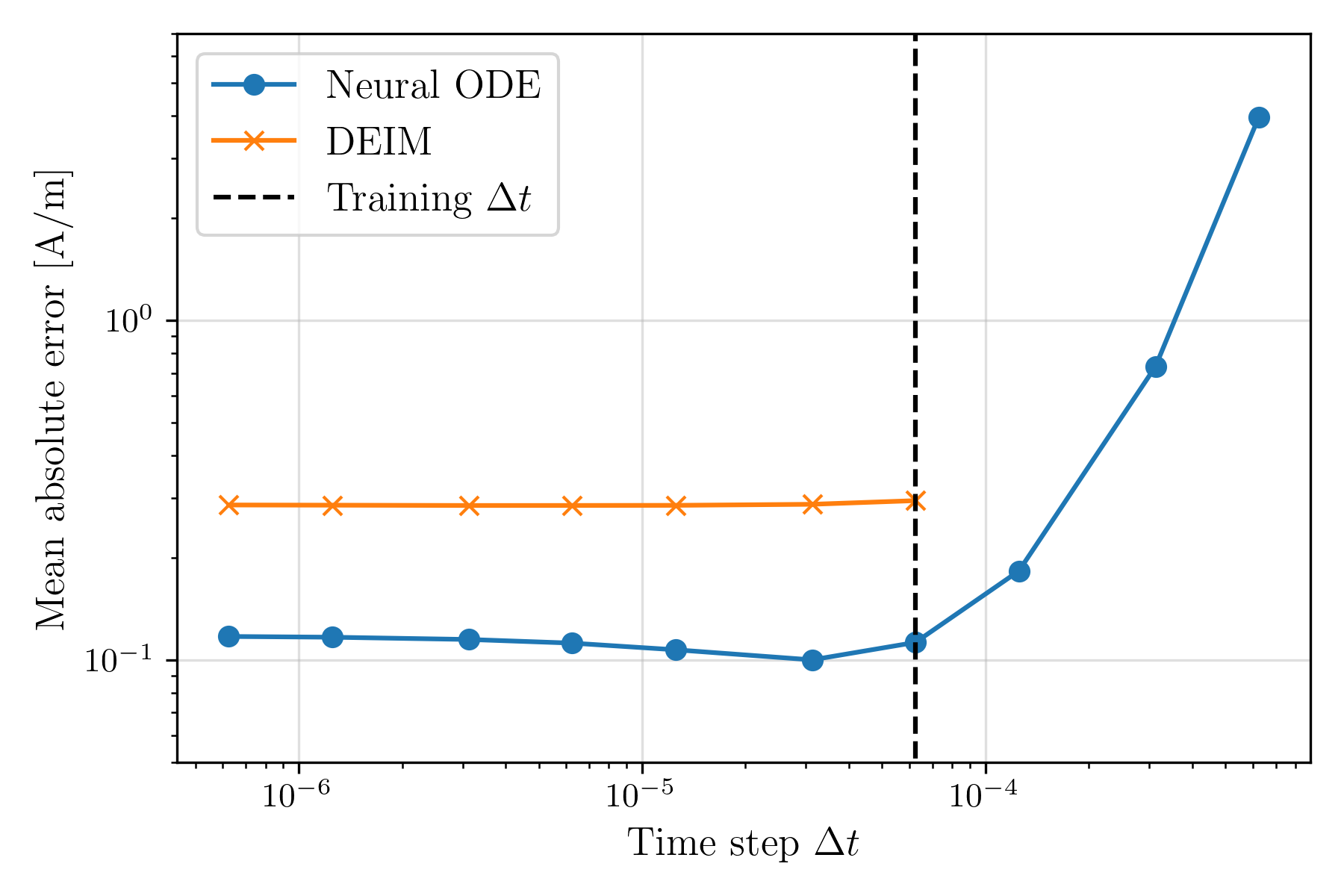}
    \caption{Time-step sensitivity study. Each marker reports the mean absolute current error averaged over space and time with respect to the \gls{fom} reference solution, so that each point summarizes the discrepancy over the full transient. The dashed line denotes the training time step $\Delta t_{\text{train}}$. The \gls{deim} curve terminates at the largest time step for which the Newton solver converges.}
    \label{fig:different_dt_error}
\end{figure}

We additionally investigate the effect of the time-step size $\Delta t$ on accuracy and computational cost. The Structured \gls{neuralode}, trained at $\Delta t_{\text{train}}$, remains accurate when deployed with smaller time steps, whereas accuracy degrades for $\Delta t > \Delta t_{\text{train}}$, consistently with~\cite{FARENGA2025107146}. The fully implicit \gls{pod}-\gls{deim} scheme is, in principle, less sensitive to $\Delta t$, but in our experiments, Newton iterations fail to converge beyond a certain $\Delta t$, limiting the admissible time-step range unless additional solver enhancements are introduced (e.g., problem-specific preconditioning strategies and improved initial guesses), which bring non-trivial algorithmic complexity. These trends are summarized in Fig.~\ref{fig:different_dt_error}.

In terms of computational cost, the Structured \gls{neuralode} exhibits an approximately constant per-step complexity (dominated by a single neural-network evaluation and reduced linear algebra), so the total transient evaluation time scales linearly with the number of time steps, i.e., proportional to $1/\Delta t$. The computational cost of the fully implicit \gls{pod}-\gls{deim} approach instead depends both on the number of time steps and on the number of Newton iterations per step. Moreover, as discussed in Section~\ref{subsec:ts}, the per-step cost of \gls{pod}-\gls{deim} with an explicit time integrator is already comparable to that of the Structured \gls{neuralode}. Therefore, even in the most favorable implicit scenario (i.e., a single Newton iteration), one should expect a higher per-step cost due to the additional assembly of the Jacobian on top of the residual evaluation. Consequently, across the tested range of $\Delta t$, the proposed method retains a substantial computational advantage while maintaining competitive accuracy for $\Delta t \le \Delta t_{\text{train}}$.

\subsection{Note on the Time-stepping Strategy}\label{subsec:ts}

\begin{figure}[!htbp]
\centering

{\includegraphics[trim=1cm 0cm 2.5cm 0cm, clip, width=0.49\textwidth]{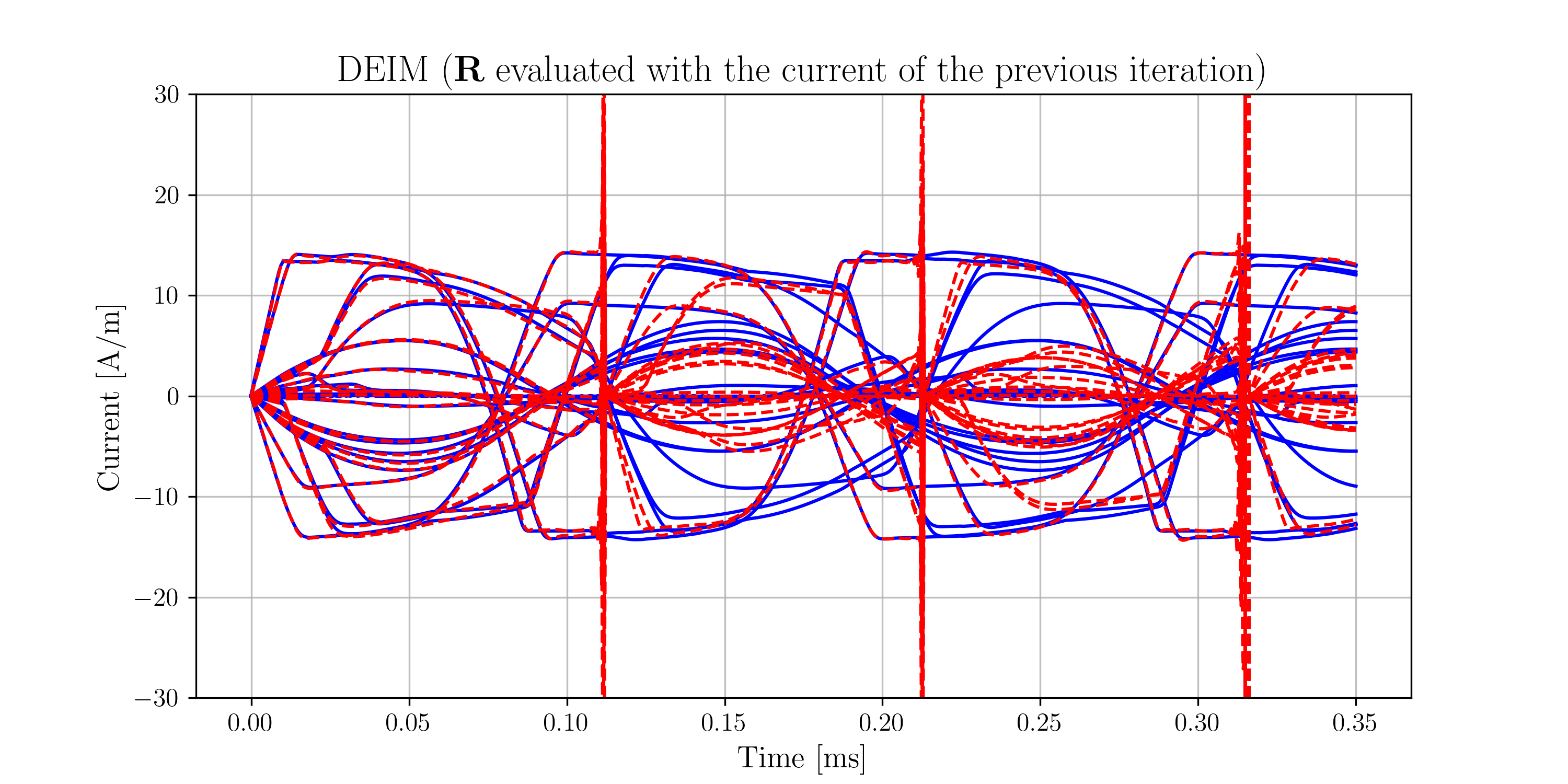}}
{\includegraphics[trim=1cm 0cm 2.5cm 0cm, clip, width=0.49\textwidth]{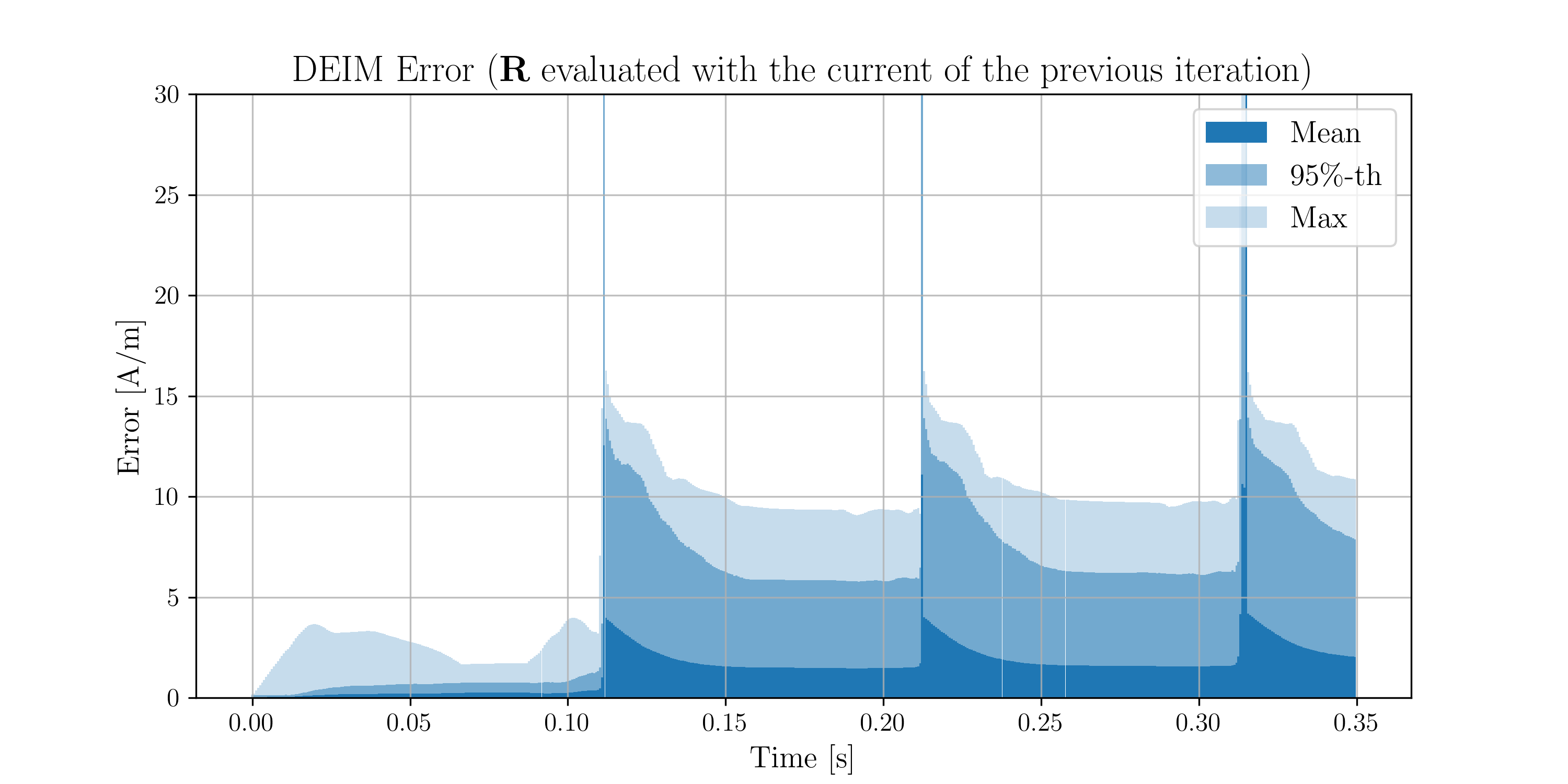}}

\caption{Evolution of selected current states in the validation transient for the \gls{pod}-\gls{deim} \gls{rom} (dashed red) compared with the \gls{fom} (solid blue). The corresponding mean, 95th-percentile, and maximum errors between the \gls{rom} and the \gls{fom} are also reported. In these simulations, the nonlinear operator is evaluated using the current state from the previous iteration. The results clearly show that this approach does not yield accurate or practically useful solutions.}
\label{fig:expl_deim}
\end{figure}

As explained in Section \ref{subsec:rom_sim}, the \gls{deim} approach could also be implemented by evaluating $\mathbf{R}_r$ at the previously computed state, in which case only a linear system would need to be solved at each time step. This would reduce the computational effort to a level comparable with that of our \gls{neuralode}. In particular, if we consider this approximation, the \gls{deim} requires approximately $1.99\times10^5$ \gls{flops} to advance the state, which is about 1.29 times fewer operations than the Structured \gls{neuralode}. However, a crucial difference lies in the fact that the neural ODE has been explicitly trained to perform this task, and the training process inherently accounts for this simplification, thereby ensuring accurate results. By contrast, due to the strong nonlinearity of the problem, applying the same type of approximation within the \gls{pod}-\gls{deim} \gls{rom} leads to unreliable and ultimately uninformative simulations, as illustrated in Figure~\ref{fig:expl_deim}.

\subsection{Post-processing: Power losses}

\begin{figure}[!htbp]
\centering

{\includegraphics[trim=1cm 0cm 2.5cm 0cm, clip, width=0.49\textwidth]{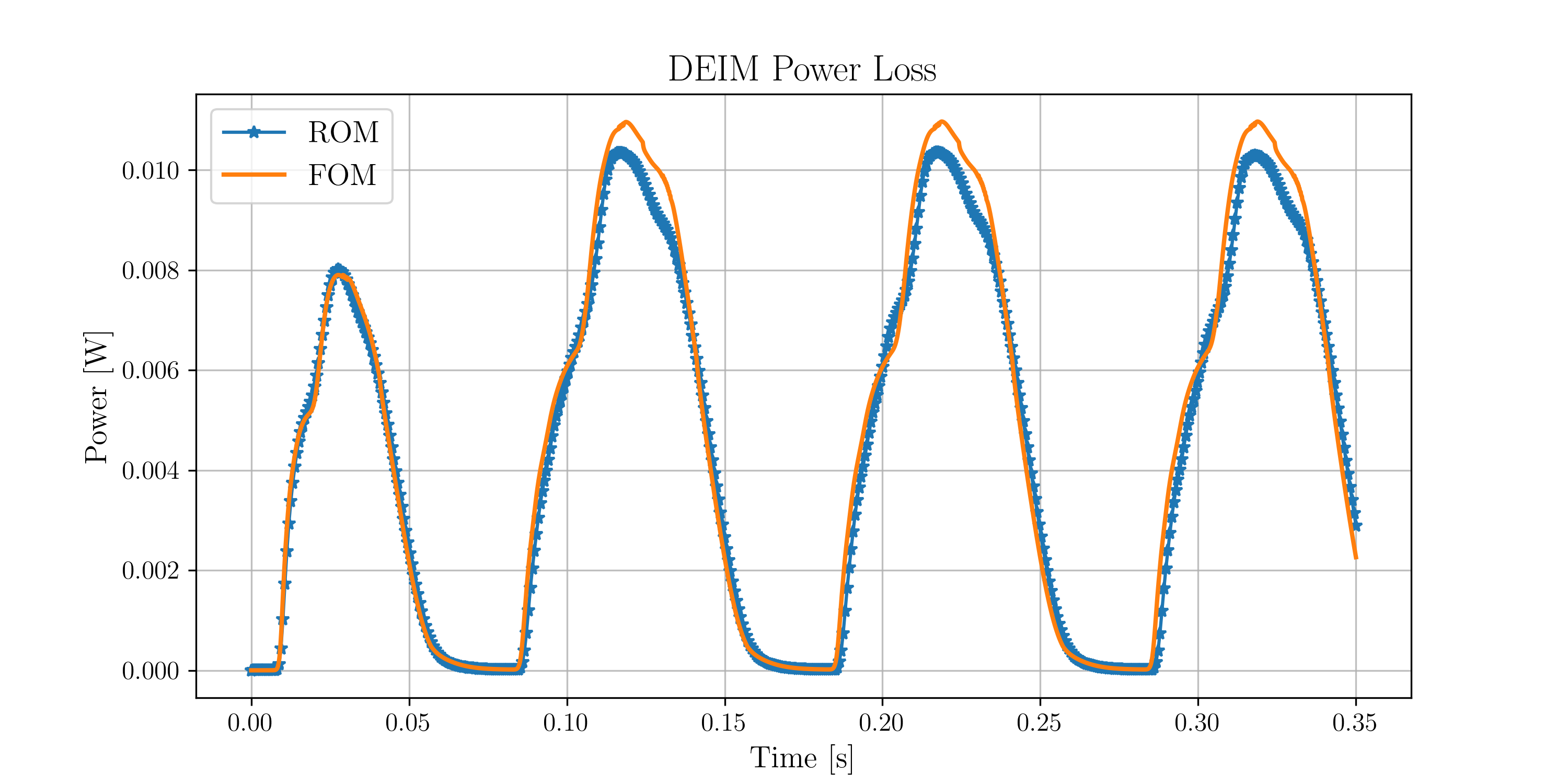}}
{\includegraphics[trim=1cm 0cm 2.5cm 0cm, clip, width=0.49\textwidth]{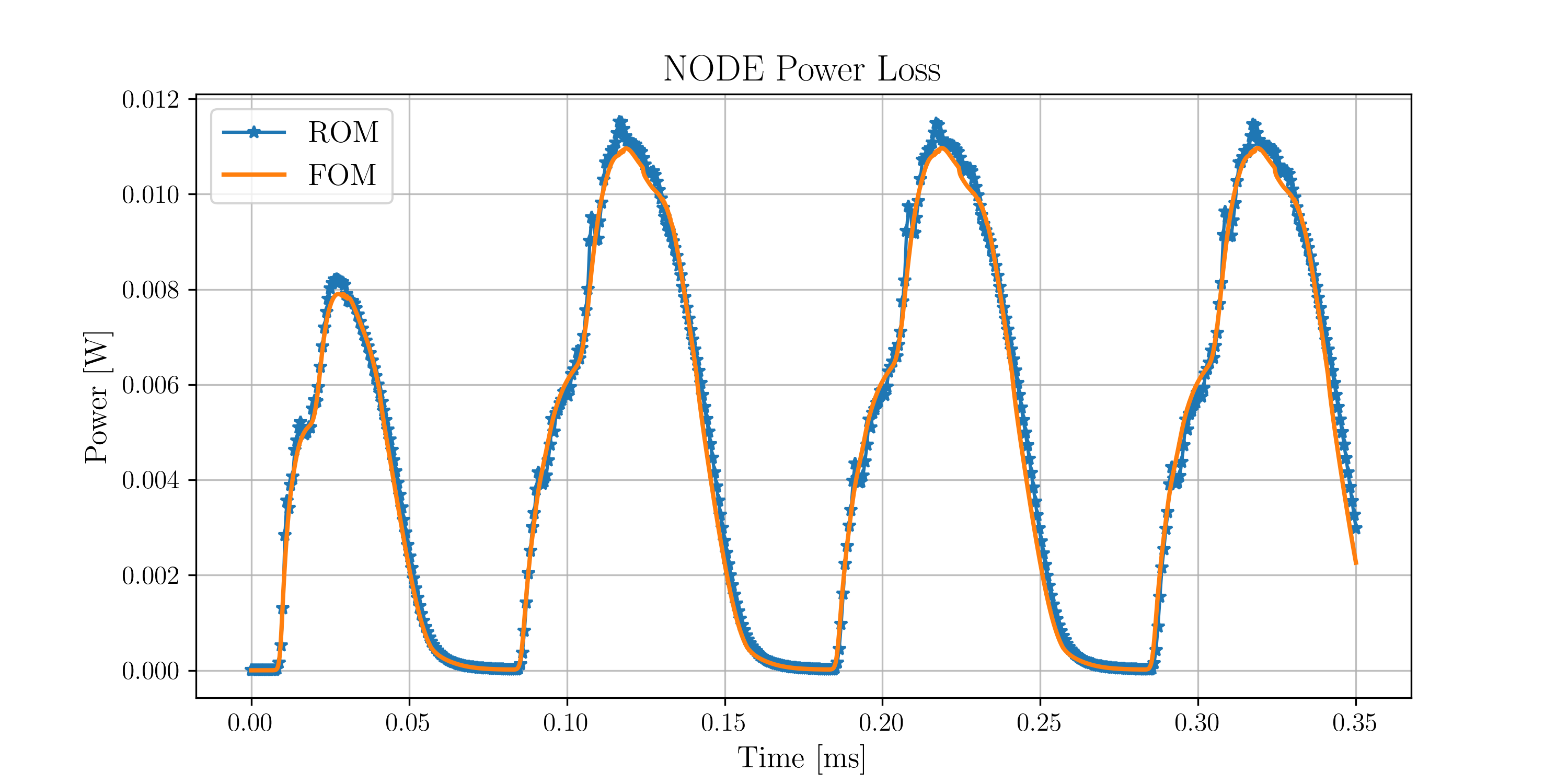}}

{\includegraphics[trim=1cm 0cm 2.5cm 0cm, clip, width=0.49\textwidth]{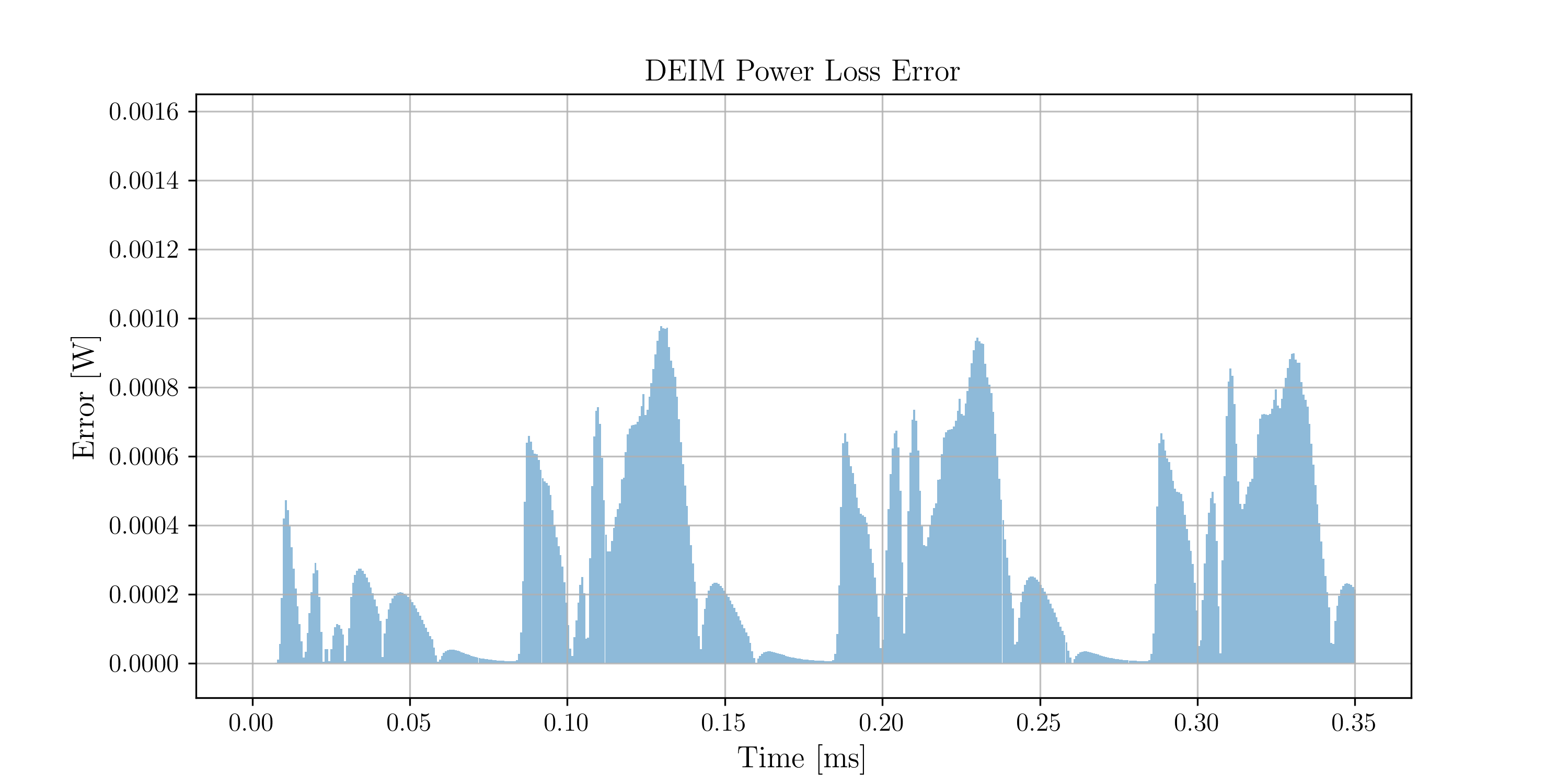}}
{\includegraphics[trim=1cm 0cm 2.5cm 0cm, clip, width=0.49\textwidth]{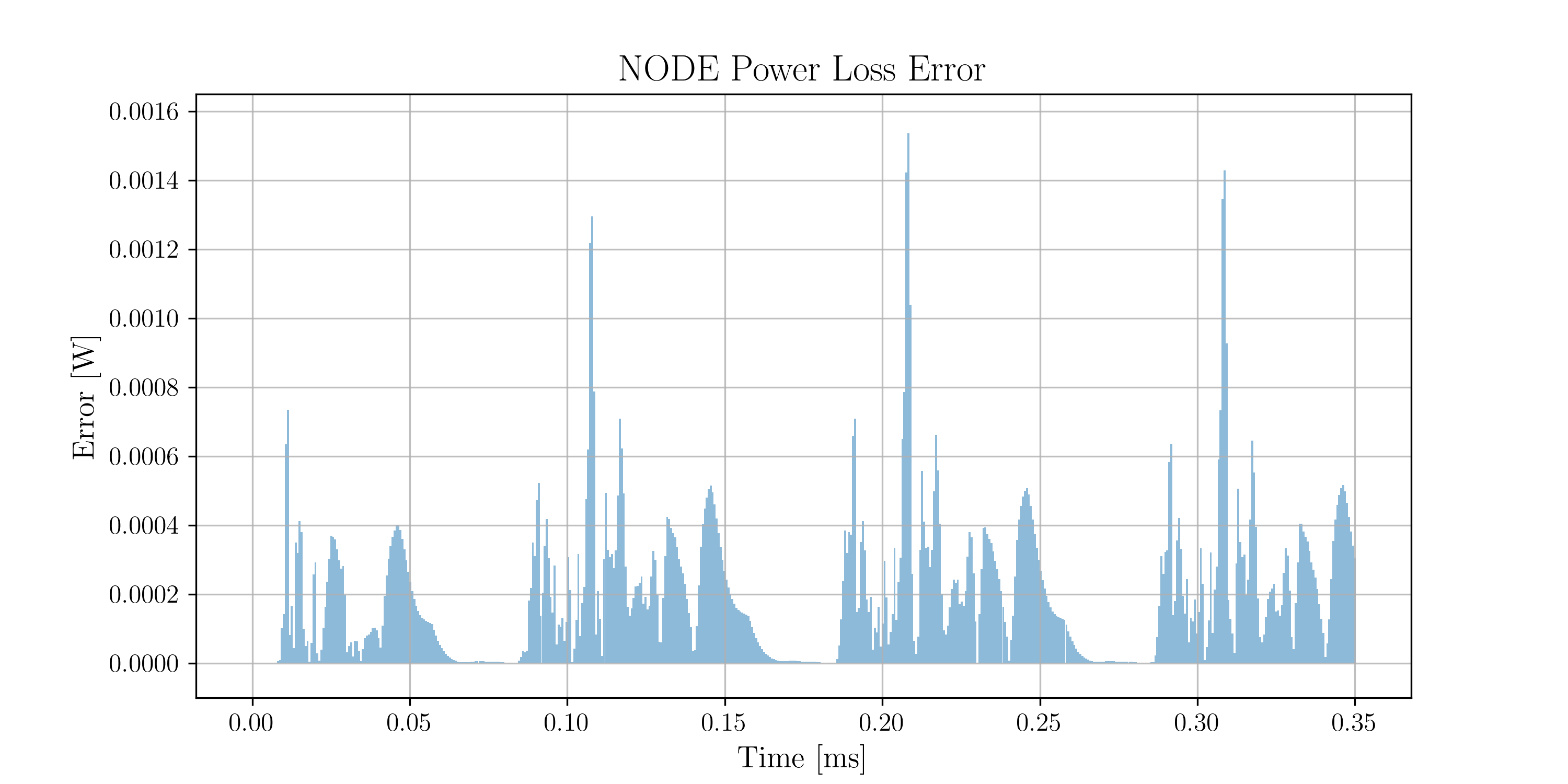}}

\caption{Comparison of power loss of the  \textit{within-the-distribution} validation transient computed from the current density obtained using the \gls{fom} and the two \gls{rom}s, including the associated error.}
\label{fig:ac_loss}
\end{figure}

As a post-processing step, we evaluate the AC losses in the \gls{corc} cable, which serve as input for a thermal model aimed at predicting the cable’s temperature distribution--a key quantity for quench prediction in both design and control. The AC losses are computed as:
\begin{equation}
p(t)=\int_{\Omega_{sc}}\mathbf{J}\cdot\mathbf{E}d\Omega = \int_{\Omega_{sc}}\rho\lVert\mathbf{J}\rVert^2d\Omega.
\end{equation}
These losses are related to the current loops in the HTS tapes, also recognized as magnetization losses \cite{yazdani2018calculation}.
The integral is performed numerically, assuming the electric field and current density are constant within each element. Power losses are evaluated using current densities obtained from both the \gls{pod}-\gls{deim} \gls{rom} and the \gls{neuralode} \gls{rom}. Figure~\ref{fig:ac_loss} shows the results of these evaluations, along with the error relative to the \gls{fom} values. The comparison indicates that both \gls{rom}s achieve comparable accuracy in predicting power loss.

\section{Discussion}\label{sec:discussion}

The comparison between the Structured \gls{neuralode} and the \gls{pod}-\gls{deim} approaches highlights a clear trade-off between accuracy, computational efficiency, intrusiveness, and training requirements.

Both \gls{rom}s achieve comparable overall accuracy, with the Structured \gls{neuralode} generally providing slightly higher precision. In particular, the Structured \gls{neuralode} yields the most accurate results for inputs within the training distribution, while the \gls{pod}-\gls{deim} \gls{rom} proves more reliable when the excitation amplitude lies outside the training range. For excitations at different frequencies, the Structured \gls{neuralode} performs remarkably well, maintaining accuracy comparable to the training frequency case and outperforming the \gls{pod}-\gls{deim} model. This behavior confirms that the proposed data-driven structure generalizes effectively across variations in frequency, provided that the system trajectories remain within regions of the state space represented during training.

In terms of computational cost, the Structured \gls{neuralode} is significantly faster because it advances the state by solving only a linear system at each time step, whereas the \gls{pod}-\gls{deim} \gls{rom} requires a high number of Newton-Raphson iterations due to the strong nonlinearity of the model, with repeated evaluations of the Jacobian at each step.

The Structured \gls{neuralode} is also considerably less intrusive, as it only requires access to the assembled linear operators and to evaluations of the nonlinear operator at selected working points. Such information can typically be extracted without modifying the underlying simulation code, making the approach compatible with commercial solvers and straightforward to integrate into existing workflows. In contrast, the \gls{pod}-\gls{deim} method is highly intrusive, since it requires direct access to and modification of the \gls{fom} assembly routines to incorporate the \gls{deim} interpolation points. This need for code-level intervention limits its applicability in commercial or legacy software environments and makes the optimization of the \gls{deim} approximation more demanding in practice.

When it comes to training, \gls{pod}-\gls{deim} is easier to set up and tune due to its linear algebra-based structure, whereas Structured \glspl{neuralode} demand careful tuning of hyperparameters and more complex training procedures.

Finally, data requirements differ: \gls{pod}-\gls{deim} typically needs less data and generalizes well across operating conditions, while Structured \glspl{neuralode} require larger, representative datasets to ensure reliable performance during operation. A concise summary of these pros and cons is provided in Table~\ref{tab:rom_comparison}.

Taken together, these results indicate that the Structured \gls{neuralode} offers clear advantages in terms of intrusiveness and online efficiency, particularly in strongly nonlinear regimes where Newton iterations dominate the runtime. \gls{pod}-\gls{deim} remains preferable primarily in data-scarce or extrapolative scenarios, whereas the Structured \gls{neuralode} is the method of choice whenever sufficient training data are available.

\begin{table*}[!htbp]
\small
\centering
\renewcommand{\arraystretch}{1.3}
\begin{tabular}{l c c}
\hline
\textbf{Aspect} & \textbf{\gls{neuralode} \gls{rom}} & \textbf{\gls{pod}-\gls{deim} \gls{rom}} \\
\hline
Accuracy & High (overall, slightly higher) & High \\
Cost & Low ($\sim$176× speedup) & High (Newton Iterations; Jacobian Evaluations)  \\
Data requirements & High (Dense Dataset) & Low (Few snapshots are enough) \\
Training procedure & Empirical and delicate  & Straightforward (Linear Algebra-based)\\
Intrusiveness & Non-intrusive & Intrusive \\
\hline
\end{tabular}
\caption{Detailed comparison between \gls{neuralode}-based \gls{rom} and \gls{pod}-\gls{deim} \gls{rom}.}
\label{tab:rom_comparison}
\renewcommand{\arraystretch}{1}
\end{table*}

\section{Conclusion}\label{sec:conclusion}

In this work, we developed and assessed reduced-order modeling strategies for nonlinear \gls{ems} simulations of \gls{hts} conductors within the \gls{iem} framework. Specifically, we presented the first application of \gls{pod}-\gls{deim} to \gls{iem}-based \gls{hts} models, and proposed a Structured \gls{neuralode} approach that learns the nonlinear dynamics directly in the reduced-order space. 

These methods were pursued to enable real-time prediction and monitoring of quench phenomena in superconducting devices. While conventional \gls{fem} or \gls{iem} remain too costly for online use, the proposed reduced-order strategies significantly improve computational efficiency without sacrificing accuracy. In particular, the Structured \gls{neuralode} demonstrated superior predictive capability compared to the standard \gls{pod}-\gls{deim} approach, achieving faster evaluations while maintaining high fidelity to the \gls{fom}.  

Future work will extend this framework toward multiphysics simulations, where the computed AC losses from the electromagnetic model will serve as inputs to thermal solvers. This coupling will allow for a more comprehensive description of quench initiation and propagation, thereby bringing reduced-order modeling closer to practical deployment in the design and operation of next-generation superconducting systems.


\bibliographystyle{unsrtnat}

\bibliography{references}

@ARTICLE{11220970,
  author={Al-Ssalih, Hasan N. H. and Clegg, Matthew and Badía-Majós, Antonio and Ruiz, Harold S.},
  journal={IEEE Access}, 
  title={{Magnetization Losses and Non-Reduced 3D Modeling of Hybrid CORC-TSTC Composite Conductors}}, 
  year={2025},
  volume={13},
  number={},
  pages={186952-186964},
  doi={10.1109/ACCESS.2025.3627080}}

@article{yazdani2018calculation,
  title={{Calculation of AC magnetizing loss of ReBCO superconducting tapes subjected to applied distorted magnetic fields}},
  author={Yazdani-Asrami, Mohammad and Gholamian, S Asghar and Mirimani, Seyyed Mehdi and Adabi, Jafar},
  journal={Journal of Superconductivity and Novel Magnetism},
  volume={31},
  number={12},
  pages={3875--3888},
  year={2018},
  publisher={Springer}
}

@ARTICLE{9462395,
  author={Shen, Boyang and Chen, Xiaoyuan and Fu, Lin and Jiang, Shan and Gou, Huayu and Sheng, Jie and Huang, Zhen and Wang, Wei and Zhai, Yujia and Yuan, Yupeng and Gao, Shuo and Wang, Sheng and Li, Chuanyue and Chen, Wei and Mu, Shuai and Zhou, Qian and Song, Wenjuan and Pei, Xiaoze and Öztürk, Yavuz and Gawith, James and Liu, Yi and Patel, Ismail and Tian, Mengyuan and Yang, Jiabin and Coombs, Tim},
  journal={IEEE Transactions on Applied Superconductivity}, 
  title={{Superconducting Conductor on Round Core (CORC) Cables: 2D or 3D Modeling?}}, 
  year={2021},
  volume={31},
  number={8},
  pages={1-5},
  doi={10.1109/TASC.2021.3091091}}

@article{ROSSI20231354360,
title = {{Particle Accelerators and Cuprate Superconductors}},
journal = {Physica C: Superconductivity and its Applications},
volume = {614},
pages = {1354360},
year = {2023},
issn = {0921-4534},
doi = {doi.org/10.1016/j.physc.2023.1354360},
author = {Lucio Rossi}
}

@article{MERRILL20152196,
title = {{Modeling an unmitigated thermal quench event in a large field magnet in a DEMO reactor}},
journal = {Fusion Engineering and Design},
volume = {98-99},
pages = {2196-2200},
year = {2015},
note = {Proceedings of the 28th Symposium On Fusion Technology (SOFT-28)},
issn = {0920-3796},
doi = {doi.org/10.1016/j.fusengdes.2015.03.007},
author = {Brad J. Merrill}
}

@ARTICLE{7393501,
  author={Takayasu, Makoto and Chiesa, Luisa and Allen, Nathaniel C. and Minervini, Joseph V.},
  journal={IEEE Transactions on Applied Superconductivity}, 
  title={{Present Status and Recent Developments of the Twisted Stacked-Tape Cable Conductor}}, 
  year={2016},
  volume={26},
  number={2},
  pages={25-34},
  doi={10.1109/TASC.2016.2521827}}

@article{Huber_2022,
doi = {10.1088/1361-6668/ac5163},
year = {2022},
month = {mar},
publisher = {IOP Publishing},
volume = {35},
number = {4},
pages = {043003},
author = {Huber, Felix and Song, Wenjuan and Zhang, Min and Grilli, Francesco},
title = {{The T-A formulation: an efficient approach to model the macroscopic electromagnetic behaviour of HTS coated conductor applications}},
journal = {Superconductor Science and Technology}
}

@ARTICLE{9097858,
  author={Shen, Boyang and Grilli, Francesco and Coombs, Tim},
  journal={IEEE Access}, 
  title={{Overview of H-Formulation: A Versatile Tool for Modeling Electromagnetics in High-Temperature Superconductor Applications}}, 
  year={2020},
  volume={8},
  number={},
  pages={100403-100414},
  doi={10.1109/ACCESS.2020.2996177}}

@ARTICLE{10897790,
  author={Lucchini, Francesco},
  journal={IEEE Transactions on Applied Superconductivity}, 
  title={{Evaluating Magnetization Losses in 3-D CORC Tapes With Integral and Finite-Element Methods}}, 
  year={2025},
  volume={35},
  number={3},
  pages={1-8},
  doi={10.1109/TASC.2025.3544512}}

@Article{instruments5030027,
AUTHOR = {Marchevsky, Maxim},
TITLE = {{Quench Detection and Protection for High-Temperature Superconductor Accelerator Magnets}},
JOURNAL = {Instruments},
VOLUME = {5},
YEAR = {2021},
NUMBER = {3},
ARTICLE-NUMBER = {27},
ISSN = {2410-390X},
DOI = {10.3390/instruments5030027}
}

@ARTICLE{10966202,
  author={Zhou, Linjie and Wang, Yihan and Yuan, Qi and Song, Xiaowei and Li, Liang and Wang, Qiuliang},
  journal={IEEE Transactions on Applied Superconductivity}, 
  title={{AC Loss Calculation of High Temperature Superconducting Coils Based on a Surrogate Model}}, 
  year={2025},
  volume={35},
  number={5},
  pages={1-5},
  doi={10.1109/TASC.2025.3561097}}

@ARTICLE{10830004,
  author={Sorti, Stefano and Balconi, Lorenzo and Rossi, Lucio and Santini, Carlo and Statera, Marco},
  journal={IEEE Transactions on Applied Superconductivity}, 
  title={{Toward Real-Time Electromagnetic Simulations of HTS Non-Insulated Coils Through Proper Orthogonal Decomposition}}, 
  year={2025},
  volume={35},
  number={5},
  pages={1-5},
  doi={10.1109/TASC.2025.3526741}}

@ARTICLE{10896621,
  author={Peng, Pai and Fu, Yutong and Peng, Weihang and Wang, Yawei},
  journal={IEEE Transactions on Applied Superconductivity}, 
  title={{A Quench Behavior Predictive Model for High Temperature Superconducting Magnet Based on Deep-Learning Neural Network}}, 
  year={2025},
  volume={35},
  number={5},
  pages={1-6},
  doi={10.1109/TASC.2025.3543793}}

@article{Zappatore_2024,
doi = {10.1088/1361-6668/ad8e8a},
year = {2024},
month = {nov},
publisher = {IOP Publishing},
volume = {37},
number = {12},
pages = {125012},
author = {Zappatore, A},
title = {{Full 3D thermal-hydraulic and electric modelling of quench propagation in HTS conductors}},
journal = {Superconductor Science and Technology}
}

@ARTICLE{10412643,
  author={Li, Kexing and Li, Ke and Zou, Liang and Fu, Yutong and Yang, Longhao and Peng, Weihang and Wang, Yawei},
  journal={IEEE Transactions on Applied Superconductivity}, 
  title={{Overcurrent Quench Detection of Parallel-Wound No-Insulation High Temperature Superconductor Coil Based on Digital Twin Method}}, 
  year={2024},
  volume={34},
  number={5},
  pages={1-6},
  doi={10.1109/TASC.2024.3357445}}

@ARTICLE{11143214,
  author={Huang, Yizhao and Yazdani-Asrami, Mohammad and Song, Wenjuan},
  journal={IEEE Access}, 
  title={{A Deep Learning-Based Ultra-Fast Surrogate Model for AC Loss Estimation in Superconductors Using COMSOL}}, 
  year={2025},
  volume={13},
  number={},
  pages={153845-153854},
  doi={10.1109/ACCESS.2025.3603953}}

@ARTICLE{9756879,
  author={Zhai, Y. and Brown, T. and Menard, J. E. and van der Laan, D. C. and Weiss, J. D. and Johnson, Z.},
  journal={IEEE Transactions on Applied Superconductivity}, 
  title={{HTS Cable Conductor for Compact Fusion Tokamak Solenoids}}, 
  year={2022},
  volume={32},
  number={6},
  pages={1-5},
  doi={10.1109/TASC.2022.3167343}}

@ARTICLE{9765384,
  author={Kang, Rui and Wang, Juan and Xu, Qingjin},
  journal={IEEE Transactions on Applied Superconductivity}, 
  title={{Detecting Quench in HTS Magnets With LTS Wires—A Theoretical and Numerical Analysis}}, 
  year={2022},
  volume={32},
  number={6},
  pages={1-5},
  doi={10.1109/TASC.2022.3171185}}

@article{van_der_Laan_2019,
doi = {10.1088/1361-6668/aafc82},
year = {2019},
month = {feb},
publisher = {IOP Publishing},
volume = {32},
number = {3},
pages = {033001},
author = {van der Laan, D C and Weiss, J D and McRae, D M},
title = {{Status of CORC® cables and wires for use in high-field magnets and power systems a decade after their introduction}},
journal = {Superconductor Science and Technology}
}

@article{Goldacker_2014,
doi = {10.1088/0953-2048/27/9/093001},
year = {2014},
month = {aug},
publisher = {IOP Publishing},
volume = {27},
number = {9},
pages = {093001},
author = {Goldacker, Wilfried and Grilli, Francesco and Pardo, Enric and Kario, Anna and Schlachter, Sonja I and Vojenčiak, Michal},
title = {{Roebel cables from REBCO coated conductors: a one-century-old concept for the superconductivity of the future}},
journal = {Superconductor Science and Technology},
}

@INPROCEEDINGS{chaturantabut_discrete_2009,
  author={Chaturantabut, Saifon and Sorensen, Danny C.},
  booktitle={Proceedings of the 48h IEEE Conference on Decision and Control (CDC) held jointly with 2009 28th Chinese Control Conference}, 
  title={{Discrete Empirical Interpolation for nonlinear model reduction}}, 
  year={2009},
  volume={},
  number={},
  pages={4316-4321},
  doi={10.1109/CDC.2009.5400045}}

@ARTICLE{lusch_deep_2017,
  author={Lusch, B. and Kutz, J.N. and Brunton, S.L.},
  journal={Nature Communications}, 
  title={{Deep learning for universal linear embeddings of nonlinear dynamics}}, 
  year={2018},
  month = {Apr.},
  volume={9},
  number={1},
  pages={4950},
  doi={0.1038/s41467-018-07210-0}}

@article{peherstorfer_data-driven_2016,
  author    = {Peherstorfer, B. and Willcox, K.},
  title     = {{Data-driven Operator Inference for Nonintrusive Projection-based Model Reduction}},
  journal   = {Computer Methods in Applied Mechanics and Engineering},
  month = {Jul.},
  volume    = {306},
  pages     = {196-215},
  year      = {2016},
  doi       = {10.1016/j.cma.2016.03.025}}

@article{kramer_nonlinear_2019,
author = {Kramer, Boris and Willcox, Karen E.},
title = {Nonlinear Model Order Reduction via Lifting Transformations and Proper Orthogonal Decomposition},
journal = {AIAA Journal},
month = {Jun.},
volume = {57},
number = {6},
pages = {2297-2307},
year = {2019},
doi = {10.2514/1.J057791}}

@article{benner_survey_2015,
  author    = {Benner, Peter and Gugercin, Serkan and Willcox, Karen},
  title     = {A Survey of Projection‑Based Model Reduction Methods for Parametric Dynamical Systems},
  journal   = {SIAM Review},
  volume    = {57},
  number    = {4},
  pages     = {483-531},
  year      = {2015},
  doi       = {10.1137/130932715}}

@inproceedings{chen_neural_2019,
 author = {Chen, Ricky T. Q. and Rubanova, Yulia and Bettencourt, Jesse and Duvenaud, David K},
 booktitle = {Advances in Neural Information Processing Systems},
 publisher = {Curran Associates, Inc.},
 title = {Neural Ordinary Differential Equations},
 volume = {31},
 year = {2018}}

@inproceedings{klambauer_self-normalizing_2017,
title = {Self-{Normalizing} {Neural} {Networks}},
volume = {30},
booktitle = {Advances in Neural Information Processing Systems},
publisher = {Curran Associates, Inc.},
author = {Klambauer, Günter and Unterthiner, Thomas and Mayr, Andreas and Hochreiter, Sepp},
year = {2017}}

@article{barrault_empirical_2004,
  author    = {M. Barrault and Y. Maday and N. C. Nguyen and A. T. Patera},
  title     = {{An ‘Empirical Interpolation’ Method: Application to Efficient Reduced-Basis Discretization of Partial Differential Equations}},
  journal   = {Comptes Rendus. Mathématique},
  month = {Nov.},
  volume    = {339},
  number    = {9},
  pages     = {667--672},
  year      = {2004},
  doi       = {10.1016/j.crma.2004.08.006}}

@book{brunton_data-driven_2021,
  author    = {Steven L. Brunton and J. Nathan Kutz},
  title     = {Data-Driven Science and Engineering: Machine Learning, Dynamical Systems, and Control},
  edition   = {2},
  publisher = {Cambridge University Press},
  place     = {Cambridge},
  year      = {2022}}

@article{Chatterjee_intro_00,
  author  = {Anindya Chatterjee},
  title   = {An Introduction to the Proper Orthogonal Decomposition},
  journal = {Current Science},
  year    = {2000},
  volume  = {78},
  number  = {7},
  pages   = {808--817}}

@article{kingma_adam_2017,
    author = {Diederik P. Kingma and Jimmy Ba},
    title = {Adam: A Method for Stochastic Optimization},
    archivePrefix={arXiv},
    eprint = {1412.6980},
    journal = {arXiv preprint arXiv:1412.6980},
    year = {2017},
    url={https://arxiv.org/abs/1412.6980}
}

@article{fast_J_phi,
  author={Lucchini, Francesco and Torchio, Riccardo and Morandi, Antonio and Dughiero, Fabrizio},
  journal={IEEE Transactions on Applied Superconductivity}, 
  title={A Fast Integral Equation $J-\varphi$ Formulation for Superconducting Structures}, 
  year={2024},
  volume={34},
  number={4},
  pages={1-8},
  doi={10.1109/TASC.2024.3366189}}

@article{sirios_comparison_2019,
  author={Sirois, Frédéric and Grilli, Francesco and Morandi, Antonio},
  journal={IEEE Transactions on Applied Superconductivity}, 
  title={Comparison of Constitutive Laws for Modeling High-Temperature Superconductors}, 
  year={2019},
  volume={29},
  number={1},
  pages={1-10},
  doi={10.1109/TASC.2018.2848219}}

@article{Wang_2019,
doi = {10.1088/1361-6668/aaf011},
year = {2019},
month = {jan},
publisher = {IOP Publishing},
volume = {32},
number = {2},
pages = {025003},
author = {Wang, Yawei and Zhang, Min and Grilli, Francesco and Zhu, Zixuan and Yuan, Weijia},
title = {{Study of the magnetization loss of CORC® cables using a 3D T-A formulation}},
journal = {Superconductor Science and Technology}}

@article{GUO2025114352,
title = {{Tensor decomposition-based DEIM for model order reduction applied to nonlinear parametric electromagnetic problems}},
journal = {Journal of Computational Physics},
volume = {542},
pages = {114352},
year = {2025},
issn = {0021-9991},
doi = {doi.org/10.1016/j.jcp.2025.114352},
author = {Ze Guo and Zuqi Tang and Zhuoxiang Ren},
}

@INPROCEEDINGS{10686274,
  author={Torchio, Riccardo and Schops, Sebastian and Lucchini, Francesco},
  booktitle={2024 IEEE International Symposium on Antennas and Propagation and INC/USNC‐URSI Radio Science Meeting (AP-S/INC-USNC-URSI)}, 
  title={{Block Structure Preserving Model Order Reduction for A-EFIE Integral Equation Method}}, 
  year={2024},
  volume={},
  number={},
  pages={977-978},
  doi={10.1109/AP-S/INC-USNC-URSI52054.2024.10686274}}

@article{MELCHERS202394,
title = {{Comparison of neural closure models for discretised PDEs}},
journal = {Computers \& Mathematics with Applications},
volume = {143},
pages = {94-107},
year = {2023},
issn = {0898-1221},
doi = {doi.org/10.1016/j.camwa.2023.04.030},
author = {Hugo Melchers and Daan Crommelin and Barry Koren and Vlado Menkovski and Benjamin Sanderse}
}

@article{RAISSI2019686,
title = {Physics-informed neural networks: A deep learning framework for solving forward and inverse problems involving nonlinear partial differential equations},
journal = {Journal of Computational Physics},
volume = {378},
pages = {686-707},
year = {2019},
issn = {0021-9991},
doi = {doi.org/10.1016/j.jcp.2018.10.045},
author = {M. Raissi and P. Perdikaris and G.E. Karniadakis},
}

@article{lu_learning_2021,
	title = {Learning nonlinear operators via {DeepONet} based on the universal approximation theorem of operators},
	volume = {3},
	issn = {2522-5839},
	doi = {10.1038/s42256-021-00302-5},
	number = {3},
	journal = {Nature Machine Intelligence},
	author = {Lu, Lu and Jin, Pengzhan and Pang, Guofei and Zhang, Zhongqiang and Karniadakis, George Em},
	month = mar,
	year = {2021},
	pages = {218--229},
}

@inproceedings{
li2021fourier,
title={{Fourier Neural Operator for Parametric Partial Differential Equations}},
author={Zongyi Li and Nikola Borislavov Kovachki and Kamyar Azizzadenesheli and Burigede liu and Kaushik Bhattacharya and Andrew Stuart and Anima Anandkumar},
booktitle={International Conference on Learning Representations},
year={2021},
}

@inproceedings{
anandkumar2019neural,
title={Neural Operator: Graph Kernel Network for Partial Differential Equations},
author={Anima Anandkumar and Kamyar Azizzadenesheli and Kaushik Bhattacharya and Nikola Kovachki and Zongyi Li and Burigede Liu and Andrew Stuart},
booktitle={ICLR 2020 Workshop on Integration of Deep Neural Models and Differential Equations},
year={2019},
}

@article{JMLR:v24:21-1524,
  author  = {Nikola Kovachki and Zongyi Li and Burigede Liu and Kamyar Azizzadenesheli and Kaushik Bhattacharya and Andrew Stuart and Anima Anandkumar},
  title   = {{Neural Operator: Learning Maps Between Function Spaces With Applications to PDEs}},
  journal = {Journal of Machine Learning Research},
  year    = {2023},
  volume  = {24},
  number  = {89},
  pages   = {1--97},
  doi = {}
}

@article{BENNER2020113433,
title = {Operator inference for non-intrusive model reduction of systems with non-polynomial nonlinear terms},
journal = {Computer Methods in Applied Mechanics and Engineering},
volume = {372},
pages = {113433},
year = {2020},
issn = {0045-7825},
doi = {doi.org/10.1016/j.cma.2020.113433},
author = {Peter Benner and Pawan Goyal and Boris Kramer and Benjamin Peherstorfer and Karen Willcox}
}

@article{FARENGA2025107146,
title = {On latent dynamics learning in nonlinear reduced order modeling},
journal = {Neural Networks},
volume = {185},
pages = {107146},
year = {2025},
issn = {0893-6080},
doi = {doi.org/10.1016/j.neunet.2025.107146},
author = {Nicola Farenga and Stefania Fresca and Simone Brivio and Andrea Manzoni},
}

@article{bonneville2024comprehensivereviewlatentspace,
      title={A Comprehensive Review of Latent Space Dynamics Identification Algorithms for Intrusive and Non-Intrusive Reduced-Order-Modeling}, 
      author={Christophe Bonneville and Xiaolong He and April Tran and Jun Sur Park and William Fries and Daniel A. Messenger and Siu Wun Cheung and Yeonjong Shin and David M. Bortz and Debojyoti Ghosh and Jiun-Shyan Chen and Jonathan Belof and Youngsoo Choi},
      year={2024},
      journal={arXiv preprint arXiv:2403.10748},
      archivePrefix={arXiv},
      primaryClass={cs.CE},
      url={https://arxiv.org/abs/2403.10748}
}

@article{loya2025structurepreservingneuralordinarydifferential,
      title={Structure-Preserving Neural Ordinary Differential Equations for Stiff Systems}, 
      author={Allen Alvarez Loya and Daniel A. Serino and J. W. Burby and Qi Tang},
      year={2025},
      eprint={2503.01775},
      journal={arXiv preprint arXiv:2503.01775},
      archivePrefix={arXiv},
      primaryClass={math.NA},
      url={https://arxiv.org/abs/2503.01775}
}

@article{1517384113,
author = {Steven L. Brunton  and Joshua L. Proctor  and J. Nathan Kutz },
title = {Discovering governing equations from data by sparse identification of nonlinear dynamical systems},
journal = {Proceedings of the National Academy of Sciences},
volume = {113},
number = {15},
pages = {3932-3937},
year = {2016},
doi = {10.1073/pnas.1517384113}
}

@article{LINOT2023111838,
title = {Stabilized neural ordinary differential equations for long-time forecasting of dynamical systems},
journal = {Journal of Computational Physics},
volume = {474},
pages = {111838},
year = {2023},
issn = {0021-9991},
doi = {https://doi.org/10.1016/j.jcp.2022.111838},
author = {Alec J. Linot and Joshua W. Burby and Qi Tang and Prasanna Balaprakash and Michael D. Graham and Romit Maulik},
}

@article{SERINO2025114090,
title = {Fast-slow neural networks for learning singularly perturbed dynamical systems},
journal = {Journal of Computational Physics},
volume = {537},
pages = {114090},
year = {2025},
issn = {0021-9991},
doi = {https://doi.org/10.1016/j.jcp.2025.114090},
author = {Daniel A. Serino and Allen {Alvarez Loya} and J.W. Burby and Ioannis G. Kevrekidis and Qi Tang},
}

@article{10.1063/5.0308144,
    author = {Pan, Xinyu and Xiao, Dunhui and Wang, Lihua and Yang, Xiaoquan},
    title = {A physics-data combined neural network-based finite volume parametric reduced order model},
    journal = {Physics of Fluids},
    volume = {37},
    number = {12},
    pages = {123602},
    year = {2025},
    month = {12},
    issn = {1070-6631},
    doi = {10.1063/5.0308144},
    eprint = {https://pubs.aip.org/aip/pof/article-pdf/doi/10.1063/5.0308144/20826109/123602_1_5.0308144.pdf},
}

@article{FuAnonlinear,
author = {Fu, R. and Xiao, D. and Navon, I.M. and Fang, F. and Yang, L. and Wang, C. and Cheng, S.},
title = {A non-linear non-intrusive reduced order model of fluid flow by auto-encoder and self-attention deep learning methods},
journal = {International Journal for Numerical Methods in Engineering},
volume = {124},
number = {13},
pages = {3087-3111},
doi = {https://doi.org/10.1002/nme.7240},
eprint = {https://onlinelibrary.wiley.com/doi/pdf/10.1002/nme.7240},
year = {2023}
}

@ARTICLE{9693316,
  author={Fareed, M. U. and Kapolka, M. and Robert, B. C. and Clegg, M. and Ruiz, H. S.},
  journal={IEEE Transactions on Applied Superconductivity}, 
  title={{3D FEM Modeling of ${\mathrm CORC}$ Commercial Cables With Bean’s Like Magnetization Currents and Its AC-Losses Behavior}}, 
  year={2022},
  volume={32},
  number={4},
  pages={1-5},
  doi={10.1109/TASC.2022.3145309}}

@ARTICLE{10591441,
  author={Wu, B. H. and Gao, S. Y. and Yang, X.S. and Jiang, J. and Zhang, H. and Zhao, Y.},
  journal={IEEE Transactions on Applied Superconductivity}, 
  title={{Magnetization Losses in HTS Coil With Different Winding Method}}, 
  year={2024},
  volume={34},
  number={8},
  pages={1-4},
  doi={10.1109/TASC.2024.3425323}}

@article{guo2025nonlinearmodelreductionprobabilistic,
author = {Guo, Jiaming and Xiao, Dunhui},
title = {Nonlinear Model Reduction by Probabilistic Manifold Decomposition},
journal = {SIAM Journal on Scientific Computing},
volume = {48},
number = {1},
pages = {A209-A235},
year = {2026},
doi = {10.1137/25M1738863},
}






\end{document}